\RequirePackage{fix-cm}
\documentclass[smallextended]{svjour3}       
\smartqed  
\usepackage{graphicx}
\usepackage{natbib}
\usepackage{amssymb}

\newcommand{\be}{\,\begin{equation}}
\newcommand{\ee}{\,\end{equation}}
\newcommand{\der}{\,\partial}

\begin{document}

\title{The Origin of Galactic Cosmic Rays
}

\titlerunning{Cosmic rays}        

\author{Pasquale Blasi}


\institute{P. Blasi \at
              INAF/Osservatorio Astrofisico di Arcetri, Firenze, Italy and \\
              INFN/Gran Sasso Science Institute, L'Aquila, Italy\\
              Tel. (Arcetri): +39 055-2752-297\\
              Fax (Arcetri): +39 055 220039\\
              \email{blasi@arcetri.astro.it}           \\
}

\date{Received: date / Accepted: date}

\maketitle

\begin{abstract}
One century ago Viktor Hess carried out several balloon flights that led him to conclude that the penetrating radiation responsible for the discharge of electroscopes was of extraterrestrial origin. One century from the discovery of this phenomenon seems to be a good time to stop and think about what we have understood about {\it Cosmic Rays}. The aim of this review is to illustrate the ideas that have been and are being explored in order to account for the observable quantities related to cosmic rays and to summarize the numerous new pieces of observation that are becoming available. In fact, despite the possible impression that development in this field is somewhat slow, the rate of new discoveries in the last decade or so has been impressive, and mainly driven by beautiful pieces of observation. At the same time scientists in this field have been able to propose new, fascinating ways to investigate particle acceleration inside the sources, making use of multifrequency observations that range from the radio, to the optical, to X-rays and gamma rays. These ideas can now be confronted with data. 

I will mostly focus on supernova remnants as the most plausible sources of Galactic cosmic rays, and I will review the main aspects of the modern theory of diffusive particle acceleration at supernova remnant shocks, with special attention for the dynamical reaction of accelerated particles on the shock and the phenomenon of magnetic field amplification at the shock. Cosmic ray escape from the sources is discussed as a necessary step to determine the spectrum of cosmic rays at the Earth. The discussion of these theoretical ideas will always proceed parallel to an account of the data being collected especially in X-ray and gamma ray astronomy. 

In the end of this review I will also discuss the phenomenon of cosmic ray acceleration at shocks propagating in partially ionized media and the implications of this phenomenon in terms of width of the Balmer line emission. This field of research has recently experienced a remarkable growth, in that $H\alpha$ lines have been found to bear information on the cosmic ray acceleration efficiency of supernova shocks. 

\keywords{Cosmic rays \and Acceleration}
\end{abstract}

\section{Introduction}
\label{intro}

In 1962 Bruno Rossi finalized the writing of his book {\it Cosmic Rays} \cite[]{rossi1964cosmic} in coincidence with the $50^{th}$ anniversary of the discovery of cosmic rays (CRs) (though the book was published in 1964). In the epilogue of the book he emphasizes how the field of CR research had become a complex combination of several fields, from Astronomy to Plasma Physics and Particle Physics. He also argues that {\it ``It is quite possible that future historians of science will close the chapter on cosmic rays with the fiftieth anniversary of Hess's discovery''}. Interestingly enough, very little of what will be discussed in the present review was known or even proposed at the time of Rossi's book: scientists in this field have been extremely active and many new ideas and new observations have changed much of what was believed in the early 60's. The purpose of this review is to provide a recount of these exciting developments, especially the ones that took place in the last decade or so. I am pretty sure that historians of science will not close the chapter on cosmic rays with the $100^{th}$ anniversary of their discovery. Too many loose ends need to be put in place. 

Cosmic rays are mainly charged particles that contribute an energy density in the Galaxy of about $1$ eV $cm^{-3}$. They are mainly protons (hydrogen nuclei) with about 10\% fraction of helium nuclei and smaller abundances of heavier elements. Despite the much lower fluxes of electrons and positrons, these particles provide us with precious information on the sources of CRs and the transport of these particles through the Galactic magnetic field. An even smaller flux of electromagnetic radiation (from radio frequencies to gamma rays) reaches the Earth from both the sources and from the interactions that CRs occasionally suffer during propagation. The models we develop for the origin of CRs are all based on an attempt to interpret these separate pieces of observations within a unified frame. 

\begin{figure}
\includegraphics[width=350pt]{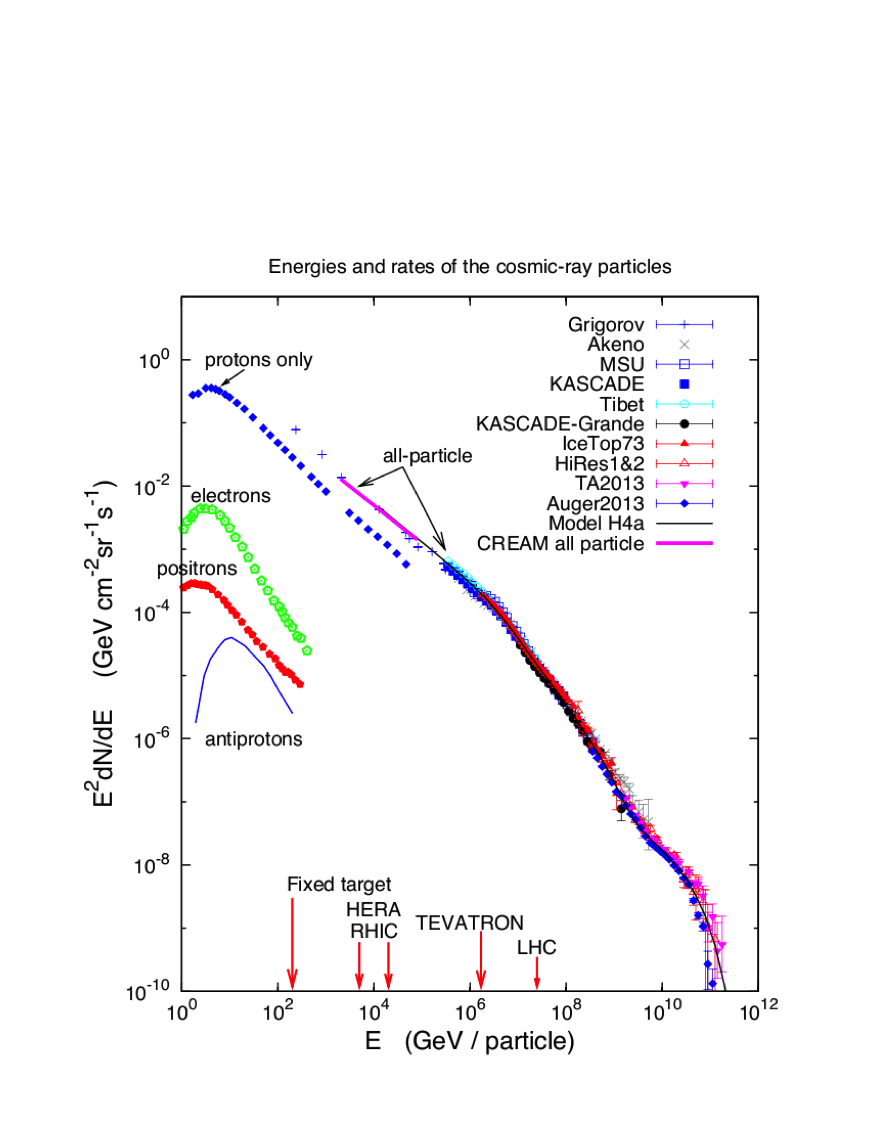}
\caption{Spectrum of cosmic rays at the Earth (courtesy Tom Gaisser). The all-particle spectrum measured by different experiments is plotted, together with the proton spectrum. The subdominant contributions from electrons, positrons and antiprotons as measured by the PAMELA experiment are shown.}
\label{fig:spectrum}       
\end{figure}

The flux of all nuclear components present in CRs (the so-called all-particle spectrum) is shown in Fig. \ref{fig:spectrum}. At low energies (below $\sim 30$ GeV) the spectral shape bends down, as a result of the modulation imposed by the presence of a magnetized wind originated from our Sun, which inhibits very low energy particles from reaching the inner solar system. The prominent steepening of the spectrum at energy $E_{K}=3\times 10^{15}$ eV is named {\it the knee}: at this point the spectral slope of the differential flux (flux of particles reaching the Earth per unit time, surface and solid angle, per unit energy interval) changes from $\sim -2.7$ to $\sim -3.1$. There is evidence that the chemical composition of CRs changes across the knee region with a trend to become increasingly more dominated by heavy nuclei at high energy \citep[see][for a review]{2006JPhCS4741H}, at least up to $\sim 10^{17}$ eV. At even higher energies the chemical composition remains matter of debate. Recent measurements carried out with KASCADE-GRANDE \cite[]{Apel:2013p3150} reveal an interesting structure in the spectrum and composition of CRs between $10^{16}$ and $10^{18}$ eV: the collaboration managed to separate the showers in electron-rich (a proxy for light chemical composition) and electron-poor (a proxy for heavy composition) showers and showed that the light component (presumably protons and He, with some contamination from CNO) has an ankle like structure at $10^{17}$ eV. The authors suggest that this feature signals the transition from Galactic to extragalactic CRs (in the light nuclei component). The spectrum of Fe-like CRs continues up to energies of $\sim 10^{18}$ eV, where the flux of Fe and the flux of light nuclei are comparable. A similar conclusion was recently reached by the ICETOP Collaboration \cite[]{Aartsen:2013p042004}. This finding does not seem in obvious agreement with the results of the Pierre Auger Observatory \cite[]{2010PhRvL104i1101A}, HiRes \cite[]{2007JPhG34401S} and Telescope Array \cite[]{2013EPJWC5206002S}, which find a chemical composition at $10^{18}$ eV that is dominated by the light chemical component. 

The presence of a knee and the change of chemical composition around it have stimulated the idea that the bulk of CRs originates within our Galaxy. The knee could for instance result from the superposition of cutoffs in the spectra of different chemicals as due to the fact that most acceleration processes are rigidity dependent: if protons are accelerated in the sources to a maximum energy $E_{p,max}\sim 5\times 10^{15}$ eV, then an iron nucleus will be accelerated to $E_{Fe,max}=26E_{p,max}\sim (1-2)\times 10^{17}$ eV (it is expected that at such high energies even iron nuclei are fully ionized, therefore the unscreened charge is $Z=26$). A knee would naturally arise as the superposition of the cutoffs in the spectra of individual elements (see for instance \cite{Horandel:2004p1543,Blasi:2012p2051,Gaisser:2013p3152}). 

The apparent regularity of the all-particle spectrum in the energy region below the knee is at odds with the recent detection of features in the spectra of individual elements, most notably protons and helium: the PAMELA satellite has provided evidence that both the proton and helium spectra harden at $230$ GV \cite[]{Adriani:2011p1893}. The spectrum of helium nuclei is also found systematically harder than the proton spectrum, through only by a small amount. The slope of the proton spectrum below $230$ GeV was measured to be $\gamma_{1}=2.89\pm 0.015$, while the slope above $230$ GeV becomes $\gamma_{2}=2.67\pm 0.03$. The slopes of protons and helium spectra at high energies as measured by PAMELA appear to be in agreement with those measured by the CREAM experiment  \cite[]{Ahn:2010p624} at supra-TeV energies. Some evidence also exists for a similar hardening in the spectra of heavier elements (see \cite{Maestro:2010p739} and references therein). 

Different explanations for the feature at 230 GV have been put forward: \cite{2012MNRAS4211209T,2013arXiv13041400T} suggested that a local source of CRs might appear in the total spectrum as a spectral hardening. On the other hand, the fact that a similar feature has been detected in the spectrum of helium nuclei (and possibly heavier nuclei) might suggest that a new physical phenomenon is showing up, probably due to CR transport. For instance, \cite{2012ApJ...752L..13T} showed that a spatially dependent diffusion coefficient may induce a spectral hardening under some assumptions on the functional shape of the function representing the diffusion coefficient (non separability between energy and space coordinates is required). \cite{Blasi:2012p2344} and \cite{2013JCAP07001A} showed that a similar feature may naturally appear if CRs can produce their own scattering centers (diffusion) through streaming instability. In the latter model, the feature appears at $\sim 200$ GeV/n as a result of the transition from self-generated diffusion and diffusion in a pre-existing turbulence. 

Very recently, some preliminary data from the AMS-02 experiment on the International Space Station have been presented \footnote{Presentation by S. Ting at the $33^{rd}$ International Cosmic Ray Conference, Rio De Janeiro, July 2013} and do not confirm the existence of the spectral breaks in the protons and helium spectra, as observed by PAMELA. Given the preliminary nature of these data and the lack of refereed publications at the time of writing of this review, I cannot comment further on their relevance.

The measurement of the ratio of fluxes of some nuclei that can only be produced by CR spallation and the flux of their parent nuclei provides the best estimate so far of the amount of matter that CRs traverse during their journey through the Galaxy. In order to account for the observed B/C ratio, CRs must travel for times that exceed the ballistic time by several orders of magnitude before escaping the Galaxy (this number decreases with energy). This is the best argument to support the ansatz that CRs travel diffusively in the Galactic magnetic field \cite[]{1972PhRvL29445J}. A similar conclusion can be drawn from the observed flux of some unstable isotopes such as $^{10}Be$ \cite[]{Simpson:1988p773}. The decrease of the B/C ratio with energy per nucleon is well described in terms of a diffusion coefficient that increases with energy. 

In principle a similar argument can be applied to the so-called positron fraction, the ratio of fluxes of positrons and electrons plus positrons, $\Phi_{e^{+}}/(\Phi_{e^{+}}+\Phi_{e^{-}})$, where however special care is needed because of the important role of energy losses for leptons. In first approximation, it is expected that positrons may only be secondary products of inelastic CR interactions that lead to the production and decays of charged pions. In this case it can be proven that the positron fraction must decrease with energy. In fact several past observations, and most recently the PAMELA measurements \cite[]{Adriani:2009p787} and the AMS-02 measurement \cite[]{Aguilar:2013p3117}, showed that the positron fraction increases with energy above $\sim 10$ GeV. This anomalous behaviour is not reflected in the flux of antiprotons \cite[]{Adriani:2008p641}: the ratio of the antiprotons to proton fluxes $\Phi_{\bar p}/\Phi_{p}$ is seen to decrease, as expected based on the standard model of diffusion. Although the rise of the positron fraction has also been linked to dark matter annihilation in the Galaxy, there are astrophysical explanations of this phenomenon that can account for the data without extreme assumptions (see the review paper by \cite{2012APh392S} for a careful description of both astrophysical models and dark matter inspired models). 

The simple interpretation of the knee as a superposition of the cutoffs in the spectra of individual elements, as discussed above, would naively lead to the conclusion that the spectrum of Galactic CRs should end at $\sim 26 E_{K}\lesssim 10^{17}$ eV. Clearly this conclusion is not straightforward: a rare type of sources that can potentially accelerate CRs to much larger energies may leave the interpretation of the knee unaffected and yet change the energy at which Galactic CRs end. This opens the very important question of where should one expect the transition to extragalactic CRs to take place. Although in the present review I will only occasionally touch upon the problem of ultra high energy cosmic rays (UHECRs), it is important to realize that the quest for their origin is intimately connected with the nature of the transition from Galactic CRs to UHECRs.

At the time of this review, there is rather convincing and yet circumstantial evidence that the bulk of CRs are accelerated in supernova remnants (SNRs) in our Galaxy, as first proposed by \cite{Baade:1934p886,Ginzburg:1961p170}. The evidence is based on several independent facts: gamma rays unambiguously associated with production of neutral pions have been detected from several SNRs close to molecular clouds \cite[]{Ackermann:2013p3110,Tavani:2010p2219}; the gamma ray emission detected from the Tycho SNR \cite[]{2012ApJ744L2G,2011ApJ730L20A} also appears to be most likely of hadronic origin \cite[]{Morlino:2011p1965,2013ApJ76314B}; the bright X-ray rims detected from virtually all young SNRs (see \cite[]{Vink:2012p2755,2006AdSpR.37.1902B} for a recent review) prove that the local magnetic field in the shock region has been substantially amplified, probably by accelerated particles themselves, due to streaming instability (for recent reviews see \cite[]{Bykov:2013p3165,Bykov:2011p1909,Schure:2012p3068}). Despite all this circumstantial evidence, no proof has been found yet that SNRs can accelerate CRs up to the knee energy.

Charged particles can be energized at a supernova shock through diffusive shock acceleration (DSA) \cite[]{1977DoSSR2341306K,1978ApJ221L29B,1977ICRC11.132A,Bell:1978p1344,Bell:1978p1342}. If SNRs are the main contributors to Galactic CRs, an efficiency of $\sim 10\%$ in particle acceleration is required (see \S \ref{sec:paradigm}). The dynamical reaction of accelerated particles at a SNR shock is large enough to change the shock structure, so as to call for a non-linear theory of DSA \cite[]{Malkov:2001p765}. Such a theory should also be able to describe the generation of magnetic field in the shock region as due to CR-driven instabilities \cite[]{Amato:2006p139,Caprioli:2008p123,Caprioli:2009p157}, although many problems still need to be solved.

The combination of DSA and diffusive propagation in the Galaxy represents what I will refer to as the {\it supernova remnant paradigm}. Much work is being done at the time of this review to find solid proofs in favor or against this paradigm. I will summarize this work here. 

The review is structured as follows: in \S \ref{sec:paradigm} I will review the basic aspects of the SNR paradigm for the origin of CRs; in \S \ref{sec:dsa} I will provide a pedagogical discussion of the mechanism of diffusive shock acceleration (DSA) at collisionless shocks and the maximum energy achievable. The non-linear version of the theory of DSA is illustrated in \S \ref{sec:nldsa}, where the dynamical reaction of accelerated particles and magnetic field amplification are discussed in depth. In \S \ref{sec:OB} I briefly discuss the issue of SN explosions in superbubbles. A discussion of several crucial pieces of the SNR paradigm (CR escape, spectra of SNRs and SNRs close to molecular clouds) are discussed in \S \ref{sec:multinu}. The phenomenon of DSA in partially ionized material is discussed in \S \ref{sec:Halpha}, with special emphasis of the implication of CR acceleration for the width of the $H\alpha$ line in Balmer dominated shocks. I conclude in \S \ref{sec:concl}.

\section{The bases of the SNR paradigm}
\label{sec:paradigm}

\begin{figure}
\includegraphics[width=300pt]{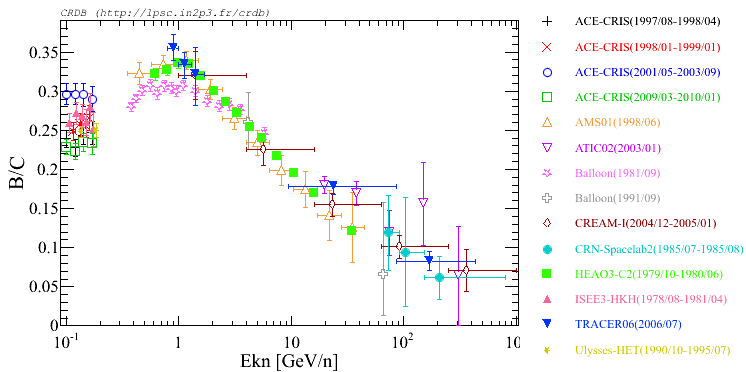}
\caption{B/C ratio as a function of energy per nucleon. Data have been extracted from the Cosmic Ray Database \cite[]{2013arXiv1302.5525M}.}
\label{fig:BC}      
\end{figure}

The abundances of some light elements such as boron, lithium and beryllium in CRs provide us with the best estimates of the time $\tau_{esc}(E)$ spent by CRs in the Galaxy before escaping. More precisely, the ratio of boron and carbon fluxes is related to the grammage traversed by CRs, $X(E)=\bar n \mu v \tau_{esc}(E)$, where $\bar n$ is the mean gas density in the confinement volume of the Galaxy (disc plus halo), $\mu$ is the mean mass of the gas, $v$ is the speed of particles. For particles with energy per nucleon of 10 GeV/n the measured B/C corresponds to $X\sim 10 g~cm^{-2}$. If the sources are located in the thin disc of the Galaxy with half thickness $h=150$ pc and the halo extends to a height $H$, the mean density can be estimated as $\bar n=n_{disc} h/H = 5\times 10^{-2}\left(\frac{n_{disc}}{1cm^{-3}}\right)\left(\frac{H}{3 kpc}\right)^{-1} cm^{-3}$. For a standard chemical composition of the ISM ($n_{He}\approx 0.15 n_{H}$) the mean mass is $\mu = (n_{H}+4n_{He})/(n_{H}+n_{He})\approx 1.4 m_{p}$. It follows that for a proton with energy $E_{*}=10$ GeV the typical escape time is 
\be
\tau_{*} \sim \frac{X(E_{*})}{\bar n \mu c} = 90 \left(\frac{H}{3 kpc}\right) Myr,
\label{eq:taustar}
\ee
which exceeds the ballistic propagation time scale by at least three orders of magnitude. This remains the strongest evidence so far for diffusive motion of CRs in the Galaxy. A diffusion coefficient can be introduced as $\tau_{esc}(E)=H^{2}/D(E)=\tau_{*}(E/E_{*})^{-\delta}$, so that at 10 GeV $D(E)\simeq 3\times 10^{28} \left(\frac{H}{3 kpc}\right) cm^{2} s^{-1}$. The grammage (and therefore the escape time) decreases with energy (or rather with rigidity) as inferred from the B/C ratio, illustrated in Fig. \ref{fig:BC}, which shows a collection of data points on the ratio of fluxes of boron and carbon, as obtained by using the data collection provided by the Cosmic Ray Database \cite[]{2013arXiv1302.5525M}. Fig. \ref{fig:BC} illustrates the level of uncertainty in the determination of the slope of the B/C ratio at high energies, which reflects on the uncertainty in the high energy behaviour of the diffusion coefficient. At low energies the uncertainty is even more severe due to the effects of solar modulation which suppresses CR fluxes in a different way during different phases of the solar activity (see \cite[]{2013LRSP.10.3P} for a recent review). The high rigidity behavior of the B/C ratio is compatible with a power law grammage $X(R)\propto R^{-\delta}$ with $\delta = 0.3-0.6$. 

\begin{figure}
\includegraphics[width=300pt]{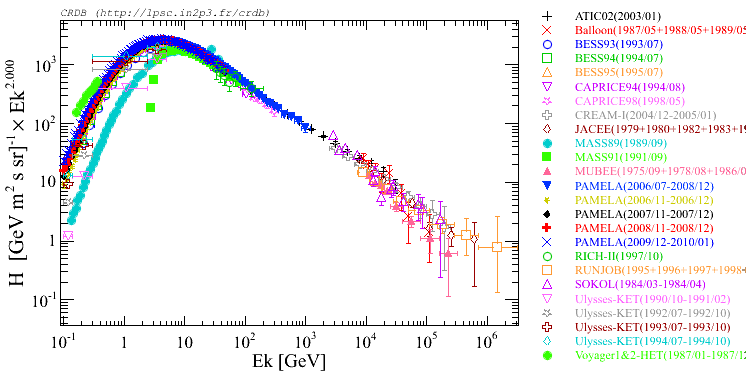}
\caption{Proton spectrum as measured by different experiments. Data are from the Cosmic Ray Database \cite[]{2013arXiv1302.5525M}.}
\label{fig:protons}      
\end{figure}

Supernovae exploding in our Galaxy at a rate ${\cal R}_{SN}$ liberate a kinetic energy in the form of moving ejecta of $E_{SN}=10^{51}E_{51}$ erg. This number is weakly dependent upon whether the SN is of type Ia or a core collapse SN, although it might be somewhat different for rare types of SNe (type Ib, Ic), possibly connected with gamma ray bursts. As I discuss below, particle acceleration in SNRs is believed to occur through diffusive shock acceleration, which leads to power law spectra of accelerated particles, and for the sake of the present discussion I assume that such an injection spectrum is in the form 
$$N(p)=\xi_{CR}\frac{E_{SN}}{m^{2}}I(\gamma)\left(\frac{p}{m}\right)^{-\gamma}~~~~~I(\gamma)\approx \frac{2(3-\gamma)(\gamma-2)}{4-\gamma},$$
where $\gamma>2$ is the slope of the differential spectrum of accelerated particles and $\xi_{CR}<1$ is the CR acceleration efficiency. Here $I(\gamma)$ is a normalization factor obtained by imposing that the total energy at the source equals $\xi_{CR}E_{SN}$. It is best to normalize the flux of CRs to the observed proton flux, since it is not expected to be affected by spallation reactions. The flux of protons observed by different experiments is shown in Fig. \ref{fig:protons} (data are from  the Cosmic Ray Database \cite[]{2013arXiv1302.5525M}). Provided we focus on sufficiently high energies, ionization losses can also be neglected and the effects of solar modulation play no role (we can also assume $E\simeq p c$). In this case the spectrum of CR protons contributed by SNRs at the Earth can be simply written as:
$$
J(E) = \frac{c}{4\pi}\frac{N(E){\cal R}_{SN}}{\pi R_{d}^{2} 2 H} \tau_{esc}(E) =
$$
\be
8\times 10^{5} \xi_{CR} I(\gamma) \left(\frac{{\cal R}_{SN}}{30 yr^{-1}} \right) \left(\frac{E}{m}\right)^{-\gamma-\delta} 
\left(\frac{E_{*}}{m}\right)^{\delta} ~ m^{-2}s^{-1}sr^{-1}GeV^{-1},
\ee
and I assumed that the disc of the Galaxy has a radius $R_{d}=10$ kpc. It is useful to notice that if the escape time is normalized to the B/C ratio at a given energy $E_{*}$ (see Eq. \ref{eq:taustar}) then the expected flux becomes independent of the size of the halo $H$. This reflects the fact that in the simple diffusion model introduced here the CR flux in the absence of losses and the grammage both scale with the ratio $H/D(E)$. This rule of thumb remains valid even in more sophisticated propagation calculations, such as GALPROP. 

Normalizing to the proton flux at $E_{*}=10$ GeV, $E_{*}^{2}J(E_{*})\approx 2\times 10^{3} GeV m^{-2}s^{-1}sr^{-1}$ (see Fig. \ref{fig:protons}), one immediately gets 
\be
\xi_{CR} \approx \frac{2.5\times 10^{-3}}{I(\gamma)} (E_{*}/m)^{(\gamma-2)} \left(\frac{{\cal R}_{SN}}{30 yr^{-1}} \right)^{-1}.
\label{eq:xi}
\ee
The required efficiency turns out to be a weak function of the slope of the injection spectrum $\gamma$ and is typically $\xi_{CR}\simeq 2-3\%$ when changing the value of $\delta$. The total CR acceleration efficiency is somewhat higher than the estimate in Eq. \ref{eq:xi} because of the contribution of nuclei heavier than hydrogen. More refined calculations provide a better estimate of the total acceleration efficiency that is between 5\% and 10\% for the bulk of SNRs, while it can be higher or smaller for individual objects, depending upon the environment in which the supernova event takes place. 

\section{The theory of diffusive shock acceleration of test particles}
\label{sec:dsa}

A supernova explosion in the interstellar medium (ISM) results in the injection of metal enriched ejecta with a total mass $M_{ej}$ moving with a velocity $V_{ej}$. If the total energy output in the form of kinetic energy is $E_{SN}=10^{51} E_{51}$ erg, then the velocity of the ejecta in the initial phases can be written as
\be
V_{ej} = 10000  E_{51}^{1/2} M_{ej,\odot}^{-1/2} \rm km/s,
\ee
where $M_{ej,\odot}$ is the mass of the ejecta in units of solar masses.

The sound speed in the ISM can be estimated as 
\be
c_{s} = \sqrt{\gamma_{g} \frac{k T}{m_{p}}} \approx 11 \left(\frac{T}{10^{4}K}\right)^{1/2} \rm km/s,
\ee
where $\gamma_{g}$ is the adiabatic index (assumed here to be $\gamma_{g}=5/3$) and $T$ is the temperature. It follows that the typical Mach number of the plasma ejected in a SN explosion is:
\be
M_{s} = \frac{V_{ej}}{c_{s}} \approx 900 E_{51}^{1/2} M_{ej,\odot}^{-1/2} \left(\frac{T}{10^{4}K}\right)^{-1/2}.
\label{eq:mach}
\ee
The motion of the ejecta is highly supersonic and drives the formation of a shock front. The motion of the shock front is heavily affected by the environment around the parent star and by the density profile in the ejecta (see \cite[]{1995PhR.256.157M} for a review). The matter accumulated behind the shock during the expansion increases the inertia of the expanding shell and eventually slows down the explosion at a time when the accumulated mass equals that of the ejecta. For an explosion in the standard ISM one can write:
\be
\frac{4}{3}\pi\rho_{ISM} R_{ST}^{3} = M_{ej} \to R_{ST} = \left( \frac{3 M_{ej}}{4\pi\rho_{ISM}}\right)^{1/3} \approx 2 ~ M_{ej,\odot}^{1/3} \left(\frac{n_{ISM}}{1 cm^{-3}}\right)^{-1/3}~\rm pc,
\ee
where $R_{ST}$ defines the radius of the expanding shell at the beginning of the so-called Sedov-Taylor (adiabatic) phase. This stage of the SNR evolution starts at the time 
\be
T_{ST}=\frac{R_{ST}}{V_{ej}}\approx 200 M_{ej,\odot}^{5/6}E_{51}^{-1/2} \left(\frac{n_{ISM}}{1 cm^{-3}}\right)^{-1/3}~\rm years.
\label{eq:sedovtime}
\ee
These estimates of the Sedov-Taylor radius and time should be considered as orders of magnitude, while the actual values depend on the conditions around the supernova explosion. For instance, for a core collapse SN explosion the material ejected by the pre-supernova star may dominate the density in the initial phases of the explosion, and the adiabatic phase may start at earlier times than indicated by Eq. \ref{eq:sedovtime}. On the other hand, in the case of a fast wind, with low density (such as would be the case for Wolf-Rayet pre-supernova star) the SN explosion might take place in an underdense bubble of hot dilute gas. In this case the adiabatic phase might start at later time. In any case, for core-collapse SN explosions the dynamics of the expanding shell is usually much more complex to describe than in the case of type Ia SN explosions in the ISM. This also reflects in the morphology of the non-thermal emission from SNRs of different types. The morphology of SNRs of core-collapse SN explosions is usually irregular and often asymmetric. This is also due to the fact that the environment in which massive stars explode through a core-collapse is often complex, with inhomogeneous distribution of gas and the presence of molecular clouds that provide the gas material for the formation of these massive, relatively short lived stars. On the other hand, type Ia SNRs are usually more regular and it is not rare to find cases of almost perfectly spherical SN shells as observed at all wavelengths. 

\begin{figure}
\includegraphics[width=175pt]{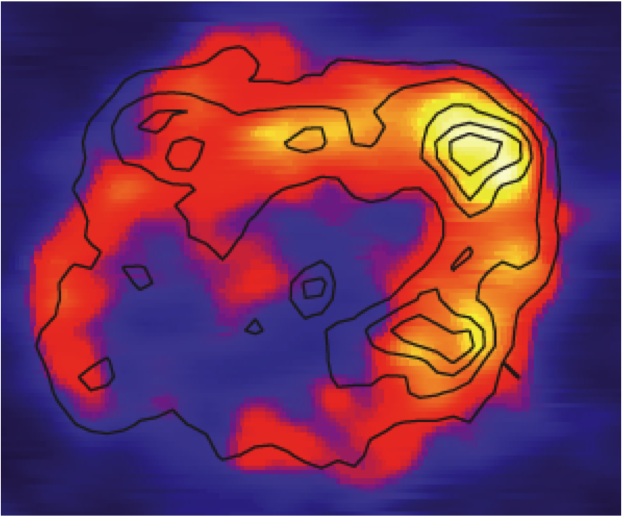}
\includegraphics[width=145pt]{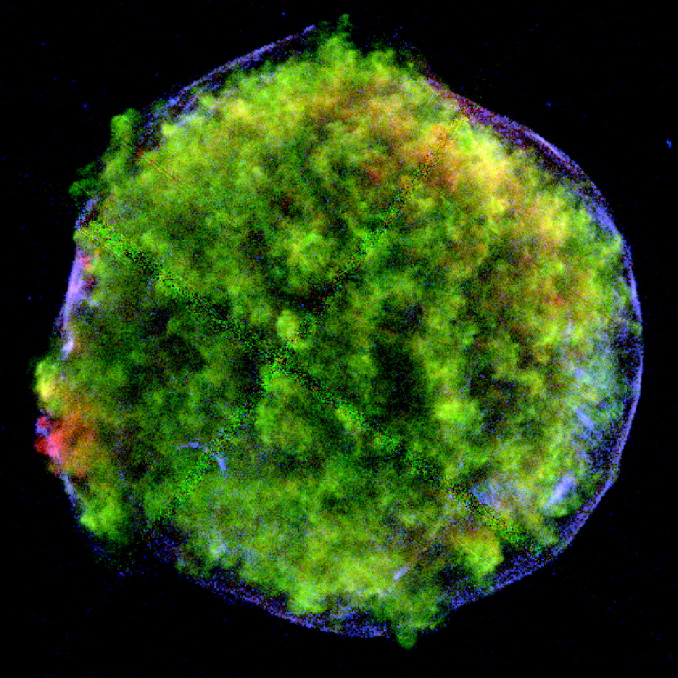}
\caption{{\it Left Panel:} Morphology of the RX J1713.7-3946. The colors illustrate the high energy gamma ray emission as measured by HESS \cite[]{2007A&A.464.235A}, while the contours show the X-ray emission in the 1-3 keV band measured by ASCA \cite[]{2002PASJ.54L.73U}. {\it Right Panel:} Morphology of the Tycho SNR as measured with Chandra \cite[]{2005ApJ.634.376W}. The three colors refer to emission in the photon energy range $0.95 - 1.26$ keV (red), $1.63-2.26$ keV (green), and $4.1-6.1$ keV (blue). The latter emission is very concentrated in a thin rim and is the result of synchrotron emission of very high energy electrons.}

\label{fig:morpho}      
\end{figure}

In Fig. \ref{fig:morpho} I show the cases of RX J1713.7-3946 (left panel, from \cite{2007A&A.464.235A}) and Tycho (right panel, from \cite{2005ApJ.634.376W}). The former is a SNR originated from a core collapse SN explosion and its gamma ray emission (color) and X-ray emission (lines) show the irregular morphology of this remnant. The Tycho SNR is the leftover of a type Ia SN exploded in 1572 at $\sim 3$ kpc distance from the solar system. The image shows its thermal X-ray emission, mainly contributed by the ejecta in the central part of the explosion region, and the non-thermal X-ray emission which has the rim-like morphology shown in the picture. In \S \ref{sec:multinu} I will discuss at length the implications of the non-thermal X-ray morphology of SNRs and of Tycho in particular. All these aspects are very important whenever the predictions of a theory of acceleration of CRs have to face observations.

As anticipated above, the supersonic motion of the ejecta of a stellar explosion leads to the formation of a shock front propagating in the ISM or in the circumstellar medium, depending on the type of SN explosion. The Mach number of the shock depends on the conditions in the region in which the explosion takes place. For instance the Mach number of the shock becomes appreciably lower than the one quoted in Eq. \ref{eq:mach} if the shock propagates in the hot tenuous gas around a core-collapse SNR. 

The first question that we have to face is however about the nature of these shock waves. In the section below I will argue that SN shocks (and in fact most astrophysical shock waves) are intrinsically different from the shock waves that we are used to in the Earth atmosphere, in that the latter are mediated by molecular collisions, while the former could not be formed based on particle-particle collisions in the ISM. The SN shocks expanding in the ordinary ISM belong to the class of {\it collisionless shocks}. Since many fundamental concepts of the physics of particle acceleration in astrophysical shocks rely on this property, I dedicate some space here to a discussion of the basic principles that regulate the formation of a collisionless shock. 

\subsection{Collisionless shocks}
\label{sec:collisionless}

Collisionless shocks are formed because of the excitation of electro-magnetic instabilities, namely collective effects generated by groups of charged particles in the background plasma. A thorough review of the theory of collisionless shock waves has recently been published by \cite{Treumann:2009p2233}, and I refer to that paper for a careful discussion of the many subtle aspects of the physics of collisionless shocks. Here I limit myself with a qualitative description of the conditions necessary for their formation, to be used as background material for some of the topics discussed in connection with the physics of particle acceleration. Moreover, since the shocks that will be discussed in this review are non-relativistic, here I will restrict the discussion to non-relativistic shocks $v\ll c$ and cases where the temperature downstream of the shock is much smaller than the electron mass, so as to avoid pair production. The requirement of a shock being non-relativistic can also be rewritten in terms of the Alfvenic Mach number:
\be
v\ll c \to M_{A}=\frac{v}{v_{A}}\ll \left(\frac{m_{p}}{m_{e}}\right)^{1/2}\frac{\omega_{p,e}}{\omega_{c,e}}=
1.3\times 10^{5} n_{cm^{-3}}^{1/2} B_{\mu G}^{-1}~,
\ee
where $v_{A}=B_{0}/\sqrt{4\pi n m_{p}}$ is the Alfv\'en velocity, $\omega_{p,e}$ and $\omega_{c,e}$ are the electron plasma frequency and cyclotron frequency. 

In an electron-proton plasma, Coulomb scattering acts in three different ways: 1) it leads the electrons to thermalize, namely to reach a Maxwellian distribution; 2) it leads protons to thermalize; 3) it leads to thermalization of electrons and protons. Typically these three processes have a well defined hierarchy: electron thermalization is the fastest process, followed by electron-proton thermalization. The slowest process is the thermalization of protons. This clearly opens several questions: first, electron-proton collisions are likely to occur when the proton distribution is not yet maxwellian; second, the time scale for electron-proton equilibration may be exceedingly long as compared with the age of the system at hand. 

The time scale for equilibration between two generic populations of particles with temperature $T_{1}$ and $T_{2}$, masses $m_{1}$ and $m_{2}$ and the same electric charge $q$ and same density $n$ is \cite[]{spitzerBook}:
\be
\tau_{eq} = \frac{3 m_{1}m_{2} k_{B}^{3/2}}{8 (2\pi)^{1/2} n q^{4} \ln\Lambda} \left( \frac{T_{1}}{m_{1}} + \frac{T_{2}}{m_{2}} \right)^{3/2},
\label{eq:equilibrate}
\ee
where $k_{B}$ is the Boltzmann constant and $\ln \Lambda\sim 10$ is the Coulomb logarithm. For instance, the equilibration time of electrons with themselves would be:
\be
\tau_{eq,ee} \approx 1200 \left(\frac{n}{1~cm^{-3}}\right)^{-1} \left(\frac{T_{e}}{10^{8} K}\right)^{3/2} ~ years,
\ee
while for protons:
\be
\tau_{eq,pp} = \approx 2.3\times 10^{6} \left(\frac{n}{1~cm^{-3}}\right)^{-1} \left(\frac{T_{p}}{10^{8} K}\right)^{3/2} ~ years.
\ee
Having in mind the case of a SNR, it is easy to envision that these equilibration times are long compared with the scale on which the shocks associated with the blast waves are actually observed, thereby raising the question of how such shocks are actually formed. On the other hand, the comparison with some plasma related quantities may be illuminating: for instance, for a velocity of $1000$ km/s and density $1 cm^{-3}$, the cyclotron radius of a particle is $m v c/e B_{0}$ which is $\sim 10^{10}$ cm for a proton in a $\mu G$ magnetic field and about 2000 times smaller for an electron. The electron plasma frequency is $\omega_{p,e}=(4\pi e^{2} n/m_{e})^{1/2}\sim 5.6\times 10^{4} n_{cm^{-3}}^{1/2}$, corresponding to a spatial scale $v/\omega_{p,e}\sim 2\times 10^{3}$ cm for a velocity $v\sim 10^{8} \rm cm/s$.

The formation of shock waves in these conditions is likely due to collective effects of charged particles. Several aspects of the physics of these collisionless shocks are all but trivial. Since the thermalization of these plasmas is directly linked to isotropization of the directions of motion of particles, it is natural to expect that the temperatures of electrons and protons immediately behind the shock front are proportional to the masses and therefore different for electrons and protons: 
\be
k T_{e} \approx \frac{3}{2} m_{e} v^{2} = \frac{m_{e}}{m_{p}} k T_{p}.
\label{eq:ratio}
\ee
Coulomb collisions between electrons and protons eventually lead them to reach the same temperature, but the time necessary to achieve this situation often exceeds the age of the source, hence the equilibration is all but guaranteed in collisionless shocks. This is especially true for young SNRs, since for typical gas densities $n\sim 0.1-1~cm^{-3}$ typical of the average ISM, the thermalization time may be of several thousands years. For instance, for a strong shock one has $T_{p}=\frac{3}{16} \frac{m_{p}V_{sh}^{2}}{k_{B}} = 5.6\times 10^{8} (V_{sh}/5000 km/s)^{2}$ and using Eq. \ref{eq:ratio} for $T_{e}$, one finds that electrons would need several hundred years to reach the same temperature as protons (even assuming that protons are thermalized in the first place). 

On the other hand, even partial equilibration between electrons and protons may produce observational signatures, such as the excitation of lines in the regime of non equilibrium ionization of heavy atoms such as Oxygen, which takes place whenever the electron temperature is above $\sim 1$ keV \cite[]{2007ApJ.661.879E}. For a shock moving with velocity $v$ the temperatures of protons and electrons immediately downstream can be estimated to be of order $k T_{p}\sim 15 v_{8}^{2}$ keV and $T_{e}\sim 80 v_{8}^{2}$ eV, where $v_{8}$ is the shock velocity in units of $10^{8}$ cm/s $=1000$ km/s.

The formation of collisionless shocks raises the important question about the mechanism for dissipation, needed in order to transform part of the kinetic energy of the plasma crossing the shock from upstream into thermal energy of the plasma downstream. The dissipation is expected to be qualitatively different depending upon the orientation of the background magnetic field. For parallel shocks (background field oriented along the normal to the shock surface) the excitation of Weibel instability leads to the generation of small scale magnetic fields which become part of the dissipation mechanism (see discussion by \cite{Treumann:2009p2233}). 

It is easy to picture how the physics of dissipation at a collisionless shock also affects the injection of particles into the acceleration cycle. Similar to the case of collisional shocks, where the thickness of the shock front is of the order of the collisional mean free path, for collisionless shocks the thickness of the front is of the order of the typical scale of the instabilities that are responsible for dissipation. As an order of magnitude one can expect that the thickness of the front is several gyration radii of the thermal particles in the plasma. While gyrating in the self-produced magnetic fields, a small fraction of particles on the tail of the distribution may end up in the upstream side of the shock that is being formed, thereby bootstrapping the injection of the first accelerated particles. Injection remains one of the most poorly known aspects of particle acceleration at astrophysical shocks. In the last few years, Particle in Cell (PIC) simulations have been instrumental in reaching a better understanding of the formation of collisionless shocks (both relativistic and non-relativistic) and the initial stages of the acceleration process \cite[]{2008ApJ.673L.39S,2008ApJ.682L.5S,2011ApJ.726.75S,2012ApJ.744.67G}.  

Independent of the specific mechanism for dissipation, after the collisionless shock has been formed one can write equations for conservation of mass, momentum and energy across the shock surface. Here I limit myself with the simple case of a plain parallel infinite shock and with accelerated particles treated as test particles, having no dynamical role. For simplicity I also assume that on the scales we are interested in the shock can be considered stationary in time. In a realistic situation, basically all of these conditions get broken to some extent, and it becomes important to always have under control the limitations of the calculations we carry out, depending on their application. 

Conservation of mass, momentum and energy across the shock read:
\be
\frac{\partial }{\partial x} \left( \rho u \right)=0,
\ee
\be
\frac{\partial }{\partial x} \left( \rho u^{2} + P_{g} \right)=0,
\ee
\be
\frac{\partial }{\partial x} \left( \frac{1}{2}\rho u^{3}+\frac{\gamma_{g}}{\gamma_{g}-1}u P_{g}  \right)=0,
\ee
where $\gamma_{g}$ is the adiabatic index, $P_{g}$ is the gas pressure and $\rho$ and $u$ are the density and velocity of the plasma as seen in the reference frame of the shock. These conservation equations have the trivial solution $\rho=constant$, $u=constant$, $P_{g}=constant$, but they also admit the discontinuous solutions:
\be
\frac{\rho_{2}}{\rho_{1}}=\frac{u_{1}}{u_{2}}=\frac{(\gamma_{g}+1)M_{1}^{2}}{(\gamma_{g}-1)M_{1}^{2}+2}
\ee
\be
\frac{P_{g,2}}{P_{g,1}}=\frac{2\gamma_{g} M_{1}^{2}}{\gamma_{g}+1}-\frac{\gamma_{g}-1}{\gamma_{g}+1}
\ee
\be
\frac{T_{2}}{T_{1}}=\frac{(2\gamma_{g} M_{1}^{2}-\gamma_{g} (\gamma_{g}-1))((\gamma_{g}-1)M_{1}^{2}+2)}{(\gamma_{g}+1)^{2}M_{1}^{2}}.
\ee
For a plasma with adiabatic index $\gamma_{g}=5/3$ and $M_{1}\gg 1$ the jump conditions simplify considerably.  I refer to this case as the strong shock limit and it is easy to show that in this asymptotic limit
\be
\frac{\rho_{2}}{\rho_{1}}=\frac{u_{1}}{u_{2}} = 4,~~~~~\frac{P_{g,2}}{P_{g,1}} = \frac{5}{4} M_{1}^{2},~~~~~\frac{T_{2}}{T_{1}} = \frac{5}{16} M_{1}^{2}.
\ee
Recalling that $M_{1}^{2}=u_{1}^{2}/c_{s,1}^{2}$ and $c_{s,1}^{2}=\gamma P_{g,1}/\rho_{1}$ one easily obtains that
\be
k T_{2} = \frac{3}{16} m_{p} u_{1}^{2}, 
\ee
namely for a strong shock a large fraction of the kinetic energy of the particles upstream is transformed into internal energy of the gas behind the shock. The downstream temperature becomes basically independent of the temperature upstream, $T_{1}$. 

The presence of non-thermal particles accelerated at the shock front, and of magnetic fields in the shock region both change the conservation equations written above, as described in \S \ref{sec:nldsa}. It is important to realize that the processes involved in the formation of a collisionless shock also determine the injection of a few particles in the acceleration cycle that may lead to CRs. At the same time CRs change the structure of the collisionless shock, thereby affecting their own injection. This complex chain of effects illustrates in a qualitative way what is known as non-linear particle acceleration. 

\subsection{Transport of charged particles in magnetic fields: basic concepts}
\label{sec:transport}

The original idea that the bulk motion of magnetized clouds could be transformed into the kinetic energy of individual charged particles was first introduced by Enrico Fermi \cite[]{1949PhRv.75.1169F,1954ApJ.119.1F} and is currently widely referred to as {\it second order Fermi acceleration}. Each interaction of a test particle with a magnetized cloud results in either an energy gain or an energy loss, depending upon the relative direction of motion at the time of the scattering. On average however, the head-on collisions dominate upon tail-on collisions and the momentum vector of the charged particle performs a random walk in momentum space, in which the length of the vector increases on average by an amount $\sim \Delta E/E=(4/3) (V/c)^{2}$, where $V/c$ is the modulus of the velocity of the clouds in units of the speed of light. The scaling with the second power of $V/c$ is the reason why the mechanism is named {\it second order} Fermi mechanism. In the ISM the role of the magnetized clouds is played by plasma waves, most notably Alfv\'en waves, which move at speed $v_{A}=B/\sqrt{4\pi \rho_{i}}=2 B_{\mu} n_{i,cm^{-3}}$ km/s, where $\rho_{i}=n_{i} m_{p}$ is the mass density of ionized material. Given the smallness of the wave velocity it is easy to understand that the role of second order Fermi acceleration is, in general, rather limited. However the revolutionary concept that it bears is still of the utmost importance: the electric field induced by the motion of the magnetized cloud (or wave) may accelerate charged particles. Given the importance of this phenomenon, not only for particle acceleration but for propagation as well, in this section I will illustrate some basic concepts that turn out to be useful in order to understand the behavior of a charged particle in a background of waves. 

\begin{figure}
\includegraphics[width=300pt]{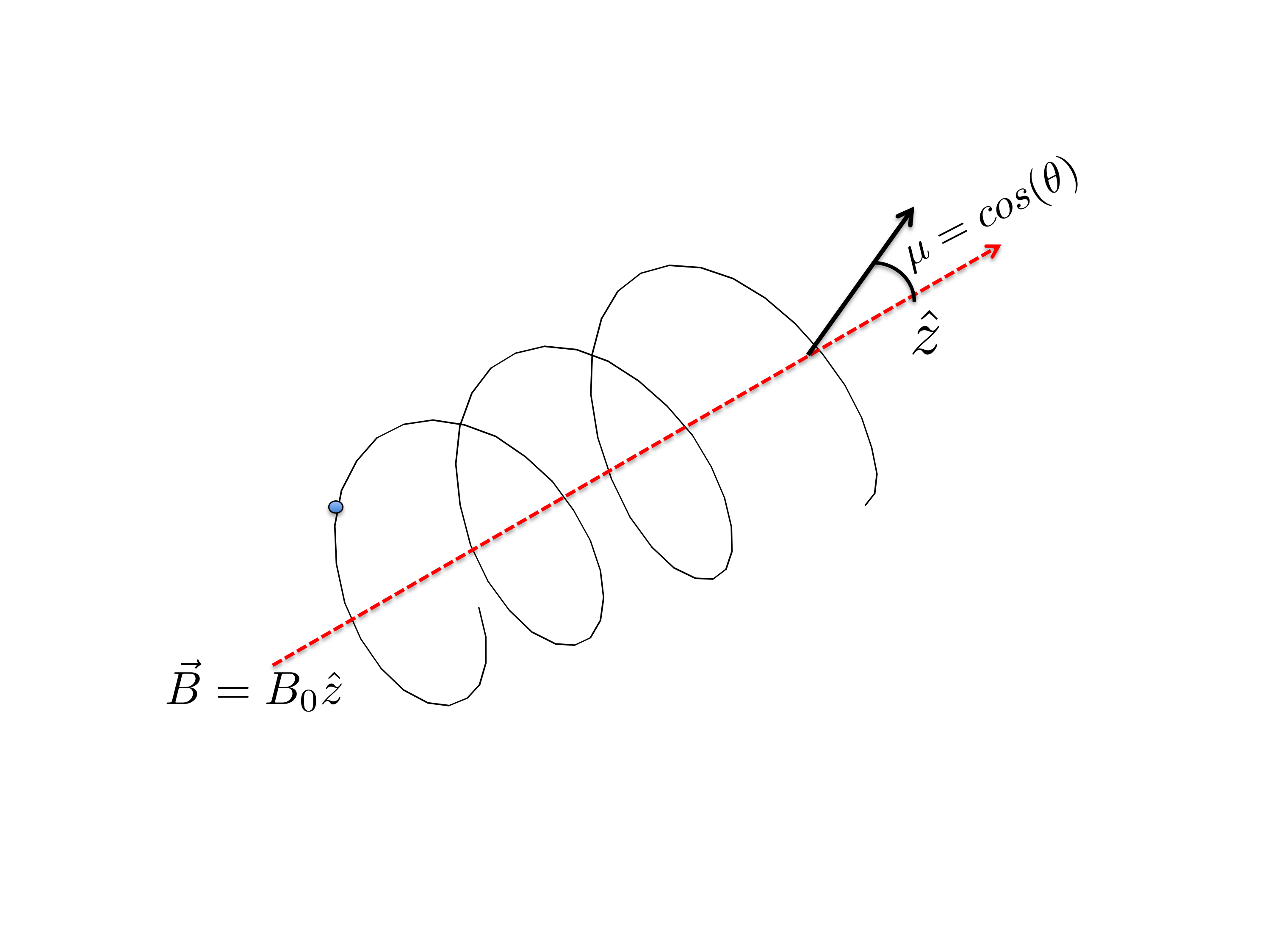}
\caption{Trajectory of a charged particle moving with a pitch angle $\theta$ with respect to an ordered magnetic field $B_{0}$, along the $\hat z$ axis.}
\label{fig:helix}      
\end{figure}

The motion of a particle moving in an ordered magnetic field $\vec B_{0} = B_{0} \hat z$ conserves the component of the momentum in the $\hat z$ direction and since the magnetic field cannot do work on a charged particle, the modulus of the momentum is also conserved. This implies that the particle trajectory consists of a rotation in the $xy$ plane perpendicular to $\hat z$, with a frequency $\Omega = q B_{0}/(m c \gamma)$ (gyration frequency) and a regular motion in the $\hat z$-direction with momentum $p_{z}= p\mu$ where $\mu$ is the cosine of the pitch angle of the particle (see Fig. \ref{fig:helix}). The velocity of the particle in the three spatial dimensions can therefore be written as:
\begin{eqnarray}
v_{x}(t)=v_{\perp} \cos\left(\Omega t+\phi\right)\\
v_{y}(t)=-v_{\perp} \sin\left(\Omega t+\phi\right)\\
v_{z}(t)=v_{\parallel}= v\mu = constant,
\end{eqnarray}
where $\phi$ is an arbitrary phase and $v_{\parallel}$ and $v_{\perp}$ are the parallel and perpendicular components of the particle velocity.

Let us assume now that on top of the background magnetic field $\vec B_{0}$ there is an oscillating magnetic field consisting of the superposition of Alfv\'en waves polarized linearly along the $x-$axis. In the reference frame of the waves ($v_{A}\ll c$) the electric field vanishes and one can write the individual Fourier modes as
\be
\delta \vec B = \delta B \hat x \sin (kz - \omega t) \approx \delta B \hat x \sin (kz),
\ee 
where the $z$ coordinate of the particle is $z=v \mu t$. The Lorentz force on the particle in the $z-$ direction is 
\be
mv\gamma \frac{d \mu}{dt} = -\frac{q}{c}\delta B v_{y} \to \frac{d\mu}{dt} = \Omega \frac{\delta B}{B_{0}} (1-\mu^{2})^{1/2} \sin\left(\Omega t+\phi\right) \sin (k v \mu t),
\ee
which can also be rewritten as 
\be
\frac{d\mu}{dt} = \frac{1}{2}\Omega \frac{\delta B}{B_{0}} (1-\mu^{2})^{1/2} \left\{ \cos\left[(\Omega-kv\mu) t+\phi\right] - \cos\left[(\Omega+kv\mu) t+\phi\right]  \right\}.
\ee
From this expression it is clear that for $\mu>0$ (particles moving in the positive direction) $\Omega+kv\mu>0$ and the cosine averages to zero on a long time scale. The first cosine also averages to zero unless $\Omega=kv\mu$, in which case the sign of $\delta\mu$ depends on $\cos(\phi)$ and it is random if the phase is random. The average over the phase also vanishes, but the mean square variation of the pitch angle does not vanish:
\be
\langle \frac{\Delta \mu \Delta \mu}{\Delta t}\rangle_{\phi} =  \pi \Omega^{2} \left(\frac{\delta B}{B_{0}}\right)^{2} \frac{(1-\mu^{2})}{\mu} \delta\left(k-\frac{\Omega}{v \mu}\right).
\ee
The linear scaling of the square of the pitch angle cosine with time is indicative of the diffusive motion of the particles. The rate of scattering in pitch angle is usually written in terms of pitch angle diffusion coefficient:
\be
\nu = \langle \frac{\Delta \theta \Delta \theta}{\Delta t}\rangle_{\phi} =  \pi \Omega^{2} \left(\frac{\delta B}{B_{0}}\right)^{2} \frac{1}{\mu} \delta\left(k-\frac{\Omega}{v \mu}\right).
\ee
If $P(k) dk$ is the wave energy density in the wave number range $dk$ at the resonant wave number $k=\Omega/v\mu$, the total scattering rate can be written as:
\be
\nu = \frac{\pi}{4} \left( \frac{kP(k)}{B_{0}^{2}/8\pi}\right) \Omega.
\ee
The time required for the particle direction to change by $\delta\theta\sim 1$ is 
\be
\tau\sim 1/\nu \sim \Omega^{-1}\left( \frac{kP(k)}{B_{0}^{2}/8\pi}\right)^{-1}
\ee 
so that the spatial diffusion coefficient can be estimated as
\be
D(p) = \frac{1}{3} v (v\tau) \simeq  \frac{1}{3} v^{2} \Omega^{-1}\left( \frac{kP(k)}{B_{0}^{2}/8\pi}\right)^{-1} = \frac{1}{3} \frac{r_{L}v}{{\cal F}},
\label{eq:diffcoeff}
\ee
where $r_{L}=v/\Omega$ is the Larmor radius of the particles and ${\cal F}=\left( \frac{kP(k)}{B_{0}^{2}/8\pi}\right)$.

It is interesting to notice that the escape time of CRs as measured from the B/C ratio and/or from unstable elements, namely a time of order $10^{7}$ years in the energy range $\sim 1$ GeV, corresponds to require $H^{2}/D(p)\sim 10^{7}$ years, where $H\sim 3$ kpc is the estimated size of the galactic halo. This implies $D\approx 10^{29} cm^{2}s^{-1}$, which corresponds to require $\delta B/B\sim 6\times 10^{-4}$ at the resonant wave number. A very small power in the form of Alfv\'en waves can easily account for the level of diffusion necessary to confine CRs in the Galaxy. The requirements become even less demanding when higher energy CRs are considered. 

The simple treatment presented here should also clarify the main physical aspects of particle scattering in the ISM, not only in terms of CR confinement in the Galaxy, but also in terms of particle transport inside the accelerators. Alfv\'en waves in proximity of a shock front can lead to a diffusive motion of particles on both sides of the shock surface. This apparently simple conclusion is the physical basis of diffusive shock acceleration, that will be discussed in the sections below. However, it is also important to realize the numerous limitations involved in the simple description illustrated above. 

First, the perturbative nature of the formalism introduced here limits its applicability to situations in which $\delta B/B\ll 1$. Second, as discussed already by \cite{Jokipii:1969p2497} and \cite{Jokipii:1969p2496}, when $\delta B/B$ becomes closer to unity, the random walk of magnetic field lines may become the most important reason for particle transport perpendicular to the background magnetic field. The combined transport of particles as due to diffusion parallel to the magnetic field and perpendicular to it is not yet fully understood, and in fact it is not completely clear that the overall motion can be described as purely diffusive. In other words, the mean square displacement $\langle z^{2} \rangle$ may not scale linearly with time (see for instance \cite[]{2013SSRv.176.73G} and references therein). The particle transport perpendicular to the background field most likely plays a very important role in terms of confinement of CRs in the Galaxy, especially when realistic models of the galactic magnetic field are taken into account \cite[]{DeMarco:2007p1952,Effenberger:2012p2499}.

Third, as discussed by \cite{Goldreich:1995p3018}, the cascade of Alfvenic turbulence from large to small spatial scales proceeds in an anisotropic way, so that at the resonant wavenumbers relevant for particle scattering, small power might be left in the parallel wavenumbers. The CR transport in these conditions might be better modeled as diffusion in a slab plus two dimensional turbulence and the diffusion of particles in such turbulence can de described by the so-called non-linear guiding center theory, first developed by \cite{2003ApJ.590L.53M}. The main physical characteristic of this theory of CR transport is that the diffusion coefficient perpendicular to the magnetic field is a non trivial function of the diffusion coefficient parallel to the field. This non-linearity makes it difficult to achieve a fully self-consistent treatment of CR propagation either in the Galaxy or in the accelerators. This point has recently been investigated in detail by \cite{Shalchi:2010p2721}. 

\subsection{DSA through the transport equation}
\label{sec:diffapproach}

Let us consider a shock front characterized by a Mach number $M_{s}$. The compression factor at the shock $r=u_{1}/u_{2}$ is then
\be
r=\frac{4 M_{s}^{2}}{M_{s}^{2}+3},
\ee
which tends to 4 in the limit of strong shocks, $M_{s}\to \infty$. A test particle diffusing in the upstream plasma does not gain or lose energy (although the second order Fermi process discussed above may be at work). 

\begin{figure}
\includegraphics[width=400pt]{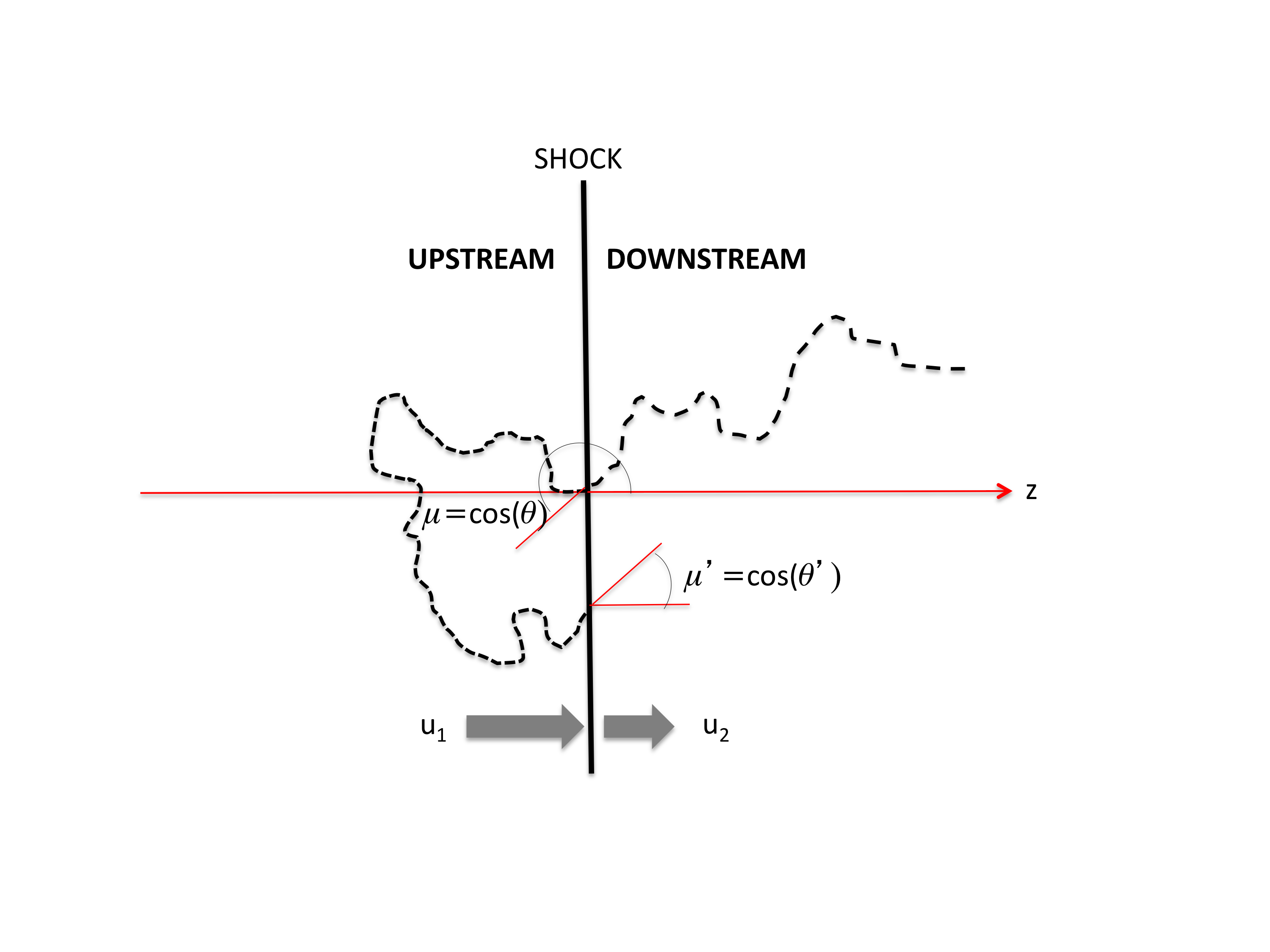}
\caption{Illustration of test-particle acceleration at a collisionless shock. In the shock frame the plasma enters from the left with velocity $u_{1}$ and exits to the right with velocity $u_{2}<u_{1}$. Here the test particle is shown to enter downstream with cosine of the pitch angle $\mu$ (as measured in the upstream plasma frame) and exit with a cosine of the pitch angle $\mu'$ (as measured in the downstream plasma frame).}
\label{fig:shock}      
\end{figure}

For a stationary parallel shock, namely a shock for which the normal to the shock is parallel to the orientation of the background magnetic field (see Fig. \ref{fig:shock}) the transport of particles is described by the diffusion-convection equation \cite[]{Skilling:1975p2166} (see \cite[]{Blandford:1987p2498} for a detailed derivation), which in the shock frame reads:
\be
u \frac{\partial f}{\partial z} = \frac{\partial}{\partial z}\left[ D \frac{\partial f}{\partial z}\right] + \frac{1}{3} \frac{du}{dz} p \frac{\partial f}{\partial p} + Q,
\label{eq:transport}
\ee
where $f(z,p)$ is the distribution function of accelerated particles, normalized in a way that the number of particles with momentum $p$ at location $z$ is $\int dp 4\pi p^{2} f(p,z)$. In Eq. \ref{eq:transport} the LHS is the convection term, the first term of the RHS is the spatial diffusion term. The second term on the RHS describes the effect of fluid compression on the accelerated particles, while $Q(x,p)$ is the injection term. 

A few comments on Eq. \ref{eq:transport} are in order: 1) the shock will appear in this equation only in terms of a boundary condition at $z=0$, and the shock is assumed to have infinitely small size along $z$. This implies that this equation cannot properly describe the thermal particles in the fluid. The distribution function of accelerated particles is continuous across the shock. 2) In a self-consistent treatment in which the acceleration process is an integral part of the processes that lead to the formation of the shock one would not need to specify an injection term. Injection would result from the microphysics of the particle motions at the shock. This ambiguity is usually faced in a phenomenological way, by adopting recipes such as the thermal leakage one \cite[]{1998PhRvE..58.4911M,2000A&A.364.911G} that allow one to relate the injection to some property of the plasma behind the shock. This aspect becomes relevant only in the case of non-linear theories of DSA, while for the test particle theory the injection term only determines the arbitrary normalization of the spectrum. However it is worth recalling that while these recipes may apply to the case of protons as injected particles, the injection of heavier nuclei may be much more complex. In fact, it has been argued that nuclei are injected at the shock following the process of sputtering of dust grains \cite[]{Meyer:1997p610,Ellison:1997p609}.

For the purpose of the present discussion I will assume that injection only takes place at the shock surface, immediately downstream of the shock, and that it only consists of particles with given momentum $p_{inj}$:
\be
Q(p,x)=\frac{\eta n_{1}u_{1}}{4\pi p_{inj}^{2}}\delta(p-p_{inj})\delta(z)=q_{0} \delta(z),
\label{eq:inj}
\ee
where $n_{1}$ and $u_{1}$ are the fluid density and fluid velocity upstream of the shock and $\eta$ is the acceleration efficiency, defined here as the fraction of the incoming number flux across the shock surface that takes part in the acceleration process. Hereafter I will use the indexes $1$ and $2$ to refer to quantities upstream and downstream respectively. 

The compression term vanishes everywhere but at the shock since $du/dz=(u_{2}-u_{1})\delta(z)$. Integration of Eq. \ref{eq:transport} around the shock surface (between $z=0^{-}$ and $z=0^{+}$) leads to:
\be
\left[D\frac{\partial f}{\partial z}\right]_{2} - \left[D\frac{\partial f}{\partial z}\right]_{1} + \frac{1}{3} (u_{2}-u_{1}) p \frac{df_{0}}{dp} + q_{0}(p) = 0,
\label{eq:boundary}
\ee
where $f_{0}(p)$ is now the distribution function of accelerated particles at the shock surface. Particle scattering downstream leads to a homogeneous distribution of particles, at least for the case of a parallel shock, so that $\left[\partial f/\partial z\right]_{2}=0$. In the upstream region, where $du/dz=0$ the transport equation reduces to:
\be
\frac{\partial}{\partial z}\left[ u f - D\frac{\partial f}{\partial z} \right]=0,
\ee
and since the quantity in parenthesis vanishes at upstream infinity, it follows that 
\be
\left[D\frac{\partial f}{\partial z}\right]_{1} = u_{1} f_{0}.
\ee
Using this result in Eq. \ref{eq:boundary} we obtain an equation for $f_{0}(p)$
\be
u_{1} f_{0} = \frac{1}{3}(u_{2}-u_{1}) p \frac{d f_{0}}{dp} + \frac{\eta n_{1}u_{1}}{4\pi p_{inj}^{2}}\delta(p-p_{inj}),
\ee
which is easily solved to give:
\be
f_{0}(p) = \frac{3r}{r-1} \frac{\eta n_{1}}{4\pi p_{inj}^{2}} \left( \frac{p}{p_{inj}}\right)^{-\frac{3r}{r-1}}.
\ee
The spectrum of accelerated particles is a power law in momentum (and not in energy as is often assumed in the literature) with a slope $\alpha$ that only depends on the compression ratio $r$:
\be
\alpha=\frac{3r}{r-1}.
\ee
The slope tends asymptotically to $\alpha=4$ in the limit $M_{s}\to \infty$ of an infinitely strong shock front. The number of particles with energy $\epsilon$ is $n(\epsilon)d\epsilon = 4\pi p^{2} f_{0}(p)(dp/d\epsilon) d\epsilon$, therefore $n(\epsilon)\propto \epsilon^{-\alpha}$ for relativistic particles and $n(\epsilon) \propto \epsilon^{(1-\alpha)/2}$ for non-relativistic particles. In the limit of strong shocks, $n(\epsilon)\propto \epsilon^{-2}$ ($n(\epsilon)\propto \epsilon^{-3/2}$) in the relativistic (non-relativistic) regime. 

Some points are worth being mentioned: the shape of the spectrum of the accelerated particles does not depend upon the diffusion coefficient. On one hand this is good news, in that the knowledge of the diffusion properties of the particles represent the greatest challenge for any theory of particle acceleration. On the other hand, this implies that the concept of maximum energy of accelerated particles is not naturally embedded in the test particle theory of DSA. In fact, the power law distribution derived above does extend (in principle) up to infinite particle energy. In the strong shock limit, such spectrum contains a divergent energy, thereby implying a failure of the test particle assumption. Clearly the absence of a maximum energy mainly derives from the assumption of stationarity of the acceleration process, which can be achieved only in the presence of effective escape of particles from the accelerator, a point which is directly connected to the issue of maximum energy, as discussed in \S \ref{sec:Emax}.

\subsection{Maximum energy: time versus space}
\label{sec:Emax}

There is some level of ambiguity in the definition of the maximum energy achieved in a SNR shock expanding in the ISM. The ambiguity arises from the fact that the maximum energy may be due to a finite time of acceleration (the age of the remnant) or to the existence of a spatial boundary, such that particles can leak out of the system when they diffuse out to such boundary. Clearly in this second case, the physical nature of such a boundary should be discussed. 

At least three different definitions of the maximum energy should be considered, and it is not always clear which definition works the best or best describes reality. The first definition consists in requiring that the acceleration time be smaller than the age of the SNR (in case of electrons as accelerated particles the age of the remnant should be replaced by the minimum between the age of the SNR and the time scale for energy losses due to synchrotron and inverse Compton scattering (ICS) radiative processes). 

A rigorous calculation of the acceleration time was carried out by \cite{1983SSRv.36.57D}, while a generalization of such a derivation in the context of the non-linear theory of DSA was presented by \cite{Blasi:2007p144}. In this section I will illustrate a simple derivation of the acceleration time based on the very essential feature of DSA, namely the fact that it proceeds through repeated shock crossings of individual particles. Let us consider a particle that from the upstream crosses the shock towards the downstream, with a pitch angle $\mu_{1}$ and an energy $E_{1}$. For simplicity let us assume that the particle is already relativistic, so that $p\simeq E$. As seen in the reference frame of the downstream plasma the particle has energy 
\be
E_{2} = \Gamma E_{1}\left( 1+\beta \mu_{1}\right) ~~~~~ 0 \leq \mu_{1} \leq 1, 
\ee
where $\beta=(u_{1}-u_{2})$ is the relative velocity between the upstream and the downstream fluid in units of the speed of light $c$, and $\Gamma=(1-\beta^{2})^{1/2}$. While in the downstream region, the particle does not gain or lose energy to first order (there are the usual second order effects that are neglected here). If the particle returns to the shock it may recross its surface with a pitch angle with cosine $-1\leq\mu_{2}\leq 0$, so that the particle energy as seen again by an observer in the upstream fluid is
\be
E_{1}' = \Gamma E_{2} \left( 1-\beta \mu_{2}\right) = \Gamma^{2} E_{1} \left( 1+\beta \mu_{1}\right) \left( 1-\beta \mu_{2}\right).
\ee
Notice that the final energy of the particle after one full cycle upstream-downstream-upstream (or downstream-upstream-downstream) is always $E_{1}'>E_{1}$, namely particles gain energy at each cycle. In the assumption that the distribution of particles is isotropized by scatterings (diffusion) both upstream and downstream, the fluxes on both sides are normalized as $2|\mu|$. In other words $\int_{0}^{1} d\mu A\mu =\int_{-1}^{0} d\mu A|\mu|=1\to A=2$. The mean value of the energy change per cycle is therefore \cite[]{Bell:1978p1344}:
\be
\langle \frac{E_{1}'-E_{1}}{E_{1}}\rangle_{\mu_{1},\mu_{2}} = -\int_{0}^{1} d\mu_{1} 2 \mu_{1} \int_{-1}^{0} d\mu_{2} 2\mu_{2} \left[ \Gamma^{2} E_{1} \left( 1+\beta \mu_{1}\right) \left( 1-\beta \mu_{2}\right) - 1\right] = \frac{4}{3} \beta.
\ee
The scaling of $\langle \frac{\Delta E}{E}\rangle$ with the first power of $\beta$ is the reason why DSA is often named first order Fermi acceleration. 

In the assumption of isotropy, the flux of particles that cross the shock from downstream to upstream is $n_{s}c/4$, which means that the upstream section is filled through a surface $\Sigma$ of the shock in one diffusion time upstream with a number of particles $n_{s}(c/4)\tau_{diff,1}\Sigma$ ($n_{s}$ is the density of accelerated particles at the shock). This number must equal the total number of particles within a diffusion length upstream $L_{1}=D_{1}/u_{1}$, namely:
\be
n_{s} \frac{c}{4} \Sigma \tau_{diff,1}= n_{s} \Sigma \frac{D_{1}}{u_{1}},
\ee
which implies for the diffusion time upstream $\tau_{diff,1}=\frac{4D_{1}}{c u_{1}}$. A similar estimate downstream leads to $\tau_{diff,2}=\frac{4D_{2}}{c u_{2}}$, so that the duration of a full cycle across the shock is $\tau_{diff}=\tau_{diff,1}+\tau_{diff,2}$. The acceleration time is now:
\be
\tau_{acc} = \frac{E}{\Delta E/\tau_{diff}}=\frac{3}{u_{1}-u_{2}}\left[ \frac{D_{1}}{u_{1}}+\frac{D_{2}}{u_{2}}\right].
\ee
This should be compared with the formally correct and more general expression \cite[]{Lagage:1983p1348,Lagage:1983p1347}:
\be
\tau_{acc} = \frac{3}{u_{1}-u_{2}} \int_{0}^{p} \frac{dp'}{p'} \left[ \frac{D_{1}(p')}{u_{1}}+\frac{D_{2}(p')}{u_{2}}\right].
\label{eq:tauacc}
\ee
The two expressions return the same order of magnitude provided $D(p)$ is an increasing function of momentum. 

Eq. \ref{eq:tauacc} effectively illustrates the fact that the acceleration time is dominated by particle diffusion in the region with less scattering (larger diffusion coefficient) which in normal conditions is the region of the upstream fluid. 

The first definition of maximum energy is that the acceleration time be smaller than the age to the SNR $\tau_{SNR}$. Using Eq. \ref{eq:diffcoeff} for the diffusion coefficient, and concentrating our attention on the upstream fluid, one can write the condition for the maximum energy as
\be
\frac{1}{3}\frac{r_{L}(p_{max}) c}{v_{s}^{2}{\cal F}(k_{min})}\approx\tau_{SNR},
\ee
where $k_{min}=1/r_{L}(p_{max})$ is the wave number resonant with particles with momentum $p_{max}$. Using the fact that for a SNR in its ejecta dominated phase $v_{s}\tau_{{SNR}}\approx R_{SNR}$, the radius of the SNR shell, the condition becomes 
\be
{\cal F}(k_{min}) \approx \frac{1}{3} \frac{c}{v_{s}}\frac{r_{L}(p_{max})}{R_{SNR}}.
\label{eq:effe}
\ee
This condition is rather interesting since at $p_{max}$, for reference values of the parameters, one has 
\be
r_{L}(p_{max}) = 1 pc \left( \frac{E}{10^{15} eV}\right) B_{\mu}^{-1}, 
\ee
which is a fraction of order $\sim 0.1$ of the size of young known SNRs in the ejecta dominated phase or early stages of the Sedov-Taylor phase. Since $c/v_{s}\sim 100$ for the same cases, one immediately infers that in order for a SNR to be a PeVatron one has to have ${\cal F}(k_{min})\gg 1$, namely the random component of the magnetic field on the scale $\sim r_{L}(p_{max})$ must be much larger than the pre-existing ordered magnetic field, $\delta B/B_{0}\gg 1$. Clearly in these conditions the calculations that led us to the expression Eq. \ref{eq:diffcoeff} for the diffusion coefficient fail since the random field can no longer be considered as a perturbation. These last few lines are sufficient to illustrate one of the problems that the field of CR research has been facing for the last several decades: for SNRs to behave as PeVatrons one has to invoke a physical mechanism that enhances the turbulent magnetic field upstream of a SNR shock by a factor $\sim 10-100$ on all scales up to $r_{L}(p_{max})$. Notice that in the absence of such a mechanism, the maximum energy achieved at a SNR shock is rather uninteresting. For instance, if the diffusion coefficient close to the shock were the same as inferred in the ISM from measurements of the B/C ratio, the maximum energy that could be achieved at $\sim 1000$ years old SNR with the shock moving at $3000$ km/s is only a fraction of GeV. 

It is important to stress that since the acceleration time is dominated by the upstream conditions, the large magnetic field amplification is needed upstream, where only accelerated particles can reach. It is therefore natural to expect, as was initially proposed by \cite{Bell:1978p1342,Bell:1978p1344} and \cite{Lagage:1983p1348,Lagage:1983p1347} that the magnetic field may be excited by the same particles that are being accelerated. This important aspect of DSA will be discussed in the context of the non linear theory of DSA in \S \ref{sec:nldsa}. 

One last point is worth being mentioned concerning Eq. \ref{eq:effe}. One might argue that increasing the radius of the SNR the condition on $\cal F$ may be relaxed, and that acceleration of very high energy CRs may take place at the late stages of the SNR. This is however not plausible for several reasons: 1) after the beginning of the Sedov-Taylor phase, the radius of the remnant increases slowly, therefore not much changes in the constraint on ${\cal F}(k_{min})$; 2) during the Sedov-Taylor phase the velocity of the shock drops with time, therefore the acceleration time starts increasing, unless the rate of magnetic field amplification gets larger, but in this case the constraint on ${\cal F}(k_{min})$ becomes even more severe. It is therefore plausible that the highest energy in a SNR is reached sometimes during the ejecta dominated phase, and most likely right before the beginning of the Sedov-Taylor phase. 

An alternative definition of the maximum energy is inspired by the possibility of free particle escape from a boundary location at some distance $z_{0}=\chi R_{sh}$, with $\chi<1$. This definition is more often used to describe the maximum energy during the Sedov-Taylor phase, when particle escape should be easier because the shock slows down, so that not only the probability for the highest energy particles to return to the shock increases (see discussion in \S \ref{sec:escape}) but also the strength of the amplified magnetic field is likely to drop. The condition for the maximum momentum in this case can be written as:
\be
\frac{D(p_{max})}{V_{sh}} \approx \chi R_{sh}.
\label{eq:2def}
\ee
Again the highest value of $p_{max}$ can be reached at the beginning of the Sedov-Taylor phase, when one can approximately estimate the SNR radius as $R_{sh}\approx V_{sh} T_{ST}$, so that Eq. \ref{eq:2def} becomes:
\be
\frac{D(p_{max})}{V_{sh}^{2}} \approx \chi T_{ST}.
\label{eq:2def1}
\ee
Recalling that $D(p)/V_{sh}^{2}$ is a rough estimate of the acceleration time, one easily realizes that the condition in Eq. \ref{eq:2def1} is somewhat more restrictive than the one based on comparing the acceleration time with the age of the SNR, since $\chi<1$. 

The third definition of the maximum energy is purely geometric in nature and should be used more as a solid upper limit rather than as an estimate of $p_{max}$. The condition, that I will only mention here, consists in requiring that the Larmor radius of the highest energy particles equal the size of the system, $r_{L}(p_{max})=R_{sh}$. Typically this condition overestimates the value of $p_{max}$ by $\sim c/V_{sh}$ with respect to the second definition discussed above. 

All estimates of the diffusion coefficient presented above are based on the framework of particle acceleration at a quasi-parallel shock. \cite{Jokipii:1982p3136} and \cite{Jokipii:1987p3084} argued that particle acceleration may be faster at oblique shocks (angle to the shock normal larger than $\sim 30^{o}$) and be the fastest at perpendicular shocks (magnetic field perpendicular to the shock normal), even for $\delta B/B<1$. At such shocks, particles can cross the shock surface several times during Larmor gyrations while moving along the magnetic field, and be thereby accelerated by the drifts associated to the electric fields that the particles experience because of the different plasma velocity upstream and downstream of the shock. The weak point of this simple scenario is that the particles get advected at the plasma speed, with the magnetic field line that they are trapped on, thereby reducing the time that they can stay in the shock region. On the other hand, the random walk of magnetic field lines may solve this problem, as discussed by \cite{2005ApJ.624.765G}. The role of particle transport perpendicular to the field lines is however not yet completely understood: the theory that currently best describes particle diffusion perpendicular to field lines was formulated by \cite{2003ApJ.590L.53M}, and shows how the perpendicular diffusion coefficient depends in a non trivial way upon the parallel diffusion coefficient, thereby creating serious problems in building a self-consistent picture of particle acceleration at perpendicular shocks. However numerical simulations have showed that particle acceleration at perpendicular shocks may be a promising mechanism to increase the maximum energy of accelerated particles beyond the limits discussed above \cite[]{2005ApJ.624.765G,2013SSRv.176.73G}.
 
The two scenarios of effective magnetic field amplification and of perpendicular shock configuration (without magnetic field amplification) are often considered as two alternative possibilities to shorten the acceleration time and lead to higher energy particles. In fact reality can be appreciably more complex than that. For instance the field at the shock can become prevalently perpendicular as a result of magnetic field amplification upstream with $\delta B/B\gg 1$, since the perturbations are likely to evolve mainly in the plane perpendicular to the pre-existing magnetic field. Moreover, as discussed by \cite{2005ApJ.624.765G}, the large scale behaviour of the magnetic field lines is likely to speed up acceleration even for the case of parallel shocks, because when the magnetic field line crosses the shock, there is a finite probability that it happens to be oblique with respect to the shock normal, so that drifts enhance the particle energy gain. This complexity and its implications for particle acceleration to the highest energies deserve much more attention than have received until now. 

\section{The non-linear theory of diffusive shock acceleration}
\label{sec:nldsa}

In the previous section I have outlined the main principles and the main limitations of the test-particle theory of CR acceleration in SNR shocks. There are three main reasons that justify the need for a non-linear theory of DSA:

\begin{itemize}
\item[1)] {\it Dynamical reaction of accelerated particles}

For the typical rate of SNRs in the Galaxy, the acceleration efficiency per supernova required to reproduce the CR energetics observed at Earth is of order $\sim 10\%$. This implies that the pressure exerted by accelerated particles on the plasma around the shock affects the shock dynamics as well as the acceleration process. The non-linearity appears through the modification of the compression factor which in turn changes the spectrum of accelerated particles in a way that in general depends upon particle rigidity.

Note also that while $\sim 10\%$ may be a reasonable estimate of the CR acceleration efficiency averaged over the entire history of the remnant, there may be stages during which the efficiency may be appreciably larger. 

\item[2)] {\it Plasma instabilities induced by accelerated particles}

As I discussed above, SNRs can be the source of the bulk of CRs in the Galaxy, up to rigidities of order $\sim 10^{6}$ GV only if substantial magnetic field amplification takes place at the shock surface. Since this process must take place upstream of the shock in order to reduce the acceleration time, it is likely that it is driven by the same accelerated particles, which would therefore determine the diffusion coefficient that describes their motion. The existence of magnetic field amplification is also the most likely explanation of the observed bright, narrow X-ray rims of non-thermal emission observed in virtually all young SNRs (see \cite[]{Vink:2012p2755,2006AdSpR.37.1902B} for recent reviews). The non-linearity here reflects in the fact that the diffusion coefficient becomes dependent upon the distribution function of accelerated particles, which is in turn determined by the diffusion coefficient in the acceleration region.

\item[3)] {\it Dynamical reaction of the amplified magnetic field}

The magnetic fields required to explain the X-ray filaments are of order $100-1000 \mu G$. The magnetic pressure is therefore still a fraction of order $10^{-2}-10^{-3}$ of the ram pressure $\rho v_{s}^{2}$ for typical values of the parameters. However, the magnetic pressure may easily become larger than the upstream thermal pressure of the incoming plasma, so as to affect the compression factor at the shock. A change in the compression factor affects the spectrum of accelerated particles which in turn determines the level of magnetic field amplification, another non-linear aspect of DSA.

\end{itemize}
While a review of non-linear DSA (NLDSA) can be found in \cite[]{Malkov:2001p765}, here I will focus on the physical aspects of relevance for the calculations of the spectrum and multifrequency appearance of SNRs. Mathematical subtleties, when present, will be pointed out but not discussed in detail.

\subsection{Dynamical reaction of accelerated particles}
\label{sec:reaction}

The dynamical reaction that accelerated particles exert on the shock is due to two different effects: 1) the pressure in accelerated particles slows down the incoming upstream plasma as seen in the shock reference frame, thereby creating a precursor. In terms of dynamics of the plasma, this leads to a compression factor that depends on the location upstream of the shock \cite[]{1981ApJ.248.344D,1982A&A.111.317A}. 2) The escape of the highest energy particles from the shock region makes the shock {\it radiative} \cite[]{1999ApJ.526.385B}, thereby inducing an increase of the compression factor between upstream infinity and downstream. Both these effects result in a modification of the spectrum of accelerated particles, which turns out to be no longer a perfect power law \cite[]{1994APh.2.215B,1997APh.7.183B,1999ApJ.526.385B,1999ApJ.511L.53M,Blasi:2002p109}. 

\begin{figure}
\includegraphics[width=410pt]{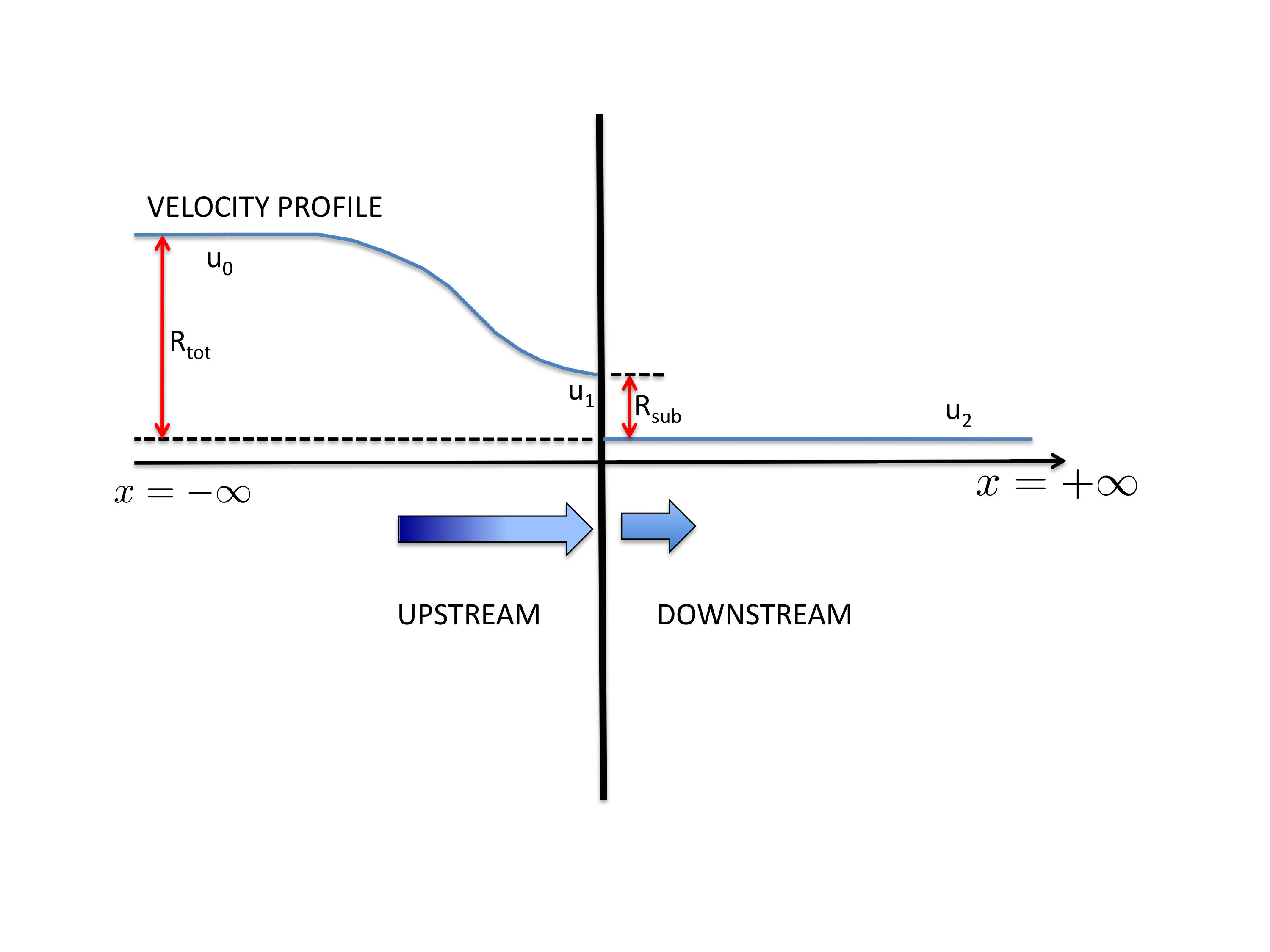}
\caption{Schematic view of a cosmic ray modified shock wave in the shock frame. Upstream infinity is on the left ($x=-\infty$), where the plasma velocity is $u_{0}$. The CR pressure slows down the inflowing plasma, so as to reduce its bulk velocity to $u_{1}<u_{0}$ immediately upstream of the subshock. The plasma in then compressed and slowed down at the subshock so that the plasma velocity downstream is $u_{2}=u_{1}/R_{sub}$.The total compression factor is $R_{tot}=u_{0}/u_{2}$.}
\label{fig:modshock}      
\end{figure}

Before embarking in outlining a theory of NLDSA, it is useful to have a feeling of the physical effects expected due to the dynamical reaction of accelerated particles on the shock. A pictorial representation of the shock modification induced by accelerated particles is reported in Fig. \ref{fig:modshock}: the plasma velocity at upstream infinity ($x=-\infty$) is $u_{0}$. While approaching the shock, a fluid element experiences an increasing pressure due to accelerated particles. This is the result of the fact that the diffusion coefficient is an increasing function of momentum, therefore at a position $z$ upstream only particles with energy $E\geq E_{min}(z)$, with $D(E_{min})/v_{s}\approx |z|$, can reach that far. The pressure of accelerated particles tends to slow down the incoming fluid, so that a precursor is created, with the gas getting slower while approaching the shock surface. Since the shock region becomes more complex in the presence of particle acceleration, the term {\it shock} is usually used to refer to the whole region between upstream infinity and downstream infinity, and is made of a precursor and a subshock, which is now the sharp discontinuity produced in the background gas. If the spectrum were $\sim E^{-2}$, the energy density would only scale logarithmically with $E_{min}$ (for a given $E_{max}$), therefore the precursor is spatially extended. For spectra steeper than $E^{-2}$, the energetics is dominated by low energies, therefore the precursor is concentrated toward the subshock. On the other hand, it will be shown later that in the presence of efficient CR acceleration, the spectrum at high energies can become appreciably harder than $E^{-2}$, so as to make the total energy in the form of accelerated particles dominated by the maximum energy. 

Although the energy density in the form of accelerated particles may become comparable with the ram pressure $\rho u^{2}$, the number density of these particles remains negligible with respect to the density of the background plasma. Therefore the equation for mass conservation is:
\be
\frac{\der \rho}{\der t} + \frac{\der}{\der z} \left( \rho u\right) = 0. 
\label{eq:mass}
\ee
The equation of motion of a fluid element under the action of a gradient in the total pressure is
\be
\rho \frac{D u}{D t} = -\frac{\der}{\der z}\left( P_{g} + P_{c}\right),
\ee
where $D/Dt=\der/\der t + u\der/\der z$ denotes the total time derivative and $P_{g}$ and $P_{c}$ are the gas and cosmic ray pressure respectively. After some simple algebra and using Eq. \ref{eq:mass} for mass conservation, one can easily rewrite this as
\be
\frac{\der}{\der t}\left( \rho u\right) = -\frac{\der}{\der z}\left[ \rho u^{2} + P_{g} + P_{c}\right],
\label{eq:momentum}
\ee
which can be viewed as the equation for momentum conservation in the presence of accelerated particles. It is useful to introduce the energy per unit mass of fluid as $\epsilon=\frac{1}{2} u^{2} + \frac{P_{g}}{\rho (\gamma_{g}-1)}$, so that the energy per unit volume is $\rho \epsilon$. The time derivative of the energy per unit volume can therefore be written as:
\be
\frac{\der}{\der t}\left(\rho\epsilon\right)  = -\frac{\der}{\der z} \left[ \left( \rho \epsilon + P_{g} \right) u \right] - u \frac{\der P_{c}}{\der z}, 
\label{eq:energy}
\ee
where I used the equations for conservation of mass and momentum and the condition that on both sides of the subshock (but not at the subshock itself) the gas evolves adiabatically:
\be
\frac{D P_{g}}{D t} = -\gamma_{g} P_{g} \frac{du}{dz}.
\label{eq:adiabat}
\ee
Eqs. \ref{eq:mass}, \ref{eq:momentum} and \ref{eq:energy} represent mass, momentum and energy conservation in a plasma in which there are accelerated particles contributing a pressure $P_{c}$. In the assumption of stationarity that is often adopted in calculations of particle acceleration at SNR shocks, the three equations read:
\begin{eqnarray}
\frac{\der}{\der z} \left( \rho u\right) = 0\\
\frac{\der}{\der z} \left( \rho u^{2} + P_{g} +P_{c} \right) = 0\\
\frac{\der}{\der z} \left( \frac{1}{2} \rho u^{3} + \frac{\gamma_{g}}{\gamma_{g}-1}u P_{g}\right) = -u\frac{\der P_{c}}{\der z}.
\end{eqnarray}
It is useful to notice that since the distribution function of accelerated particles is continuous across the subshock, $P_{c}(z=0^{-})=P_{c}(z=0^{+})$, the conservation equations at the subshock are those of an ordinary gas shock. The effect of accelerated particles only reflects in the fact that the fluid velocity immediately upstream of the subshock is different from the one at upstream infinity. In this sense, the subshock is a standard gaseous shock, while the overall structure of the shock region may be heavily affected by cosmic rays.

The dynamics of accelerated particles is defined by the transport equation, which I report here in its time dependent form:
\be
\frac{\der f}{\der t} + u \frac{\partial f}{\partial z} = \frac{\partial}{\partial z}\left[ D \frac{\partial f}{\partial z}\right] + \frac{1}{3} \frac{du}{dz} p \frac{\partial f}{\partial p} + Q.
\label{eq:transport-time}
\ee
If $T(p)$ is the kinetic energy of particles with momentum $p$, the energy density and pressure of accelerated particles can be written as
\be
E_{c} (z)= \int_{0}^{\infty} dp ~4\pi p^{2} T(p) f(p,z)
\label{eq:Ec}
\ee 
\be
P_{c} (z) = \frac{1}{3} \int_{0}^{\infty} dp~ 4\pi p^{3} v(p) f(p,z).
\label{eq:Pc}
\ee
Integrating Eq. \ref{eq:transport-time} in momentum space, and neglecting the small energy input at the shock as due to injection, one gets: 
\be
\frac{\der E_{c}}{\der t} + u \frac{\der E_{c}}{\der z} = \frac{\der}{\der z}\left[ \bar D \frac{\der E_{c}}{\der z}\right] - P_{c} \frac{d u}{d z} + \frac{1}{3} \left(\frac{d u}{d z}\right)\left[ 4\pi p^{3} T(p) f(p,z)\right]_{p=0}^{p=\infty},
\label{eq:CRstate}
\ee
where I introduced the mean diffusion coefficient defined as:
\be
\bar D(z) = \frac{\int_{0}^{\infty} dp~4\pi p^{2} T(p) D(p) \frac{\der f}{\der z}}{\int_{0}^{\infty} dp~4\pi p^{2} T(p) \frac{\der f}{\der z}}.
\ee
The last term in Eq. \ref{eq:CRstate} requires some comments: in test-particle theory, the transport equation Eq. \ref{eq:transport-time} has a time dependent solution with a steadily increasing maximum momentum (if the shock velocity remains constant), namely there is no stationary solution of that equation. A stationary solution would correspond to a power law extending to infinite energy, and for a strong shock this would lead to the last term in Eq. \ref{eq:CRstate} being finite. In the context of a non-linear theory of particle acceleration, the situation is even worse since spectra can become harder than $p^{-4}$, thereby making the same term diverging. Clearly the system would be destroyed by CR pressure before reaching that situation. A meaningful
stationary solution (or a quasi-stationary solution) can only be obtained by assuming the existence of a physical boundary at a finite location $z_{0}$ upstream, where particles escape from the acceleration region. This corresponds to requiring $f(z_{0},p)=0$, so as to have an escape flux proportional to the space derivative of the distribution function in $z_{0}$ (which does not vanish). Within this framework the distribution function has a strong suppression at $p_{max}$ (see below) and the last term in Eq. \ref{eq:CRstate} vanishes. Hence Eq. \ref{eq:CRstate} becomes:
\be
\frac{\der E_{c}}{\der t} + \frac{\der}{\der z} \left[ \frac{\gamma_{c}}{\gamma_{c}-1}u P_{c}\right] = \frac{\der}{\der z}\left[ \bar D \frac{\der E_{c}}{\der z}\right] + u \frac{\der P_{c}}{\der z},
\label{eq:CRstate1}
\ee
where I introduced the adiabatic index of accelerated particles as $E_{c}=P_{c}/(\gamma_{c}-1)$. Eq. \ref{eq:CRstate1} can be used to derive $u\der P_{c}/\der z$, to be substituted in Eq. \ref{eq:energy} \cite[]{Caprioli:2009p145}:
\be
\frac{\der}{\der t} \left[ \frac{1}{2} \rho u^{3} +\frac{P_{g}}{\gamma_{g}-1}+E_{c}\right] = - \frac{\der}{\der z} \left[ \frac{1}{2} \rho u^{3} + \frac{\gamma_{g}}{\gamma_{g}-1}u P_{g} + \frac{\gamma_{c}}{\gamma_{c}-1}u P_{c}  \right] + \frac{\der}{\der z} \left[\bar D \frac{\der E_{c}}{\der z} \right]. 
\label{eq:energy1}
\ee
In the stationary regime the compression factor at the subshock can be written as a function of the Mach number $M_{1}$ of the fluid immediately upstream of the subshock in the usual way:
\be
R_{sub}=\frac{u_{1}}{u_{2}}=\frac{\rho_{2}}{\rho_{1}}=\frac{(\gamma_{g}+1) M_{1}^{2}}{(\gamma_{g}-1) M_{1}^{2}+2},
\ee
which can be obtained by integrating the equations of conservation of mass and momentum around the subshock. Integrating the same equations between immediately upstream ($z=0^{-}$) and far upstream ($z=z_{0}$) one also derives 
\be
R_{tot}=\frac{u_{0}}{u_{1}}=M_{0}^{\frac{2}{\gamma_{g}+1}}\left[\frac{(\gamma_{g}+1)R_{sub}^{\gamma_{g}}-(\gamma_{g}-1)R_{sub}^{\gamma_{g}+1}}{2} \right]^{\frac{1}{\gamma_{g}+1}},
\ee
where I used the condition of adiabaticity of the upstream gas: $M_{1}^{2}=M_{0}^{2}\left(\frac{R_{sub}}{R_{tot}} \right)^{\gamma_{g}+1}$. The total compression factor changes in case of non-adiabatic heating of the precursor, for instance due to the damping of waves induced by accelerated particles (see for instance \cite[]{1999ApJ.526.385B}). 

Finally, Eq. \ref{eq:energy1} can be used to determine $F_{esc}=\bar D \frac{\der E_{c}}{\der z}|_{z=z_{0}}$ which has the meaning of an escape flux of energy in the form of accelerated particles. These equations illustrate very clearly the formation of a cosmic ray induced precursor: for instance in the limit in which the gas pressure upstream remains negligible compared with $\rho u^{2}$, which is always true for strong shocks, the equation of conservation of momentum can be written as
\be
\xi_{CR}(z)\approx \frac{P_{c} (z)}{\rho_{0}u_{0}^{2}}\approx 1-\frac{u(z)}{u_{0}},
\ee
where $u(z)$ is the gas velocity at the position $z$ upstream. Immediately upstream of the shock the gas feels the largest CR pressure $\xi_{CR}(0) = 1-\frac{u_{1}}{u_{0}}$. In other words the upstream gas is slowed down by the CR pressure by an amount which is directly related to the fraction of the ram pressure $\rho_{0}u_{0}^{2}$ that gets converted to accelerated particles. 

Since the subshock is a gaseous shock (namely its dynamics is not affected by the presence of accelerated particles), its compression factor is bound to be $R_{sub}<4$, while the total compression factor can potentially become large. In the absence of particle escape, the net effect of the accelerated particles would be to change the adiabatic index toward $\sim 4/3$ (appropriate for a relativistic gas), therefore $R_{tot}\sim 7$. However, the escape of particles makes the shock radiative-like, so that $R_{tot}$ can become larger than 7, although I will show below that in all realistic calculations of CR modified shocks both $R_{sub}$ and $R_{tot}$ stay rather close to $4$, as a consequence of additional non-linear effects that reduce the CR reaction. 

The formation of a precursor upstream implies that the spectra of accelerated particles are no longer power laws. Physically this is easy to understand: particles with momentum $p$ diffuse upstream by a distance that is proportional to the diffusion coefficient $D(p)$, that is usually a growing function of momentum. This implies that particles with low momentum experience a compression factor closer to $R_{sub}<4$, while higher momentum particles trace a compression factor closer to $R_{tot}>4$. As a consequence the spectrum is expected to be steeper than $p^{-4}$ at low momenta and harder than $p^{-4}$ at high momenta, with the transition typically taking place around a few $GeV/c$. From the mathematical point of view, the spectrum can be calculated by solving together the non linear CR transport equation, and the equations for conservation of mass, momentum and energy. This has been done in at least three different ways: 1) finite schemes of numerical integration \cite[]{1997APh.7.183B,2000A&A.357.283B,2012APh.39.12Z}, 2) Monte Carlo methods \cite[]{1984ApJ.286.691E,1996ApJ.458.641K,2008ApJ.688.1084V} and 3) semi-analytical methods \cite[]{1999ApJ.511L.53M,1997ApJ.485.638M,Blasi:2002p109,Blasi:2004p116,Amato:2005p112,Amato:2006p139}. Each method has its pros and cons: calculations of the CR transport based on finite schemes of integration are best in tracking the temporal evolution of the whole system. Monte Carlo methods could in principle be used to investigate non diffusive effects of CRs close to the maximum momentum. Both these methods are rather time-consuming and in general it is problematic to use them together with hydrodynamical simulations of a supernova evolution. Semi-analytical methods are computationally very fast and easy to implement in more complex calculations involving simulations of supernova evolution. The quasi-stationary solutions derived with quasi-analytical methods are excellent approximations to the time-dependent solutions for given parameters, as discussed by \cite{2010MNRAS.407.1773C}.

\begin{figure}
\includegraphics[width=210pt]{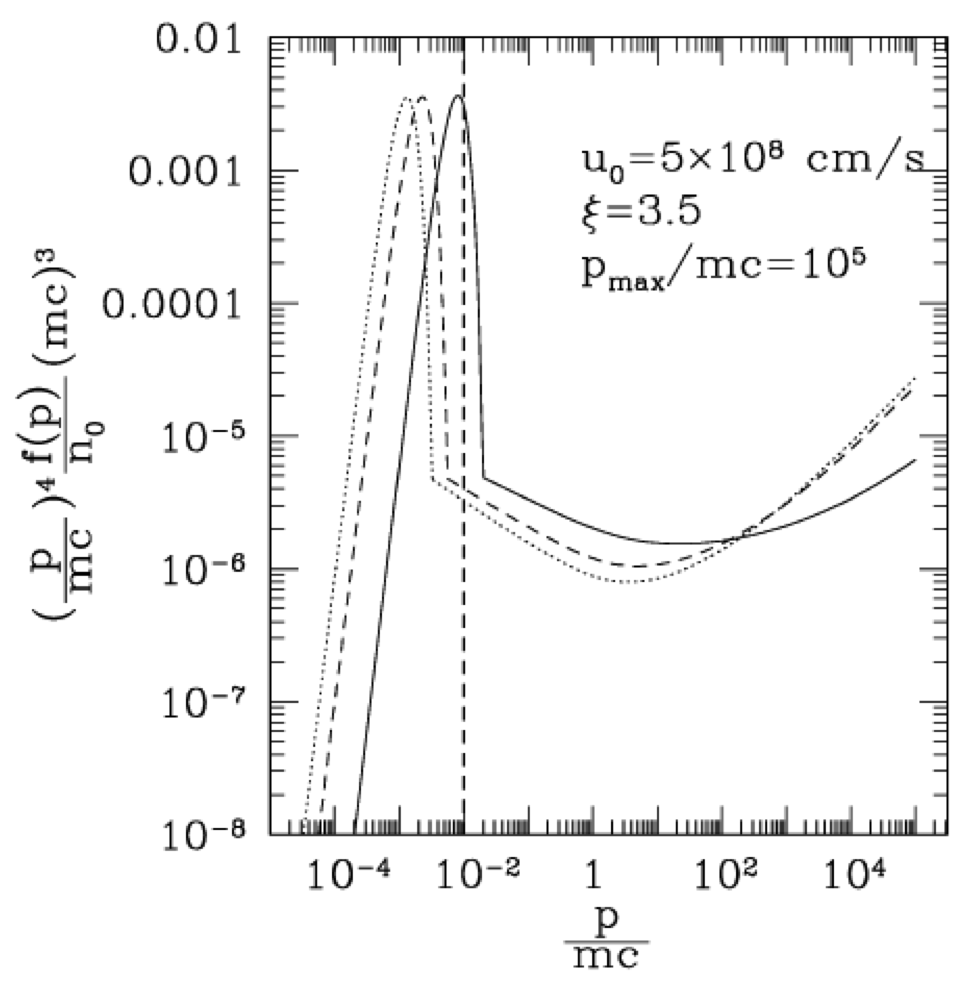}
\caption{Particle spectra (thermal plus non-thermal) at a CR modified shock with Mach number $M_{0}=10$ (solid line), $M_{0}=50$ (dashed line) and $M_{0}=100$ (dotted line). The vertical dashed line is the location of the thermal peak as expected for an ordinary shock with no particle acceleration (this value depends very weakly on the Mach number, for strong shocks). The plasma velocity at upstream infinity is $u_{0}=5\times 10^{8}$ $cm/s$, $p_{max}=10^{5} m_{p}c$ and the injection parameter is $\xi=3.5$ \cite[]{Blasi:2005p107}.}
\label{fig:modshockspec}      
\end{figure}

The encouraging agreement among these different methods of calculations of the CR dynamical reaction allows us to deduce some general conclusions on these non-linear effects: 1) the spectra of particles accelerated at a shock in the non-linear regime are not perfect power laws. 2) Since a fraction $\xi_{CR}$ of the ram pressure $\rho_{0}u_{0}^{2}$ is channelled into accelerated particles, the thermal energy of particles downstream of the shock is less than would have been found in the absence of particle acceleration. Both these effects are well illustrated in Fig. \ref{fig:modshockspec} (from \cite[]{Blasi:2005p107}), where the distribution function of particles (thermal plus accelerated) is plotted (multiplied by $p^{4}$). The three curves are obtained by changing the Mach number of the shock by lowering the temperature of the upstream gas (the shock velocity is fixed at $u_{0}=5\times 10^{8}$ km $s^{-1}$. Increasing the Mach number causes the CR acceleration to increase (the value of the maximum momentum is fixed at $p_{max}=10^{5}$ GeV/c) and the spectra become increasingly more concave so as to reflect a more pronounced CR-induced shock modification. Moreover, while increasing the CR acceleration efficiency, the temperature of the downstream plasma drops, reflecting in the peak of the Maxwellian distributions in Fig. \ref{fig:modshockspec} moving leftward. In \S \ref{sec:Halpha} I will discuss the implications of this phenomenon on the width of the broad Balmer line emission in shocks where CR acceleration is efficient.

The curvature in the spectrum is directly related to the formation of a precursor upstream of the shock. The plasma compression in the precursor is directly related not only to the pressure in the form of accelerated particles but also to any form of non-adiabatic heating possibly associated with the presence of accelerated particles. Non Adiabatic heating leads in general to a weakening of the precursor and in turn to a reduction of the concavity in the spectra of accelerated particles. Since the most natural source of non-adiabatic heating upstream is due to damping of the turbulent component of magnetic fields, this phenomenon is related to the magnetic field generation, discussed in the next section. 

\subsection{Magnetic field amplification}
\label{sec:Bfield}

The phenomenon of magnetic field amplification is probably the most important manifestation of the non-linearity of DSA. This role is related to both observational and phenomenological reasons. From the observational point of view, the main evidence for large magnetic fields in the shock region is represented by the observation of narrow filaments of non-thermal X-ray radiation in virtually all young SNRs \cite[]{Vink:2012p2755,2006AdSpR.37.1902B}. This radiation is the result of synchrotron emission from electrons with energy $E_{e}\approx 8 \left( \frac{E_{\gamma}}{100 eV} \right)^{1/2} B_{100}^{-1/2}~$ TeV, where $E_{\gamma}$ is the energy of the synchrotron photons and $B_{100}$ is the magnetic field in units of 100 $\mu G$. One can clearly see that only electrons in the $\sim 10$ TeV energy range can produce the X-rays observed from the rims. Assuming Bohm diffusion for simplicity, the acceleration time can be estimated as 
\be
\tau_{acc}\approx 3.3\times 10^{7} E_{TeV}B_{100}^{-1}V_{sh,8}^{-2}~s,
\ee
where $E_{TeV}$ is the electron energy in TeV and $V_{sh,8}=V_{sh}/(10^{8}~cm/s)$. The synchrotron loss time is
\be
\tau_{syn} = 4\times 10^{10} B_{100}^{-2} E_{TeV}^{-1}~s.
\ee
Therefore the maximum electron energy is 
\be
E_{e,max} \approx 34 B_{100}^{-1/2} V_{sh,8} ~ TeV
\ee
and the energy of the synchrotron photons reads
\be
E_{\gamma,max} \approx 1.7 V_{sh,8}^{2} ~ keV,
\ee
independent of the strength of the local magnetic field. The independence of $E_{\gamma,max}$ on the value of $B_{100}$ is a consequence of having assumed Bohm diffusion, and is not a general result. In the same approximation of Bohm diffusion, the distance covered by electrons with energy close to $E_{e,max}$ before losing their energy can be estimated as
\be
\sqrt{D(E_{e,max})\tau_{syn}} \approx 3.7 \times 10^{-2} B_{100}^{-3/2}~pc.
\ee
At the distance of the young SNRs in which the bright X-ray rims have been observed, the thickness of the rims corresponds to a physical scale of $\sim 10^{-2}$ pc, thereby implying the presence of a magnetic field of order several hundred $\mu G$, to be compared with the $1-6\mu G$ typically found in the ISM. The filaments are the best evidence so far that magnetic fields in the shock region have been amplified by a factor $\sim 10$ with respect to the interstellar magnetic field compressed at the shock. 

Establishing the nature of this phenomenon is of the utmost importance. Magnetic field amplification could be produced by the shock corrugation, through a sort of Richtmyer-Meshkov instability \cite[]{Giacalone:2007p962,2012ApJ.758.126S} or could be induced by the streaming of accelerated particles (see \cite[]{Schure:2012p3068} for a recent review), thereby representing a different aspect of the non-linear reaction of CRs on the shock. There is a qualitative, extremely important difference between these two scenarios: in the former, the field is wrapped around and its strength enhanced in the downstream region only, while in the latter case the amplification only occurs upstream of the shock, and the field is further compressed at the shock surface. These two possibilities have profoundly different implications in terms of particle acceleration, as discussed below.

Besides being needed in order to explain the thickness of the X-ray rims, magnetic field amplification is also required as a crucial aspect of the SNR paradigm. Particle acceleration as due to DSA requires effective diffusive confinement of CRs close to the shock surface in order to make it possible for the maximum energy to rise up to $\sim 10^{15}-10^{16}$ eV, as required by observations of CRs at Earth. This need is well illustrated by the following simple estimate. If the diffusion coefficient relevant for particle acceleration at SNR shocks were the one derived in the ISM from the measurement of the B/C ratio, $D(E)\approx 3\times 10^{28}(E/10 GeV)^{\delta} ~cm^{2}s^{-1}$, with $\delta\approx 0.3-0.6$, the acceleration time would be 
\be
\tau_{acc}(E) \sim \frac{D(E)}{V_{sh}^{2}} \approx 10^{5} \left(\frac{E}{10 GeV}\right)^{\delta} V_{sh,8}^{-2} ~ years,
\ee
which exceeds the typical duration of the free-expansion phase of a SNR in the ISM even for low energies (for any reasonable value of $\delta$). During the Sedov-Taylor phase the velocity of the expanding shock decreases, so that it is unlikely that the maximum energy can appreciably increase during such stage, unless the magnetic field increases with time during this phase, which is not expected to happen. 

This simple argument shows that CR acceleration in SNRs requires that magnetic field is disordered and amplified in the proximity of the shock so as to shorten the acceleration time. For instance, if the diffusion coefficient were Bohm-like with a strength of $100\mu G$, as suggested by X-ray observations, then the acceleration time would read:
\be
\tau_{acc}(E) \sim \frac{D(E)}{V_{sh}^{2}} \approx 3.3\times 10^{4} E(GeV) V_{sh,8}^{-2} B_{100}^{-1}~ s.
\ee
Comparing this time with the duration of the ejecta dominated phase of a supernova, $T_{s}$, which for typical parameters is of order a few hundred years, one easily obtains:
\be
E_{max} \approx 3\times 10^{5} GeV B_{100}\left( \frac{T_{s}}{300~years}\right) \left( \frac{V_{sh}}{1000~ km~ s^{-1}}\right)^{2}.
\label{eq:Emax}
\ee
Clearly faster shocks help reaching higher values of $E_{max}$ by decreasing the advection time $\sim D(E)/V_{sh}^{2}$, although it is worth keeping in mind that this also implies that there is less time available for magnetic field amplification. 

More realistic estimates of the maximum energy usually return somewhat lower values. Eq. \ref{eq:Emax} illustrates in a simple way the difficulty in reaching the energy of the knee in SNR shocks. All parameters have to be chosen in the most optimistic way so as to maximize $E_{max}$. 

As mentioned above, magnetic field amplification can also be due to plasma related phenomena rather than to the presence of accelerated particles. One implementation of this idea was illustrated by \cite{Giacalone:2007p962}: the shock propagates in an inhomogeneous medium with density fluctuations $\delta\rho/\rho\sim 1$. While crossing the shock surface these inhomogeneities lead to shock corrugation and to the development of eddies in which magnetic field is frozen. The twist of the eddie may lead to magnetic field amplification on time scales $\sim L_{c}/u_{2}$, where $L_{c}$ is the spatial size of these regions with larger density and $u_{2}$ is the plasma speed downstream of the shock. Smaller scales also grow so as to form a power spectrum downstream. This phenomenon could well be able to account for the observed thin X-ray filaments. The acceleration time for particles at the shock is however not necessarily appreciably reduced in that no field amplification occurs upstream. It turns out that this mechanism may be effective in accelerating particles in the cases where the initial magnetic field is perpendicular to the shock normal. In fact in this case the particles' return from the upstream region is geometrically easier and may potentially occur in just one Larmor gyration. It seems unlikely that this scenario, so strongly dependent upon the geometry of the system, may lead to a general solution of how to reach the highest energies in Galactic CRs, but this possibility definitely deserves more attention. 

It has been known for quite some time that the super-Alfv\'enic streaming of charged particles in a plasma leads to the excitation of an instability \cite[]{Skilling:1975p2165}. The role of this instability in the process of particle acceleration in SNR shocks was recognized and its implications were discussed by many authors, most notably \cite{Zweibel:1979p2308} and \cite{Achterberg:1983p2080}. The initial investigation of this instability led to identify as crucial the growth of resonant waves with wavenumber $k=1/r_{L}$, where $r_{L}$ is the Larmor radius of the particles generating the instability. The waves are therefore generated through a collective effect of the streaming of CRs but can be resonantly absorbed by individual particles thereby leading to their pitch angle diffusion. The resonance condition, taken at face value, would lead to expect that the growth stops when the turbulent magnetic field becomes of the same order as the pre-existing ordered magnetic field $\delta B\sim B_{0}$, so that the saturation level of this instability has often been assumed to occur when $\delta B/B\sim 1$. \cite{Lagage:1983p1347,Lagage:1983p1348} used this fact to conclude that the maximum energy that can possibly be reached in SNRs when the accelerated particles generate their own scattering centers is $\lesssim 10^{4}-10^{5}$ GeV/n, well below the energy of the knee. Hence, though the streaming instability leads to an appealing self-generation of the waves responsible for particle diffusion, the intrinsic resonant nature of the instability would inhibit the possibility to reach sufficiently high energy. It is important to notice that the problem with this instability is not the time scale, but again the resonant nature that forces $\delta B/B\sim 1$. In fact, the growth rate of the streaming instability can easily be found to be (see \S \ref{sec:resonant}):
\be
\Gamma_{CR}(k) = \frac{\pi}{8} \Omega_{p}^{*} \frac{V_{sh}}{V_{A}}\frac{n_{CR}(p>p_{res}(k))}{n_{i}},
\label{eq:rateofgrowth}
\ee
where $\Omega_{p}^{*}$ is the proton cyclotron frequency, $n_{CR}(p>p_{res}(k))$ is the CR density with momentum $p>p_{res}(k)$, where $p_{res}(k)=\Omega_{p}^{*}m_{p}/k$ is the minimum momentum of particles that can resonate with waves with wavenumber $k$ and $n_{i}$ is the density of ionized gas in the background plasma (here it was assumed that $V_{sh}\gg V_{A}$). 

If we introduce the power per unit logarithmic wavenumber ${\cal F}(k)$, the diffusion coefficient that rules particle acceleration is $D(p)\simeq \frac{1}{3}r_{L}(p) v(p) \frac{1}{{\cal F}(k)}$, and ${\cal F}(k)$ satisfies the advection equation
\be
V_{sh} \frac{\partial {\cal F}}{\partial z} = \sigma (k) {\cal F}(z,k),
\label{eq:growth_adv}
\ee
where $\sigma(k)=2\Gamma_{CR}(k)$ is the growth rate for the quantity $\delta B^{2}$.
It is easy to solve this equation analytically if we assume that the spectrum of accelerated particles is the standard $\propto p^{-4}$, so as to obtain that the power spectrum at the location of the shock is
\be
{\cal F}_{0}(k)=\frac{\pi}{4} \xi_{CR} \frac{V_{sh}}{V_{A}}\frac{1}{\Lambda},
\label{eq:satur}
\ee
where $\xi_{CR}$ is the fraction of $\rho V_{sh}^{2}$ that is converted to accelerated particles and $\Lambda=\log(p_{max}/m_{p}c)$. Eq. \ref{eq:growth_adv} reflects the fact that waves grow upstream of the shock while advecting towards the shock. In other words, the time available for wave growth is the advection time of a fluid element in the upstream, which is of order $D(p)/V_{sh}^{2}$. This is a sort of upper limit to the growth of waves, in that saturation might intervene at earlier times because of damping or, as mentioned above, because the growth rate gets suppressed when ${\cal F}\sim 1$. For canonical values of the parameters in Eq. \ref{eq:satur} ($\xi_{CR}=0.1$, $V_{sh}=5000$ km/s, $V_{A}=3$ km/s, $\Lambda\approx 10$), one can see that ${\cal F}_{0}\gg 1$, hence the CR induced streaming instability may potentially play a crucial role in amplifying the magnetic field upstream of the shock and enhance the scattering of accelerating particles. Moreover, for spectrum $n_{CR}(p)\sim p^{-4}$ the power spectrum ${\cal F}_{0}(k)$ is independent of $k$, thereby implying that the diffusion coefficient is Bohm-like $D(p)\propto v(p) p$. 

This qualitative conclusion is however challenged by numerous theoretical and practical difficulties: first, when ${\cal F}>1$ one qualitatively expects that the resonance condition be broken, which considerably reduces the impact of this instability; second, as I show in next section, for acceleration efficiencies $\xi_{CR}\sim 10\%$ or larger the growth rate is profoundly changed, the excited waves are no longer Alfv\'en waves and the saturation level is considerably reduced. 

\subsubsection{Resonant streaming instability induced by accelerated particles}
\label{sec:resonant}

In the reference frame of the shock the distribution of accelerated particles is approximately isotropic, while the background upstream plasma (made of protons and electrons) moves towards the shock with velocity $V_{sh}$. The condition that the total charge density vanishes at any point upstream reads:
\be
n_{CR} + n_{i} =n_{e},
\ee
where $n_{CR}$, $n_{i}$ and $n_{e}$ are the density of accelerated particles, ions and electrons respectively. Moreover the total electric current must also vanish, which implies 
\be
n_{i}v_{i} = n_{e} v_{e}.
\ee
Since $m_{p}\gg m_{e}$ we can make the assumption that electrons react more promptly than protons, so that $v_{i}\approx V_{sh}$ and
\be
v_{e} = V_{sh}\frac{n_{i}}{n_{CR}+n_{i}} \approx V_{sh}\left( 1 - \frac{n_{CR}}{n_{i}}\right),
\ee
where I also assumed $n_{CR}\ll n_{i}$, which is usually the case. The background plasma reacts to CRs moving with the shock by creating a current (relative drift between electrons and ions) that cancels the excess positive charge contributed by CRs, assumed here to be all protons. The dispersion relation of waves with wavenumber $k$ and frequency $\omega$ allowed in a system made of CRs, background ions and electrons can be written in a general form as:
\be
\frac{c^{2} k^{2}}{\omega^{2}} = 1 + \sum_{\alpha}\frac{4\pi^{2} q_{\alpha}^{2}}{\omega} \int dp \int d\mu \frac{p^{2} v(p) (1-\mu^{2})}{\omega - k v \mu \pm \Omega_{\alpha}} \left[ \frac{\partial f_{0,\alpha}}{\partial p} + \frac{1}{p} \left( \frac{v k}{\omega}-\mu\right)\frac{\partial f_{0,\alpha}}{\partial \mu}\right],
\label{eq:dispersion}
\ee
where $f_{0,\alpha}(p,\mu)$ is the unperturbed distribution function of particles of type $\alpha=CR,~i,~e$. Here $\Omega_{\alpha} = q_{\alpha}B_{0}/m_{\alpha}c$ is the cyclotron frequency of the species $\alpha$. 

Here we first consider the solutions of the dispersion relation in the regime of low frequency waves, $\omega\ll k V_{sh}$. The resulting frequency is in general a complex function of $k$, and the sign of the imaginary part of the frequency provides information on the growth or damping of the waves. The real part of the frequency describes the type of waves that get excited. 

For simplicity let us consider the case of a spectrum of accelerated particles coincident with the canonical DSA spectrum $f_{CR,0}(p)\propto p^{-4}$ for $\gamma_{min} \leq p/m_{p}c \leq \gamma_{max}$. In the case of small CR efficiency, namely when the condition
\be
\frac{n_{CR}}{n_{i}} \ll \frac{v_{A}^{2}}{V_{sh}c}
\label{eq:condition}
\ee
is fulfilled \cite[]{Zweibel:1979p2308,Achterberg:1983p2080}, it is easy to show that Alfv\'en waves are excited (namely $Re\left[\omega\right]\approx k v_{A}$) and their growth rate is:
\be
Im\left[\omega\right](k)  \equiv \omega_{I} (k)= \frac{\pi}{8} \Omega_{p}^{*} \frac{V_{sh}}{v_{A}}\frac{n_{CR}(p>p_{res}(k))}{n_{i}}.
\label{eq:Im}
\ee
This is the same growth rate as quoted in the previous section and used for the estimates of the maximum energy of accelerated particles (the factor 2 difference between Eq. \ref{eq:Im} and Eq. \ref{eq:rateofgrowth} is due to the fact that $\sigma = 2 \omega_{I}$). The same expression can also be used to estimate the growth rate of Alfv\'en waves excited in the Galaxy during propagation of CRs, if $V_{sh}$ is replaced with $\sim 2 v_{A}$. It is however very important to notice that for the usual nominal values of parameters, the condition in Eq. \ref{eq:condition} reads $\frac{n_{CR}}{n_{i}}\ll 10^{-7}$. As an order of magnitude the density of CRs can be related to the efficiency as $\frac{n_{CR}}{n_{i}} \approx \frac{3\xi_{CR}}{\gamma_{min}\Lambda}\left( \frac{V_{sh}}{c}\right)^{2}$.
Therefore Eq. \ref{eq:condition} becomes
\be
\xi_{CR} \ll \frac{\gamma_{min}\Lambda}{3} \left( \frac{v_{A}}{V_{sh}}\right)^{2}\frac{c}{V_{sh}}\approx 8\times 10^{-4} \left( \frac{V_{sh}}{5\times 10^{8} cm/s}\right)^{-3},
\label{eq:condition1}
\ee
which is typically much smaller than the value $\xi_{CR}\sim 10\%$ which is required of SNRs to be the sources of the bulk of Galactic CRs. It follows that in phases in which the SNR accelerates CRs most effectively the growth rate proceeds in a different regime. 

This regime was already investigated in the pioneering papers by \cite{Zweibel:1979p2308} and \cite{Achterberg:1983p2080} where it is referred to as {\it cosmic ray modified regime}. Two important effects come into play: 1) the excited waves are no longer pure Alfv\'en waves, in that imaginary and real part of the frequency become comparable,  and 2) their growth rate acquires different scalings with the physical quantities involved in the problem. 

In this regime, that occurs when Eq. \ref{eq:condition} is not fulfilled, the solution of the dispersion relation for $k r_{L,0}\leq 1$, namely for waves that can resonate with particles in the spectrum of accelerated particles ($\gamma\geq \gamma_{min}$) becomes:
\be
\omega_{I} \approx \omega_{R} =  \left[ \frac{\pi}{8} \Omega_{p}^{*} k V_{sh} \frac{n_{CR}(p>p_{res}(k))}{n_{i}} \right]^{1/2}.
\ee
Since $n_{CR}(p>p_{res}(k))\propto p_{res}^{-1}\sim k$, it follows that $\omega\propto k$ for $k r_{L,0}\leq 1$, but the phase velocity of the waves $v_{\phi}=\omega_{R}/k\gg v_{A}$. The fact that the phase velocity of these waves exceeds the Alfv\'en speed may affect the spectrum of particles accelerated at the shock.

One can repeat the calculation of the saturation of the turbulent magnetic field as due to advection alone, upstream of the shock, as done above (see Eq. \ref{eq:growth_adv}), but using now the growth rate appropriate for the case of efficient CR acceleration at the shock. It is easy to calculate the power spectrum at the shock location:
\be
{\cal F}_{0}(k) = \left(\frac{\pi}{6}\right)^{1/2} \left( \frac{\xi_{CR}}{\Lambda} \right)^{1/2} \left( \frac{c}{V_{sh}} \right)^{1/2}. 
\ee
For the usual canonical values of the parameters, one finds ${\cal F}_{0}\lesssim 1$, hence the effect of efficient CR acceleration is such as to reduce the growth of the waves and limit the value of the self generated magnetic field to the same order of magnitude as the pre-existing magnetic field. Magnetic field damping may possibly make the problem even more severe. 

\subsubsection{Non-resonant small-scale modes from streaming instability induced by accelerated particles}

The solution of the dispersion equation, Eq. \ref{eq:dispersion} contains more modes than the resonant ones discussed above. \cite{Bell:2004p737,2005MNRAS.358.181B} noticed that when the condition in Eq. \ref{eq:condition} is violated, namely when 
\be
\xi_{CR} > \frac{\gamma_{min}\Lambda}{3} \left( \frac{v_{A}}{V_{sh}}\right)^{2}\frac{c}{V_{sh}},
\ee
the right hand polarized mode develops a non-resonant branch for $kr_{L,0}>1$ (spatial scales smaller than the Larmor radius of all the particles in the spectrum of accelerated particles), with a growth rate that keeps increasing proportional to $k^{1/2}$ and reaches a maximum for 
\be
k_{*}r_{L,0} = \frac{3\xi_{CR}\gamma_{min}}{\Lambda}\left( \frac{V_{sh}}{v_{A}}\right)^{2}\frac{V_{sh}}{c}>1,
\ee
which is a factor $(k_{*}r_{L,0})^{1/2}$ larger than the growth rate of the resonant mode at $k r_{L,0}=1$. This non-resonant mode has several interesting aspects: first, it is current driven, but the current that is responsible for the appearance of this mode is the return current induced in the background plasma by the CR current. The fact that the return current is made of electrons moving with respect to protons is the physical reason for these modes developing on small scales (electrons in the background plasma have very low energy) and right-hand polarized. Second, the growth of these modes, when they exist, is very fast for high speed shocks, however they cannot resonate with CR particles because their scale is much smaller than the Larmor radius of any particles at the shock. On the other hand, it was shown that the growth of these modes leads to the formation of complex structures: flux tubes form, that appear to be organized on large spatial scales \cite[]{2012MNRAS.419.2433R} and ions are expelled from these tubes thereby inducing the formation of density perturbations. At present it is not clear whether the growth may lead to the formation of magnetic perturbations on scales relevant for scattering of CRs with energy $\geq 10^{5}$ GeV (see discussion in \S \ref{sec:fila}). 

\begin{figure}
\includegraphics[width=300pt]{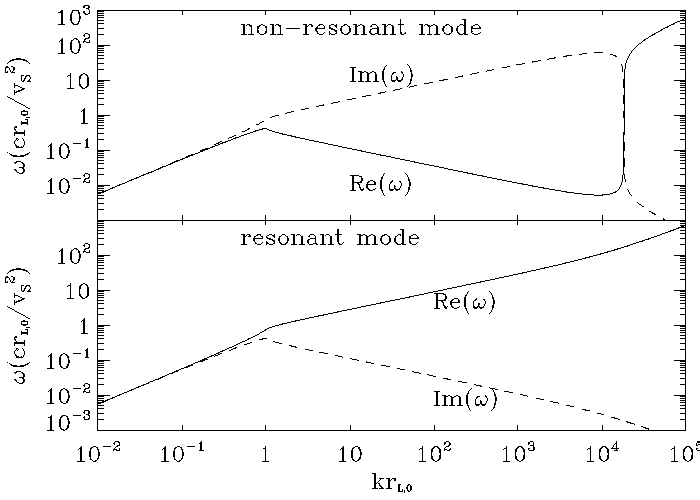}
\caption{Real and imaginary parts of the frequency as a function of wavenumber for the resonant (top panel) and non-resonant (bottom panel) modes, as calculated in \cite[]{Amato:2009p122}. Wavenumbers are in units of $1/r_{L,0}$, while frequencies are in units of $V^{2}_{sh}/(c r_{L,0})$. In each panel, the solid (dashed) curve represents the real (imaginary) part of the frequency. The values of the parameters are as follows: $V_{sh}=10^{9} cm s^{-1}$, $B_{0} = 1\mu G$, $n=1~cm^{-3}$, $\xi_{CR}=10\%$ and $p_{max}=10^{5} m_{p}c$.}
\label{fig:growth}      
\end{figure}

The situation described above is well illustrated in Fig. \ref{fig:growth}, taken from a paper by \cite{Amato:2009p122}. The top (bottom) panel refers to the left-hand (right-hand) polarized mode for a case with strong CR modification of the waves ($\xi_{CR}=10\%$). In both plots the real and imaginary part of the frequency are plotted as a solid and dashed line respectively. In this plot, the concept of resonant and non-resonant should be understood in terms of left-hand and right-hand polarization of the waves. In fact one can see that the resonant part of the dispersion relation ($k r_{L,0}\leq 1$) is present in both panels, namely these modes are excited irrespective of the polarization (this would not be true in the case of low acceleration efficiency, in which only left-hand modes are excited). In addition to the waves that can resonate with protons, the right-hand branch also has modes that grow for $k r_{L,0}\geq 1$, as discussed above. For the set of parameters used in Fig. \ref{fig:growth}, the maximum growth occurs for $k_{*} r_{L,0}\sim 10^{4}$. One can also notice how the real part of the frequency of the fast growing modes is very small: these modes are basically almost purely growing. 

Finally, it is worth recalling that damping considerably reduces the region of parameter space where the Bell's modes may effectively grow and give rise to the strongly non-linear phase of development of the instability \cite[]{Zweibel:2010p2965}.

The problem of particle acceleration at SNR shocks in the presence of small scale turbulence generated by the growth of the non-resonant mode was studied numerically by \cite{2008ApJ.678.255Z}, where maximum energies of the order to $10^{5}$ GeV were found, as a result of the fact that at the highest energies the scattering proceeds in the small deflection angle regime $D(p)\propto p^{2}$. This finding reflects the difficulty of small scale waves to resonate with particles, irrespective of how fast the modes grow. 

Recently \cite{Bykov:2009p3106,2011MNRAS.410.39B} proposed that the growth of the fast non-resonant mode may in fact also enhance the growth of waves with $k r_{L,0}<1$. If this process were confirmed by numerical calculations of the instability (current calculations are all carried out in the quasi-linear regime), it might provide a way to overcome the problem of inefficient scattering of accelerated particles off the existing turbulence around SNR shocks. 

\subsubsection{Filamentation instability}
\label{sec:fila}

Recent work has shown that the non-linear development of CR induced magnetic field amplification is more complex than illustrated above. There is no doubt that the small scale non-resonant instability \cite[]{Bell:2004p737} is very fast, provided the acceleration efficiency is large enough. The question is what happens to these modes while they grow. Both MHD simulations \cite[]{Bell:2004p737} and Particle-in-Cell simulations of this instability carried out by \cite{2009ApJ.694.626R} show how the growth leads to the development of modes on larger spatial scales. In recent numerical work \cite[]{2012MNRAS.419.2433R,2013ApJ.765L.20C} it has been shown that the current of CRs escaping the system induces the formation of filaments: the background plasma inside such filaments gets expelled from the filaments because of the $\vec J \times \vec B$ force. Different filaments attract each other as two currents would and give rise to filaments with larger cross section. Interestingly this instability, that might be a natural development of the Bell's instability to a strongly non-linear regime, leads to magnetic field amplification on a spatial scale comparable with the Larmor radius of particles in the CR current. However, since the current is made of particles that are trying to escape the system, the instability leads to a sort of self-confinement. The picture that seems to be arising consists in a possibly self-consistent scenario in which the highest energy particles (whichever that may be) generate turbulence on the scale of their own Larmor radius, thereby allowing particles of the same energy to return to the shock and sustain DSA \cite[]{2013MNRAS.431.415B,2013MNRAS.430.2873R}.

\cite{2013MNRAS.431.415B} have recently discussed the importance of the filamentation instability in achieving PeV energies in young SNRs. The authors estimated the current of particles escaping at $p_{max}$ as a function of the shock velocity and concluded that the rate of growth of the instability is such as to allow young SNRs to reach $\sim 200$ TeV energies for shock velocity $V_{sh}\sim 5000$ km/s (typical of SNRs such as Tycho), falling short of the knee by about one order of magnitude. A possible conclusion of this study might be that SNRs with an even larger velocity (therefore much younger) may be responsible for acceleration of PeV CRs. The issue of whether such young SNRs may have plowed enough material (and therefore accelerated enough particles) to account for the actual fluxes of CRs observed at Earth remains to be addressed. It is worth recalling that the argument discussed above, if applied to scenarios involving SNe type Ib,c where it has been speculated that the maximum rigidity may be as high as $\sim 10^{17}$ V \cite[]{Ptuskin:2010p1025}, imply considerably lower maximum energies. Future detection of CR protons of Galactic origin in such high energy region would be hardly reconcilable with DSA in SNRs of any type. 

\subsubsection{Non resonant large scale streaming instability induced by accelerated particles}

In addition to the resonant and non-resonant modes discussed above, the dispersion relation Eq. \ref{eq:dispersion} also returns a large scale non-resonant mode, basically a firehose instability. This instability excites waves with wavenumber smaller than the $1/r_{L,max}$, where $r_{L,max}$ is the Larmor radius of particles at some maximum momentum $p_{max}$. The instability is excited due to the anisotropy of the distribution function of accelerated particles, similar to the standard firehose instability that requires an anisotropic pressure. Interestingly the relevant anisotropy is the quadrupole rather than the dipole anisotropy (see the review paper by \cite{Bykov:2013p3165} for a discussion of this issue). The growth rate of the firehose instability can be written as
\be
\Gamma_{FH} (k) \simeq \xi_{CR}^{1/2} \frac{V_{sh}^{2} k}{c}.
\ee
Since $k\ll 1/r_{L,max}$ and $\tau_{adv}(p_{max})=r_{L,max}c/V_{sh}^{2}$ can be used as an estimate of the advection time of particles at $p_{max}$, it follows that $\Gamma_{FH}\tau_{adv}(p_{max}) \ll \xi_{CR}^{1/2}<1$, namely the instability is unlikely to have enough time to grow to a level that can be important for particles at $p_{max}$. On the other hand, the distribution of particles escaping the system could be much more anisotropic than what is implied by the diffusive approximation and hence enhance the effectiveness of the firehose instability. 

\subsection{The dynamical reaction of amplified magnetic fields on the shock}
\label{sec:Breaction}

A third aspect of the non-linearity of CR acceleration at shocks consists of the dynamical reaction of magnetic fields produced by CRs upstream on the shock itself. The theoretical aspects of this phenomenon at CR modified shocks were developed by \cite{Caprioli:2008p123} and \cite{Caprioli:2009p157}. I will refer to these papers for mathematical details, that basically represent the generalization of the conservation equations discussed in \S \ref{sec:reaction} to the case where magnetic fields are present. The conservation of mass and momentum read:
\begin{eqnarray}
\frac{\der}{\der z} \left( \rho u\right) = 0\\
\frac{\der}{\der z} \left( \rho u^{2} + P_{g} +P_{c} + P_{w}\right) = 0\\
\end{eqnarray}
where $P_{w}$ is the pressure in the form of waves. As discussed by \cite{Caprioli:2008p123}, the dynamical reaction of the amplified magnetic field can be understood by focusing on what happens at the subshock, where energy conservation reads:
\be
\left[ \frac{1}{2}\rho u^{3} + \frac{\gamma_{g}}{\gamma_{g}-1} u P_{g} + F_{w} \right]_{1}^{2} = 0,
\ee
where I used the continuity of the CR distribution function (and therefore pressure) across the subshock. As usual the indexes 1 and 2 denote quantities immediately upstream and downstream of the subshock respectively. Here $F_{w}$ is the flux of waves with pressure $P_{w}$. The connection between $P_{w}$ and $F_{w}$ is specific of the type of waves that are being studies, which unfortunately limits the applicability of the conclusions to the same cases. \cite{Caprioli:2008p123,Caprioli:2009p157} only considered the case of Alfv\'en waves, for which 
\be
P_{w} = \frac{1}{8\pi} \left(\sum_{i} \delta B_{i} \right)^{2}~~~~~F_{w} = \sum_{i} \frac{\delta B_{i}^{2}}{4\pi} (u + H_{c,i}v_{A}) + P_{w} u,
\ee
where $H=\pm 1$ is the wave helicity. The calculations illustrated in \S \ref{sec:reaction} can be repeated including the effect of waves, so as to obtain the expression connecting $R_{sub}$ (compression factor at the subshock) and $R_{tot}$ (total compression factor):
\be
R_{tot}^{\gamma_{g}+1} = \frac{M_{0}^{2}R_{sub}^{\gamma_{g}}}{2} \left[ \frac{\gamma_{g}+1-R_{sub}(\gamma_{g}-1)}{1+\Lambda_{B}} \right],
\label{eq:RtotMod}
\ee
where
\be
\Lambda_{B} = W \left[ 1+R_{sub}(2/\gamma_{g}-1) \right],~~~~~W = \frac{P_{w,1}}{P_{g,1}}.
\ee
The dynamical reaction of the amplified magnetic field is regulated by the quantity $\Lambda_{B}$, which in turn is determined by the ratio $W$ between the pressure in the form of waves and the thermal pressure immediately upstream of the subshock. If $W\ll 1$ the dynamical reaction of the magnetic field is negligible, while for $W\gtrsim 1$ the total compression factor gets reduced (Eq. \ref{eq:RtotMod}): the effect of the magnetic field is that of reducing the plasma compressibility when the magnetic pressure becomes comparable with the thermal pressure of the upstream gas, thereby acting in the direction opposite to that of CRs, which lead to larger values of $R_{tot}$. This is the reason why taking into account the effect of magnetic fields on the shock dynamics leads to predict less modified shocks, and correspondingly less concave spectra of accelerated particles \cite[]{Caprioli:2009p157}. The values of magnetic fields inferred from the thickness of the X-ray rims typically corresponds to $W\sim 1-10$, if the field is interpreted as CR induced. Hence the magnetic dynamical reaction described above is certainly important and it has been shown to have a considerable impact on the spectra of accelerated particles, making them closer to power laws. 

\subsection{A critical summary of magnetic field amplification mechanisms}

The X-ray filaments observed in virtually all young SNRs are the strongest evidence so far that magnetic field amplification takes place close to the shock. Is this the same magnetic field that is responsible for particle acceleration up to the knee? 

In the standard theory of diffusive particle transport at shocks, scattering occurs efficiently at resonance, namely when the particle encounters a wave with wavenumber $k\simeq 1/r_{L}$. This requires that sufficient power exists in the magnetic spectrum at the resonant wavenumber, so as to lead to the required scattering frequency. In the sections above I have discussed several nuances of the excitation of resonant instabilities and for all of them the case can be made that they grow too slowly. In general the strength of the magnetic field only grows to $\delta B\sim B$ for waves excited by the CRs when they are efficiently accelerated ($\xi_{CR}$ larger than few percent). Clearly if the instability led to $\delta B>B$ one could argue that the resonance condition would be ill defined. In this case the perturbative approach intrinsic in the quasi-linear theory would reveal itself as being utterly inadequate. On the other hand, the non-resonant mode first discussed by \cite{Bell:2004p737} (but see also \cite[]{2000MNRAS.314.65L,2001MNRAS.321.433B}) has a growth rate which can be much larger than any other unstable mode, and can certainly lead to large magnetic fields at the shock. However, the scales that get excited by the instability are very small compared with the gyration radii of accelerated particles and although their growth also leads to power transfer to larger scales (a sort of inverse cascade \cite[]{Bell:2004p737}), it is unlikely that this process may continue up to the scales comparable with the larmor radius of particles of $10^{5}-10^{6}$ GeV, because of the limited time available for the process to occur (roughly one advection time). Moreover, the current that induces the instability is dominated by low energy particles (say GeV particles), hence it is not easy to envision a mechanism that should move power to scales much larger than the Larmor radius of the particles representing the bulk of the current. 

In addition to the CR induced instabilities discussed above, there are also fluid instabilities (e.g. see \cite{Giacalone:2007p962}) that only amplify magnetic field downstream of the shock if a density inhomogeneity is present upstream on suitably chosen scales. In this case the scattering of particles upstream of the shock is not affected by the amplification process. 

We could speculate that the instabilities discussed above, and more specifically the non-resonant modes first found by \cite{Bell:2004p737}, play a crucial role in the production of the magnetic field as inferred from the X-ray morphology, while the same instabilities might be less important to warrant the necessary level of particle scattering to reach high energies. What would then be the mechanism to energize CRs to the knee energy? Clearly this question is still open and it may be worth keeping an open mind on how to address it. As discussed above, a possible way out might come from the investigation of the filamentation instability excited by particles escaping the acceleration region.  

A very important role in understanding the role of the different types of CR-induced instabilities in SNR shocks is being played by hybrid numerical simulations, in which the protons in the background plasma are treated by using a kinetic approach, while electrons are treated as a fluid. This approach allows one to take into account a larger range of spatial scales with respect to Particle-in-cell (PIC) simulations, which are more appropriate for the investigation of the initial stages of particle acceleration (injection). Hybrid simulations have recently been used to investigate the role of shock obliquity in the process of particle acceleration and magnetic field amplification \cite[]{2012ApJ.744.67G,2013ApJ.765L.20C}. Unfortunately, even with hybrid simulations it is, at present, difficult to describe the complex interplay between large and small scales that is so important in astrophysical sources of high energy particles: for instance, the dynamics of the shock is often dominated by the highest energy particles, that diffuse further away from the shock and probably play a crucial role in seeding magnetic instabilities (see for instance \cite{2013MNRAS.431.415B}), but these scales may be very large compared with the computation box. Another instance is in the random walk of magnetic field lines on very large scales (comparable with the size of a SNR) that facilitates the process of particles' return to the shock surface in oblique shocks, and that would not be properly described in current hybrid simulations. 

In the section below I also discuss a more mondane possibility that has been often discussed in the literature and yet received less attention than it deserved, namely the possibility that the bulk of Galactic CRs is accelerated in superbubbles excavated in the ISM by repeated SN explosions, rather than in isolated SNRs. These regions are very active in that several SNRs occur in a relatively short period of time (a few tens million years), and conditions might be better suited for particle acceleration to higher energies.

\section{The superbubble hypothesis}
\label{sec:OB}

Massive stars form mainly in the cores of dense molecular clouds in a time span that is only a few million years long. This short time inhibits the stars from acquiring a peculiar velocity larger than $\sim 2$ km/s, so that these stars explode basically within a few tens of parsecs from the place where they were born. Stars of type O and B are typically characterized by intense stellar winds with an energy injection which is of the same order of magnitude as the energy liberated at the time of the supernova event associated with the end of the nuclear reactions in the parent stars. It has been estimated that $\sim 85\%$ of the core collapse SNe in the Galaxy occur in these superbubbles (\cite{Higdon:2005p846} and references therein), excavated by the collective action of the stellar winds of O and B stars. 

The launching of the stellar winds pollutes the circumstellar region with heavy elements synthesized in the stellar interior due to nuclear reactions, therefore it may be expected that the SN explosion due to the core collapse of the parent stars take place in a metal enriched medium. It has been advocated that this may explain some anomalies in the chemical composition of CRs, most notably the overabundance of refractory elements and the $^{22}Ne$ abundance \cite[]{Higdon:2005p846,Higdon:2006p868,Higdon:2013p3116}. 

It is easy to realize that the environment in which the OB association is located is profoundly changed by the collective action of the stellar winds and the SN explosions, all within a few million years time span. In principle particle acceleration may be taking place in this environment due to several different processes, from shock acceleration in the winds, to shock acceleration at shocks formed during supernova explosions, to second order acceleration in the turbulent magnetic field deriving from merging winds and SN ejecta. These processes have been studied for instance by \cite{2001AstL.27.625B} and \cite{2004A&A.424.747P}, and the calculations seem to show a general trend to very hard spectra of accelerated particles. It has also been proposed that the maximum energy that can be achieved is higher than in isolated SNR, although these estimates are somewhat based on simple arguments that may fail to properly represent reality. Nonetheless, as a qualitative statement, it is clear that a place with enhanced background turbulence may in principle be better suited to make acceleration faster, thereby allowing us to infer higher values of the maximum energy. The problem of how to reconcile the hard injection spectra with those observed at the Earth remains to be properly addressed. 

Recently the Fermi-LAT telescope has found the first direct evidence for gamma ray emission that can be attributed to freshly accelerated CRs in the Cygnus region \cite[]{Ackermann:2011p3159}, an OB association at 1.4 kpc distance from the Sun. The spectrum of the gamma radiation is appreciably harder than the average Galactic gamma ray spectrum, again supporting the hypothesis that the parent CRs have been produced at a location close to the emission region. 

\section{Indirect evidence for CR acceleration in SNRs}
\label{sec:multinu}

There is no doubt that SNRs are sites of cosmic ray acceleration. The subject of the debate is whether all CRs are accelerated in SNRs, and which SNRs or which phases of a SNR may possibly allow for CR acceleration up to the energy of the knee. This confidence is based on direct observation of the radiation produced by CRs while being accelerated inside the sources. SNRs have long been known as radio and X-ray sources, while gamma ray emission extending to $>TeV$ energies has been detected more recently.

Radio emission is associated with synchrotron emission of non-thermal electrons, accelerated at the SNR shock. Electrons with energy $E$ would radiate at frequency $\nu \simeq 3.7 MHz B_{\mu} E(GeV)^{2}$. It is easy to see how the phenomenon of magnetic field amplification affects very profoundly the radio emission, in two ways: 1) if the field is amplified to values of, say, 100 $\mu G$, the electrons responsible for GHz radio waves have energy $E\sim 1-2$ GeV, while if the magnetic field were not amplified the corresponding electron energy would be $\sim 10-20$ GeV. The electron spectra in these two energy regions might carry information on the acceleration process: for instance in the theory of NLDSA with strong dynamical reaction of accelerated particles the spectrum is somewhat steeper (softer) at $\sim$ GeV energies than it is at $\sim 10$ GeV, which might reflect into a similar hardening in the spectrum of radio emission. This effect is more pronounced when comparing the spectrum of GeV electrons with that of particles responsible for synchrotron X-rays. X-ray radiation at 1 keV requires electrons with energy $\sim 20-30$ TeV for a 100 $\mu G$ magnetic field, therefore the concavity might be visible if one considers together radio and X-ray emission. 2) Moreover, the strong dependence of synchrotron losses from magnetic field strength implies that at given photon frequency less electrons are needed in order to explain the synchrotron emission. This reflects in a smaller value of the ratio between electrons and protons in the GeV range, what is usually referred to as $K_{ep}$. A general feature of NLDSA is to require very low values of this ratio, $K_{ep}\sim 10^{-3}-10^{-4}$ as a consequence of magnetic field amplification. The value of $K_{ep}$ measured at the Earth in the GeV energy region, where energy losses during propagation do not play an important role, is $\sim 10^{-2}$, which is a reason for concern if one wants to associate the origin of CR electrons to SNRs as well. One should however exercise some caution here, in that the effective spectrum of CRs injected by a SNR is the integral over time of the particles escaping the remnant at different times. The problem of escape of CRs from their sources is of central importance to the origin of CRs and is also one of the most uncertain aspects of the whole SNR paradigm (see \S \ref{sec:escape} below). The value of $K_{ep}$ as inferred from multiwavelength studies in the sources reflects the instantaneous ratio of densities of electrons and protons, while the value of $K_{ep}$ as measured at Earth is the result of the integration over time of the escape flux and the overlap of potentially different numerous sources. This is not a justification of the discrepancy, but rather an assessment of the complexity that lies behind the simple nature of the SNR paradigm. 

Another instance of this complexity is represented by the spectra of accelerated particles in a SNR (see \S \ref{sec:spectra} below). The basic prediction of DSA in its linear or non-linear version is that the spectra of accelerated particles at sufficiently high energies (above few GeV) should be close to $\sim E^{-2}$ or harder if the efficiency of acceleration is high enough to drive a strong dynamical reaction on the shock. As discussed below, this simple expectation is in conflict both with measurements of CR anisotropy at Earth and with measurements of the gamma ray spectrum from selected SNRs. Whether this represents a symptom of new physical effects of particle acceleration or a byproduct of the environment in which the acceleration process takes place remains to be understood. 

In the following I will try to address the strong and weak points of the SNR paradigm for the origin of CRs, stressing, whenever possible, which observational strategy could help improving our understanding. 

\subsection{Escape}
\label{sec:escape}

In an ideal plane infinite shock, the return probability of CRs from upstream of the shock is unity, namely all CRs return to the shock and are eventually advected downstream. If this were the end of the story, CRs would all be confined inside a SNR until the shock would eventually dissipate away and the particles would be able to escape into the ISM and become CRs. The adiabatic energy losses suffered by particles during the SN expansion would imply that the highest energy CRs (say with energy close to the knee) would lose part of their energy and the requirements in terms of maximum energy at the source would be even more severe than they already are. More important, one would not expect any gamma ray emission in situations in which a molecular cloud is {\it illuminated} by the CR escaping from a nearby SNR, or at least this phenomenon would appear only when CRs are left free to escape since the shock is no longer able to confine them inside the expanding shell. 

Many physical phenomena intervene in a more realistic shock wave: 1) the shock slows down due to mass accumulation, more so during the Sedov-Taylor phase. In this phase, the shock radius changes in time as $R_{sh}\propto t^{2/5}$ (if the expansion takes place in a homogeneous ISM), while the diffusion front of CRs moving with the shock expands with respect to the shock as $\propto t^{1/2}$. It seems unavoidable that more particles will diffuse away from the shock and the probability that they may return to the shock from upstream is reduced. 2) The shock may be {\it broken}, so as to allow for particles' escape to some extent. In this instance, the spectrum and density of escaping particles would depend on details of the environment in which the shock expands, making this scenario rather unappealing but not necessarily less realistic. 3) If particles can produce their own scattering centers through the collective excitation of streaming instability, it is reasonable to imagine that at some distance from the shock the particle density drops, so as to make the scattering frequency too low to warrant their return to the shock. 

A careful description of the numerous problems involved in the description of the escape of particles from a SNR shock can be found in a recent paper by \cite{2011MNRAS.415.1807D}.

Historically, in the absence of a physical theory of particle escape, this phenomenon has been modeled by introducing a spatial boundary (the same for particles of any energy) at which particles are left free to escape the system. This condition is usually implemented by solving the diffusion-convection equation with the boundary condition that $f(p,z_{0})=0$, where $z_{0}$ is the location of the escape boundary. The idea behind this boundary condition is that when self-confinement becomes inefficient, the particle density drops as a result of a transition to a sort of ballistic motion. Clearly, even this description is rather simplistic in that even the escaping particles move diffusively, but with a larger scattering length, probably closer to the one they experience while diffusing in the Galaxy. In other words, what is changing is the value of the diffusion coefficient, which increases from the small, self-generated one in the shock proximity, to the larger one present in the Galaxy.

The position of the free escape boundary is usually assumed to be located at a given fraction (of order $\sim 10\%$) of the shock radius. In this case, the solution of the transport equation can be simply found to be

\be
f(z,p)=f_{0}(p) \frac{\exp\left(\frac{u z}{D(p)}\right)-\exp\left(\frac{u z_{0}}{D(p)}\right)}{1-\exp\left(\frac{u z_{0}}{D(p)}\right)},
\ee
in the assumption that the diffusion coefficient $D(p)$ does not depend upon the spatial coordinate $x$. As usual, I assume that downstream of the shock the particle distribution is homogeneous, namely $\partial f/\partial x|_{2}=0$. The flux of particles escaping the accelerator at $x_{0}$ is then
\be
F(z_{0},p) = - D(p) \frac{\partial f}{\partial z}|_{z=z_{0}} = - \frac{u_{1}f_{0}(p)}{1-\exp\left(\frac{u z_{0}}{D(p)}\right)}\exp\left(\frac{u z_{0}}{D(p)}\right).
\label{eq:EscFlux}
\ee
The fact that $F(z_{0},p)<0$ simply expresses the fact that the particles are escaping from the system. As a function of momentum, Eq. \ref{eq:EscFlux} vanishes for $p\to 0$ and for $p\to \infty$, while it has a peak around the momentum for which $D(p_{*})/u_{1}\simeq x_{0}$, which can be used as an estimate of the maximum momentum. 

In other words, for a given location of the escape boundary, only particles in a narrow region around the maximum momentum can escape the system, so that the spectrum of escaping particles as seen from the point of view of an observer outside the system appears to be centered around the momentum $p_{*}$. On the other hand, during the Sedov-Taylor phase of a SNR the shock velocity drops, the radius of the shell increases and the magnetic field amplification causes the magnetic field to decrease with time. The spectrum of particle escaping the system is then the result of integration over time of the peaked spectra escaping at any given time. Calculating this spectrum is a useful exercise and can be done very easily \cite[]{Caprioli:2010p133,Ptuskin:2010p1025}. Let us assume that the maximum momentum reached at the beginning of the Sedov phase, $T_{s}$, is $p_{max,s}$, and that then it drops with time as $p_{max}(t)\propto (t/T_{s})^{-\alpha}$, with $\alpha>0$. The energy in the escaping particles of momentum $p$ is 
\be
d\epsilon = 4\pi p^{2} dp p c N_{esc}(p) = \xi_{esc} \frac{1}{2} \rho v_{sh}^{3} 4\pi R_{sh}^{2} dt,
\label{eq:simple}
\ee
where $\xi_{esc}(t)$ is the fraction of the income flux $\frac{1}{2} \rho v_{sh}^{3} 4\pi R_{sh}^{2}$ that is converted into escaping flux.

During the Sedov-Taylor phase in a homogeneous medium one has $R_{sh}\propto t^{2/5}$ and $V_{sh}\propto t^{-3/5}$, therefore from Eq. \ref{eq:simple}:
\be
N_{esc}(p) \propto \xi_{esc}(t) p^{-3} t^{-1} \frac{dt}{dp_{max}} \propto p^{-4} \xi_{esc}(t).
\ee
What I obtained is that the spectrum of escaping particles integrated over the Sedov-Taylor phase of the SNR is $p^{-4}$ if the fraction $\xi_{esc}$ does not depend on time. It is worth stressing that this $p^{-4}$ has nothing to do with the standard result of the DSA in the test-particle regime, neither it depends on the detailed evolution in time of the maximum momentum. It solely depends on having assumed that particles escape the SNR during the adiabatic (self-similar) phase. Notice also that in realistic calculations of the escape $\xi_{esc}$ usually decreases with time, leading to a spectrum of escaping particles which is even harder than $p^{-4}$. The total spectrum of particles injected into the ISM by an individual SNR is the sum of the escape flux and the flux of particles escaping the SNR after the shock dissipates and allows for the release of the particles accelerated throughout the history of the SNR and trapped in the expanding shell.  

This simple picture does not change qualitatively once the non-linear effects of particle acceleration are included: \cite{Caprioli:2010p133} calculated the spectrum of CRs injected by a SNR in detail in the context of the NLDSA. These calculations raise many problems, when compared with observations, as discussed below.
 
\subsection{Spectra}
\label{sec:spectra}

\begin{figure}
\includegraphics[width=300pt]{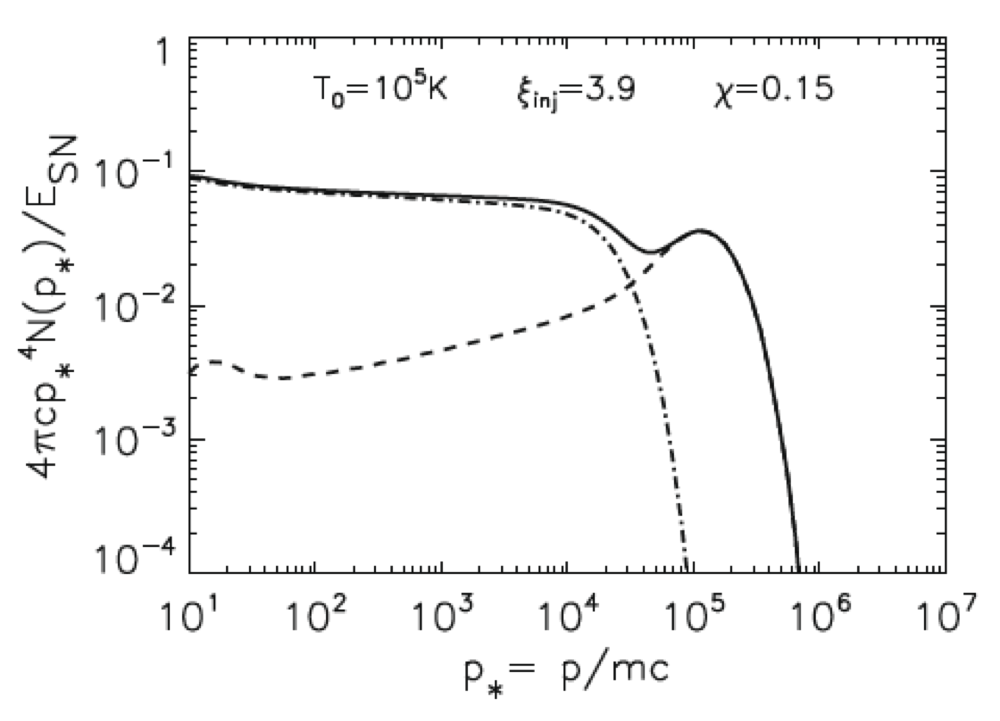}
\caption{ CR spectrum injected in the ISM by a SNR expanding in a medium with density $n_{0}=0.1~cm^{-3}$, temperature $T_{0}=10^{5}$ K and injection parameter $\xi_{inj}=3.9$ (from \cite[]{Caprioli:2010p133}). The dashed line shows the escape of particles from upstream, the dash-dotted line is the spectrum of particles escaping at the end of the evolution. The solid line is the sum of the two. The escape boundary is located at $0.15 R_{sh}$.}
\label{fig:escapeflux}      
\end{figure}

The spectrum of CRs injected by a SNR into the ISM during the few tens thousands years of its evolution is extremely complex to calculate since it requires the knowledge of the instantaneous spectrum of accelerated particles at any time, of the temporal evolution of the maximum energy, of the mechanism that leads to particle escape (see discussion above), and the entire calculation depends on the type of SN and the environment in which it explodes. The most one can do at the present time is to consider different scenarios and achieve a quantitative estimate of the amount of changes in the overall CR spectrum. Several possibilities were investigated by \cite{Caprioli:2010p133}, but a pretty general conclusion of these calculations is that the spectrum is typically very close to $E^{-2}$ at high energies if not harder, mainly as a result of the dynamical reaction of accelerated particles, and the contribution from the flux of particles escaping at any given time, which is typically harder than $E^{-2}$, as discussed above. A typical spectrum obtained from these calculations is reported in Fig. \ref{fig:escapeflux} (from the work of \cite{Caprioli:2010p133}) for a shock expanding in a uniform medium with temperature $T_{0}=10^{5}$ K and injection parameter $\xi_{inj}=3.9$. The dashed curve shows the spectrum of particles escaping through the boundary, located at $\chi R_{sh}$ (with $\chi=0.15$) from the shock, at any time. The dash-dotted line shows the spectrum of particles that leave the SNR at the end of its evolution. The maximum energy in this latter component is clearly lower, since higher energy particles escaped at earlier times through the boundary. The solid line shows the total spectrum contributed by the SNR after the end of its evolution. The bump-like structure at the highest energies is due to the hard escape flux dominating there. Notice that the escape flux as calculated in NLDSA is harder than than the naive estimate $\sim E^{-2}$ derived in \S \ref{sec:escape}, and its concavity reflects the temporal evolution of the non-linear dynamical reaction of accelerated particles on the shock. Notice also that in the absence of an escape flux from the SNR the spectrum of CRs contributed by SNRs (dash-dotted line) would exhibit a pronouned cutoff at energies much lower than the knee, as a result of adiabatic energy losses. 

The spectrum illustrated in Fig. \ref{fig:escapeflux} is troublesome in at least two ways: 1) it is harder than the spectra observed in gamma rays in several SNRs, as pointed out by \cite{Caprioli:2011p2134}; 2) if the CR spectrum injected by an individual SNR is that hard, the diffusion coefficient required in the Galaxy to fit the spectra observed at Earth is $D(E)\propto E^{0.7}$ (see also \cite[]{Berezhko:2007p1010}), which is known to result in exceedingly large CR anisotropy \cite[]{Ptuskin:2006p666,Blasi:2012p2024}. 

It is worth noticing that this discrepancy is not a consequence of the non-linear theory of DSA, in that the predictions of the test particle theory are also plagued by the same problem. 

It has been argued by \cite{Ptuskin:2010p1025,Caprioli:2010p133} that one possible reason for softer spectra might be the presence of fast moving scattering centers around the shock: as was first pointed out by \cite{Bell:1978p1344}, the compression factor that enters the calculation of the spectrum of accelerated particles is the ratio of the upstream and downstream velocity of scattering centers. In the case of ordinary Alfv\'en waves, $v_{A}\ll V_{sh}$ and the effect is weak, namely the velocity of the scattering centers (in the shock frame) is very close to the plasma velocity. On the other hand, in the case of strong magnetic field amplification it may be speculated that the speed of waves may be a sizeable fraction of the shock speed \footnote{This does not need to be so: for instance in the case of the non-resonant instability discussed by \cite{Bell:2004p737,2005MNRAS.358.181B}, magnetic field amplification may be effective, and yet the modes are almost purely growing, namely with very low phase velocity}. In this case the spectrum of accelerated particles becomes $N(E)dE\propto E^{-\alpha} dE$ with \cite{Caprioli:2012p2411}:
\be
\alpha=\frac{\tilde r+2}{\tilde r - 1}~~~~~ \tilde r = \frac{u_{1}\pm v_{W,1}}{u_{2}\pm v_{W,2}}.
\ee
While it is customary to assume that waves get isotropized downstream ($v_{W,2}=0$), the compression factor can be either decreased or increases depending on the helicity of waves upstream. This reflects in either softer or harder spectra of accelerated particles. 

Another possibility to obtain softer spectra has been discussed by \cite{Schure:2013p3169}: the authors claim that in case of a mainly perpendicular shock geometry, the return probability of particles from downstream can become smaller, thereby leading to steeper spectra. 

It is rather disappointing that both these effects rely on details of the theory, and one is left to wander if observations may actually allow us to find the correct explanation for this rather serious discrepancy between theory and observational evidence. 

\subsection{Gamma ray emission from isolated SNRs}
\label{sec:isolated}

The best chance of testing our theories of the origin of CRs in SNRs is in the modeling of the multifrequency spectrum and morphology of selected SNRs. The purpose of this section is however not that of listing the individual SNRs that have been detected in gamma rays, but rather to choose a few cases of SNRs that are sufficiently isolated so as to be modeled as individual sources, and use them to illustrate the type of information that we can gather by comparing observations with theory. 

The first clear detection of TeV gamma ray emission from a SNR came from the SNR RXJ1713.7-3946 \cite[]{2004Natur.432.75A,2006A&A.449.223A,2007A&A.464.235A}, later followed by the detection of the same remnant in the GeV energy range with the Fermi-LAT telescope \cite[]{2011ApJ.734.28A}. Here I will briefly discuss this case because it is instructive of how the comparison of theoretical predictions with data can drive our understanding of the acceleration environment. 

A discussion of the implications of the TeV data, together with the X-ray data on spectrum and morphology was presented by \cite{Morlino:2009p140}. A hadronic origin of the gamma ray emission would easily account for the bright X-ray rims (requiring a magnetic field of $\sim 160\mu G$), as well as for the gamma ray spectrum.  If electrons were to share the same temperature as protons, the model would predict a powerful thermal X-ray emission, which is not detected. Rather than disproving this possibility, this finding might be the confirmation of the expectation that at fast collisionless shocks electrons fail to reach thermal equilibrium with protons. In fact, the Coulomb collision time scale for this remnant turns out to exceed its age. On the other hand, it was pointed out by \cite{2010ApJ.712.287E} that even a slow rate of Coulomb scattering would be able to heat electrons to a temperature $\gtrsim 1$ keV, so that oxygen lines would be excited and they would dominate the thermal emission. These lines are not observed, thereby leading to a severe upper limit on the density of gas in the shock region, that would result in a too small pion production. \cite{2010ApJ.712.287E} concluded that the emission is of leptonic origin. This interpretation appears to be confirmed by the more recent Fermi-LAT data, that show a very hard gamma ray spectrum, incompatible with an origin related to pion production and decay. Clearly this does not mean that CRs are not efficiently accelerated in this remnant. It simply implies that the gas density is too low for efficient pp scattering. 

However, it should be pointed out that models based on ICS of high energy electrons are not problem free: first, as pointed out by \cite{Morlino:2009p140}, the density of IR light necessary to explain the HESS data as the result of ICS is $\sim 25$ times larger than expected. Second, the ICS interpretation requires a weak magnetic field of order $\sim 10 \mu G$, incompatible with the observed X-ray rims. Finally, recent data on the distribution of atomic and molecular hydrogen around SNR RXJ1713.7-3946 \cite[]{2012ApJ.746.82F} suggest a rather good spatial correlation between the distribution of this gas and the TeV gamma ray emission, which would be easier to explain if gamma rays were the result of $pp$ scattering. In conclusion, despite the fact that the shape of the spectrum of gamma rays would suggest a leptonic origin, the case of SNR RXJ1713.7-3946 will probably turn out to be one of those cases in which the complexity of the environment around the remnant plays a crucial role in determining the observed spectrum. Future high resolution gamma ray observations, possibly with the Cherenkov telescope array (CTA), will contribute to clarify this situation. 

A somewhat clearer case is that of the Tycho SNR, the leftover of a SN type Ia exploded in a roughly homogeneous ISM, as confirmed by the regular circular shape of the remnant. Tycho is one of the historical SNRs, as it was observed by Tycho Brahe in 1572. The multifrequency spectrum of Tycho extends from the radio band to gamma rays, and a thin X-ray rim is observed all around the remnant (see the right panel of Fig. \ref{fig:morpho}). It has been argued that the spectrum of gamma rays observed by Fermi-LAT \cite[]{2012ApJ744L2G} in the GeV range and by VERITAS \cite[]{2011ApJ730L20A} in the TeV range can only be compatible with a hadronic origin \cite[]{Morlino:2012p2243}. The morphology of the X-ray emission, resulting from synchrotron radiation of electrons in the magnetic field at the shock, is consistent with a magnetic field of $\sim 300 \mu G$, which implies a maximum energy of accelerated protons of $\sim 500$ TeV. A hadronic origin of the gamma ray emission has also been claimed by \cite{2013ApJ76314B}, where however the steep gamma ray spectrum measured from Tycho is attributed to an environmental effect: the gamma ray flux is assumed to be made of two components: one due to gamma ray production in a roughly homogeneous medium and another due to gamma ray production in denser, compact clumps where the maximum energy of CRs is lower. In the calculations of \cite{Morlino:2012p2243} the steep spectrum is instead explained as a result of NLDSA in the presence of waves moving with the Alfv\'en velocity calculated in the amplified magnetic field. In this latter case the shape of the spectrum is related, though in a model dependent way, to the strength of the amplified magnetic field, which is the same quantity relevant to determine the X-ray morphology. In the former model the steep spectrum might not be found in another SNR in the same conditions, in the absence of the small scale density perturbations assumed by the authors. 

\begin{figure}
\includegraphics[width=165pt]{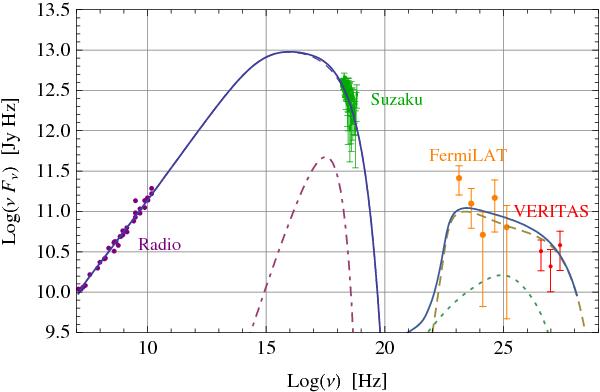}
\includegraphics[width=160pt]{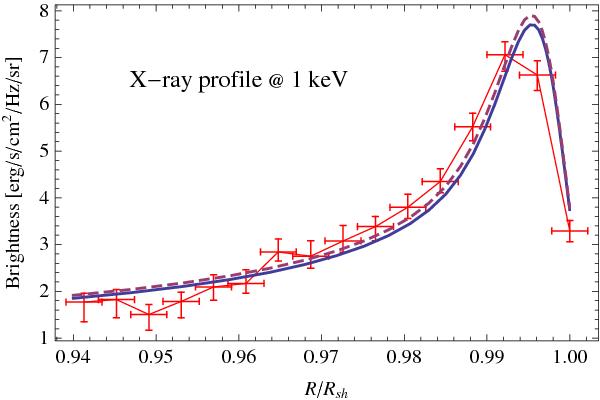}
\caption{{\it Left Panel:} Spatially integrated spectral energy distribution of Tycho. The curves show synchrotron emission, thermal electron bremsstrahlung and pion decay as calculated by \cite{Morlino:2012p2243}. Gamma ray data from Fermi-LAT \cite[]{2012ApJ744L2G} and VERITAS \cite[]{2011ApJ730L20A} are shown. {\it Right Panel:} Projected X-ray brightness at 1 keV. Data points are from \cite[]{2007ApJ...665..315C}. The solid line shows the result of the calculations by \cite{Morlino:2012p2243} after convolution with the Chandra point spread function.}
\label{fig:tycho}      
\end{figure}

The multifrequency spectrum of Tycho (left) and the X-ray brightness of its rims (right) are shown in Fig. \ref{fig:tycho} (from \cite[]{Morlino:2012p2243}). The dash-dotted line in the left panel shows the thermal emission from the downstream gas (here the electron temperature is assumed to be related to the proton temperature as $T_{e}=(m_{e}/m_{p}) T_{p}$ immediately behind the shock, and increases with time solely due to Coulomb scattering, that couples electrons with the warmer protons), the short-dashed line shows the ICS contribution to the gamma ray flux, while the dashed line refers to gamma rays from pion decays. The solid lines show the total flux. The figure shows rather impressively how the magnetic field necessary to describe the radio and X-ray radiation as synchrotron emission also describes the thickness of the X-ray rims (right panel) and pushes the maximum energy of accelerated particles to $\sim 500$ TeV (in the assumption of Bohm diffusion). 

The case of Tycho is instructive as an illustration of the level of credibility of calculations based on the theory of NLDSA: the different techniques agree fairly well (see \cite[]{2010MNRAS.407.1773C} for a discussion of this point) as long as only the dynamical reaction of accelerated particles on the shock is included. When magnetic effects are taken into account, the situation becomes more complex: in the calculations based on the semi-analytical description of \cite{Amato:2006p139} the field is estimated from the growth rate and the dynamical reaction of the magnetic field on the shock is taken into account \cite[]{Caprioli:2008p123,Caprioli:2009p157}. Similar assumptions are adopted by \cite{2008ApJ.688.1084V}, although the technique is profoundly different. Similar considerations hold for \cite{Ptuskin:2010p1025}. On the other hand, \cite{2013ApJ76314B} take the magnetic field as a parameter of the problem, chosen to fit the observations, and its dynamical reaction is not included in the calculations. The magnetic backreaction, as discussed by \cite{Caprioli:2008p123,Caprioli:2009p157} comes into play when the magnetic pressure exceeds the thermal pressure upstream, and leads to a reduction of the compression factor at the subshock, namely less concave spectra. Even softer spectra are obtained if one introduces a recipe for the velocity of the scattering centers \cite[]{Ptuskin:2010p1025,Caprioli:2010p133,Morlino:2012p2243}. This, yet speculative, effect is not included in any of the other approaches.

Even more pronounced differences arise when environmental effects are included. The case of Tycho is again useful in this respect: the predictions of the standard NLDSA theory would not be able to explain the observed gamma ray spectrum from this SNR. But assuming the existence of {\it ad hoc} density fluctuations, may change the volume integrated gamma ray spectrum as to make it similar to the observed one \cite[]{2013ApJ76314B}. Space resolved gamma ray observations would help clarify the role of these environmental effects in forging the gamma ray spectrum of a SNR. 

\subsection{SNRs near molecular clouds}
\label{sec:clouds}

There is no lack of evidence of CR proton acceleration in SNRs close to molecular clouds (MC), that act as a target for hadronic interactions resulting in pion production. Recently the AGILE \cite[]{2011ApJ.742L.30G,2010A&A.516L11G,2011ApJ.742L.30G} and Fermi-LAT \cite[]{2010ApJ.718.348A,Ackermann:2013p3110,2010ApJ.712.459A,2010Sci.327.1103A,2009ApJ.706L.1A} collaborations claimed the detection of the much sought-after pion bump in the gamma ray spectrum. This spectral feature confirms that the bulk of the gamma ray emission in these objects is due to $pp\to \pi^{0} \to 2\gamma$. 

\begin{figure}
\includegraphics[width=320pt]{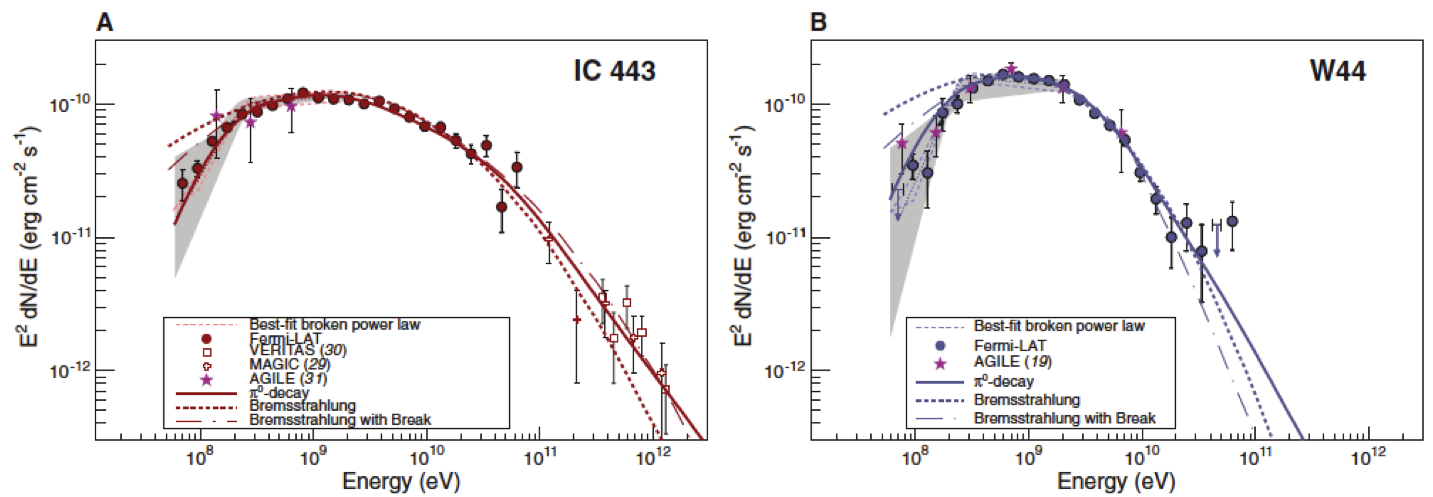}
\caption{Pion bump in the gamma ray emission of SNRs IC 443 and W44 as measured by Fermi-LAT and reported by \cite{Ackermann:2013p3110}.}
\label{fig:pionbump}      
\end{figure}

Fig. \ref{fig:pionbump} (from \cite{Ackermann:2013p3110}) shows the gamma ray spectra of SNRs IC443 (left panel) and W44 (right panel), where the pion bump is well visible. The steep gamma ray spectrum at high energies suggests that the acceleration process is no longer very active, as one may qualitatively have expected for old SNRs. 

SNRs close to molecular clouds are very interesting astrophysical objects, not so much in terms of investigating CR acceleration (as these are old objects in which one would not expect acceleration to very high energies), but rather as laboratories to investigate CR propagation around sources and escape from sources. In this respect, it is useful to separate the SNR-MC associations in two types: 1) the ones in which the shock is directly propagating inside the cloud, and 2) the ones in which the MC is {\it illuminated} by CRs propagating out of a nearby SNR, which is however at some distance from the cloud. 

In the first instance, several new effects intervene: for a density of molecular gas $n=10^{3}~cm^{-3}$, the interaction length between molecules, assuming a geometric cross section of $\sigma\sim 10^{-14}~cm^{2}$, becomes $\lambda\sim 1/n\sigma \sim 10^{11}$ cm. Moreover, the typical fraction of ionized gas in a molecular gas is so small that collisionless processes of formation of a shock wave may be less important than the ones associated with molecular collisions. The SNR shock impacting a molecular gas might become collisional, thereby leading to heating of the molecular gas on a scale $\sim \lambda$ downstream. This picture appears to be supported by the presence of maser emission from behind such shocks \cite[]{2009ApJ.706L.270H}, that prove the presence of heated molecular gas. The possibility that such shocks may accelerate particles is all but demonstrated. In fact the gamma ray emission from a MC in these conditions might be the result of the streaming of particles accelerated at previous times at the collisionless SNR shock and liberated once the shock impacts the MC. 

The second scenario has received more attention (see for instance \cite[]{2007Ap&SS.309.365G,2009MNRAS.396.1629G,2008ApJ...689..213R}). The propagation of escaping CRs from a SNR shock to a MC in its vicinity is a rather complex phenomenon to describe and model: the spectrum of CRs reaching the MC is in general time dependent, in that it is affected by both the time dependence of the escape flux (see discussion in \S \ref{sec:escape}) and by the finite time that CRs have to diffuse out to the distance of the MC, $R_{MC}$. Several authors have argued that a low energy cutoff can be expected in the CR spectrum, at the energy for which $\left[D(E) \tau_{SNR}\right]^{1/2}\simeq  R_{MC}$. This reflects the fact that higher energy particles diffuse faster, thereby reaching the MC when lower energy particles are still lagging behind. It is important to notice that a low energy cutoff in the spectrum of CRs reaching the MC at a given time does not reflect in a cutoff in the gamma ray spectrum: the cross section for pion production from a proton of given energy scales approximately as $1/E_{\pi}$, so that low energy gamma rays are expected to have a spectrum approximately $\propto E_{\gamma}^{-1}$, a signature of a low energy cut in the CR spectrum at the MC location. Possible indications of this phenomenon might have been already detected in the SNR W28 \cite[]{2010A&A.516L11G}, where two clouds at different distances from the SNR appear to be illuminated in a different way (different flux of CRs) and to be characterized by a low energy spectral break that starts at higher energies for the most distant MC, as one would expect if the break is related to CR propagation. 

Two phenomena add to the complexity of the picture presented above: 1) for isotropic diffusion, the density of CRs from the SNR dominates upon the Galactic CR spectrum for distances of a few tens of parsecs (see discussion in \cite[]{Blasi:2012p2051}). This may imply that the diffusion properties of CRs inside such distance are self-produced by the diffusing CRs, therefore possibly very different from the average conditions inside the Galaxy at large. In case of dominant parellel diffusion, this effect becomes even more important. 2) If there is a dominant orientation of the background Galactic magnetic field where the SNR and the MC are located, one can expect anisotropic diffusive effects to play a prominent role. Below I briefly discuss these issues, which might represent major sources of interesting discoveries in the near future. 

As I pointed out several times throughout this review, CRs play a crucial role in determining the diffusion properties of the medium in which they propagate. This is equally true at SNR shocks, in the Galaxy while CRs propagate, and near sources due to the CR gradient that is established there. A self-consistent solution of the propagation of CRs near their sources has recently been presented by \cite{2013ApJ.768.73M}, where effects of diffusion parallel and perpendicular to the local magnetic field have also been discussed. 

The expected pattern of diffusion mainly parallel to the background local magnetic field reflects in a spatial distribution of CRs which is elongated in the direction of the field \cite[]{Nava:2013p3140,2013PhRvD.88b3010G} at least for a time smaller than the diffusion time over a scale of the order of the coherence scale $L_{c}\sim 50-100$ pc of the magnetic field. When CRs diffuse farther than $L_{c}$ they start feeling the random walk of magnetic field lines and their distribution spreads in three spatial dimensions. If a nearby MC is located along the direction of the magnetic field it gets eventually illuminated by CRs escaping the SNR. If on the other hand the MC is not connected to the SNR by a flux tube, it is unlikely to be illuminated by CRs (because perpendicular diffusion is suppressed on these scales), and virtually no gamma ray emission is expected. This picture is strikingly more complex and richer of information than the simple picture of CRs escaping a SNR isotropically that is usually adopted in studying MCs. 

\section{$H\alpha$ line as a cosmic ray calorimeter in SNRs}
\label{sec:Halpha}

$H\alpha$ optical emission from Balmer dominated SNR shocks is a powerful indicator of the conditions around the shock \cite[]{1978ApJ.225L.27C,1980ApJ.235.186C} including the presence of accelerated particles (see \cite[]{2010PASA.27.23H} for a review). The $H\alpha$ line is produced when neutral hydrogen is present in the shock region, and it gets excited by collisions with thermal ions and electrons to the level $n=3$ and decays to $n=2$. In the following I describe the basic physics aspects of this phenomenon and how it can be used to gather information on the CR energy content at the shock. 

A collisionless shock propagating in a partially ionized background goes through several interesting new phenomena: first, neutral atoms cross the shock surface without suffering any direct heating, due to the collisionless nature of the shock (all interactions are of electro-magnetic nature, therefore the energy and momentum of neutral hydrogen cannot be changed). However, a neutral atom has a finite probability of undergoing either ionization or a charge exchange  reaction, whenever there is a net velocity difference between ions and atoms. Behind the shock, ions are slowed down (their bulk motion velocity drops down) and heated up, while neutral atoms remain colder and faster. The reactions of charge exchange lead to formation of a population of hot atoms (a hot ion downstream catches an electron from a fast neutral), which also have a finite probability of getting excited. The Balmer line emission from this population corresponds to a Doppler broadened line with a width that reflects the temperature of the hot ions downstream. Measurements of the width of the broad Balmer line have often been used to estimate the temperature of protons behind the shock, and in fact it is basically the only method to do so, since at collisionless shocks electrons (which are responsible for the continuum X-ray emission) have typically a lower temperature than protons. Equilibration between the two populations of particles (electrons and protons) may eventually occur either collisionally (through Coulomb scattering) or through collective processes. 
The broad Balmer line is produced by hydrogen atoms that suffer at least one charge exchange reaction downstream of the shock. The atoms that enter downstream and are excited before suffering a charge exchange also contribute to the $H\alpha$ line, but the width of the line reflects the gas temperature upstream, and is therefore narrow (for a temperature of $10^{4}$ K, the width is $21$ km/s).  In summary, the propagation of a collisionless shock through a partially ionized medium leads to $H\alpha$ emission, consisting of a broad and a narrow line (see the recent review by \cite{2013SSRv.tmp.75G}). 

When CRs are efficiently accelerated, two phenomena occur, as discussed in \S \ref{sec:nldsa}: 1) the temperature of the gas downstream of the shock is lower than in the absence of accelerated particles. 2) A precursor is formed upstream, as a result of the pressure exerted by accelerated particles. 

Both these phenomena have an impact on the shape and brightness of the Balmer line emission. The lower temperature of the downstream gas leads to a narrower broad Balmer line, whose width bears now information on the pressure of accelerated particles, through the conservation equations at the shock. 

The CR-induced precursor slows down the upstream ionized gas with respect to the hydrogen atoms, which again do not feel the precursor but through charge exchange. If ions are heated in the precursor (not only adiabatically, but also because of turbulent heating) the charge exchange reactions transfer some of the internal energy to neutral hydrogen, thereby heating it. This phenomenon results in a broadening of the narrow Balmer line. 

A narrower broad Balmer line and a broader narrow Balmer line are both signatures of CR acceleration at SNR shocks \cite[]{2010PASA.27.23H}. The theory of CR acceleration at collisionless SNR shocks in the presence of neutral hydrogen has only recently been formulated \cite[]{2012ApJ.755.121B,2012p3105,2013ApJ.768.148M} and has led to the prediction of several new interesting phenomena, discussed below.

\subsection{Acceleration of test particles at shocks in partially ionized media}

The presence of neutrals in the shock region changes the structure of the shock even in the absence of appreciable amounts of accelerated particles, due to the phenomenon of neutral return flux \cite[]{2012ApJ.755.121B}. A neutral atom that crosses the shock and suffers a charge exchange reaction downstream gives rise to a new neutral atom moving with high bulk velocity. There is a sizeable probability (dependent upon the shock velocity) that the resulting atom moves towards the shock and crosses it towards upstream. A new reaction of either charge exchange or ionization upstream leads the atom to deposit energy and momentum in the upstream plasma, within a distance of the order of its collision length. On the same distance scale, the upstream plasma get heated up and slows down slightly, thereby resulting in a reduction of the plasma Mach number immediately upstream of the shock (within a few pathlengths of charge exchange and/or ionization). This implies that the shock strength drops, namely its compression factor becomes less than 4 (even for strong shocks). 

This neutral return flux \cite[]{2012ApJ.755.121B} plays a very important role in the shock dynamics for velocity $V_{sh}\lesssim 3000$ km/s. For faster shocks, the cross section for charge exchange drops rather rapidly and ionization is more likely to occur downstream. This reduces the neutral return flux and the shock modification it produces. 

The consequences of the neutral return flux both on the process of particle acceleration and on the shape of the Balmer line are very serious: some hydrogen atoms undergo charge exchange immediately upstream of the shock, with ions that have been heated by the neutral return flux. These atoms give rise to a Balmer line emission corresponding to the temperature of the ions immediately upstream of the shock. As demonstrated by \cite{2012p3105} this contribution consists of an intermediate Balmer line, with a typical width of $\sim 100-300$ km/s. Some tentative evidence of this intermediate line might have already been found in existing data ({\it e.g.} see \cite[]{2000ApJ.535.266G}).

\begin{figure}
\includegraphics[width=320pt]{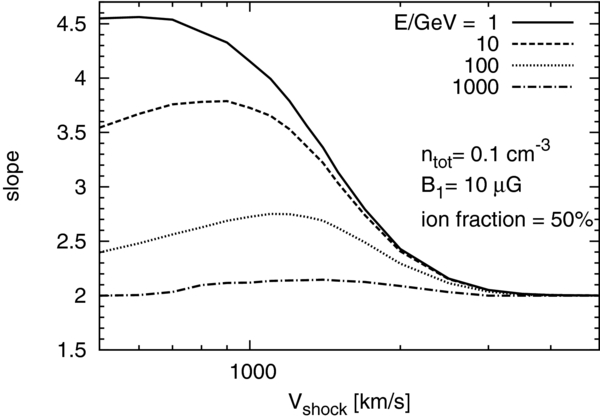}
\caption{Slope of the differential spectrum of test particles accelerated at a shock propagating in a partially ionized medium, with density $0.1~cm^{-3}$, magnetic field $10\mu G$ and ionized fraction of $50\%$, as a function of the shock velocity. The line shows the slope for particles at different energies, as indicated. The figure is taken from the paper by \cite{2012ApJ.755.121B}.}
\label{fig:slope}      
\end{figure}

The most striking consequence of the neutral return flux is however the steepening of the spectrum of test particles accelerated at the shock, first discussed by \cite{2012ApJ.755.121B}. The effect is caused by the reduction of the compression factor of the shock, which reflects on the fact that the slope of the spectrum of accelerated particles gets softer. This effect is however limited to particles that diffuse upstream of the shock out to a distance of order a few collision lengths of charge exchange/ionization upstream. It follows that the steepening of the spectrum is limited to particle energies low enough as to make their diffusion length shorter than the pathlength for charge exchange and ionization. In Fig. \ref{fig:slope} (from \cite{2012ApJ.755.121B}) I show the spectral slope as a function of shock velocity for particles with energy 1, 10, 100, 1000 GeV, as labelled (background gas density, magnetic field and ionization fraction are as indicated). One can see that the standard slope $\sim 2$ is recovered only for shock velocities $>3000$ km/s. For shocks with velocity $\sim 1000$ km/s the effect may make the spectra extremely steep, to the point that the energy content may be dominated by the injection energy, rather than, as it usually is, by the particle mass. This situation, for all practical purposes, corresponds to not having particle acceleration but rather a strong modification of the distribution of thermal particles. For milder neutral induced shock modifications, the effect is that of making the spectra of accelerated particles softer. It is possible that this effect may play a role in reconciling the predicted CR spectra with those inferred from gamma ray observations (see \cite[]{Caprioli:2011p2134} and \S \ref{sec:spectra} for a discussion of this problem), although the effect is expected to be prominent only for shocks slower than $\sim 3000$ km s$^{-1}$.

\subsection{NLDSA in partially ionized media}

The theory of NLDSA in the presence of partially ionized media was fully developed by \cite{2013ApJ.768.148M}, using the kinetic formalism introduced by \cite{2012ApJ.755.121B} to account for the fact that neutral atoms do not behave as a fluid, and their distribution in phase space can hardly be approximated as being a maxwellian. The theory describes the physics of particle acceleration, taking into account the shock modification induced by accelerated particles as well as neutrals, and magnetic field amplification. The theory is based on a mixed technique in which neutrals are treated through a Boltzmann equation while ions are treated as a fluid. The collision term in the Boltzmann equation is represented by the interaction rates of hydrogen atoms due to charge exchange with ions and ionization, at any given location. The Boltzmann equation for neutrals, the fluid equations for ions and the non-linear partial differential equation for accelerated particles are coupled together and solved by using an iterative method. The calculation returns the spectrum of accelerated particles at any location, all thermodynamical quantities of the background plasma (density, temperature, pressure) at any location, the magnetic field distribution, and the distribution function of neutral hydrogen in phase space at any location from far upstream to far downstream. 

\begin{figure}
\includegraphics[width=330pt]{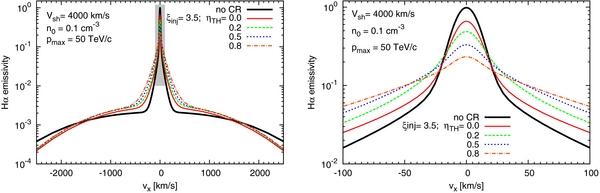}
\caption{{\it Left Panel:} Shape of the Balmer line emission for a shock moving with velocity $V_{sh}=4000$ km/s in a medium with density $0.1~cm^{-3}$, as calculated by \cite{2013ApJ.768.148M}. The thick (black) solid line shows the result in the absence of particle acceleration. The other lines show the broadening of the narrow component and the narrowing of the broad component when CR are accelerated with an injection parameter $\xi_{inj}=3.5$ and different levels of turbulent heating ($\eta_{TH}$) as indicated. {\it Right Panel:} Zoom in of the left panel on the region of the narrow Balmer line, in order to emphasize the broadening of the narrow component in the case of efficient particle acceleration.}
\label{fig:balmerline}      
\end{figure}

These quantities can then be used to infer the Balmer line emission from the shock region, taking into account the excitation probabilities to the different atomic levels in hydrogen. An instance of such calculation is shown in Fig. \ref{fig:balmerline}, where I show the shape of the Balmer line for a shock moving with velocity $V_{sh}=4000$ km/s in a medium with density $0.1~cm^{-3}$ with a maximum momentum of accelerated particles $p_{max}=50$ TeV/c. The left panel shows the whole structure of the line, including the narrow and broad components, while the right panel shows a zoom-in on the narrow Balmer line region (gray shadowed region in the left panel). The black line is the Balmer line emission in the absence of accelerated particles. Allowing for particle acceleration to occur leads to a narrower broad Balmer line (left panel) and to a broadening of the narrow component (right panel). The latter is rather sensitive however to the level of turbulent heating in the upstream plasma, namely the amount of energy that is damped by waves into thermal energy of the background plasma. In fact turbulent heating is also responsible for a more evident intermediate Balmer line (better visible in the left panel) with a width of few hundred km/s. It is worth recalling that observations of the Balmer line width are usually aimed at either the narrow or the broad component, but usually not both, because of the very different velocity resolution necessary for measuring the two lines. Therefore the intermediate line is usually absorbed in either the broad or the narrow component, depending on which component is being measured. This implies that an assessment of the observability of the intermediate Balmer component requires a proper convolution of the predictions with the velocity resolution of the instrument. 

At the time of this review, an anomalous shape of the broad Balmer line has been reliably measured in a couple of SNRs, namely SNR 0509-67.5 \cite[]{Helder:2010p618,Helder:2011p2386} and SNR RCW86 \cite[]{Helder:2009p660}. As I discuss below, the main problem in making a case for CR acceleration is the uncertainty in the knowledge of the shock velocity and the degree of electron-ion equilibration downstream of the shock. The ratio of the electron and proton temperatures downstream is indicated here as $\beta_{down}=T_{e}/T_{p}$. The other parameters of the problem have a lesser impact on the inferred value of the CR acceleration efficiency. 

The SNR 0509-67.5 is located in the Large Magellanic Cloud (LMC), therefore its distance is very well known, $50\pm 1$ kpc. \cite{Helder:2010p618,Helder:2011p2386} carried out a measurement of the broad component of the $H\alpha$ line emission in two different regions of the blast wave of SNR 0509-67.5, located in the southwest (SW) and northeast (NE) rim, obtaining a FWHM of $2680 \pm 70$ km/s and $3900 \pm 800$ km/s, respectively. The shock velocity was estimated to be $V_{sh} = 6000 \pm 300$ km/s when averaged over the entire remnant, and $6600\pm 400$ km/s in the NE part, while a value of 5000 km/s was used by \cite{Helder:2010p618,Helder:2011p2386} for the SW rim. The width of the broad Balmer line was claimed by the authors to be suggestive of efficient CR acceleration. In order to infer the CR acceleration efficiency the authors made use of the calculations by \cite{2008ApJ.689.1089V}, that, as discussed by \cite{2013arXiv1306.6454M}, adopt some assumptions on the distribution function of neutral hydrogen that may lead to a serious overestimate of the acceleration efficiency for fast shocks. Moreover, a closer look at the morphology of this SNR, reveals that the SW rim might be moving with a lower velocity than assumed by \cite{Helder:2010p618,Helder:2011p2386}, possibly as low as $\sim 4000$ km/s. Both these facts have the effect of implying a lower CR acceleration efficiency, as found by \cite{2013arXiv1306.6762M}. 

\begin{figure}
\includegraphics[width=330pt]{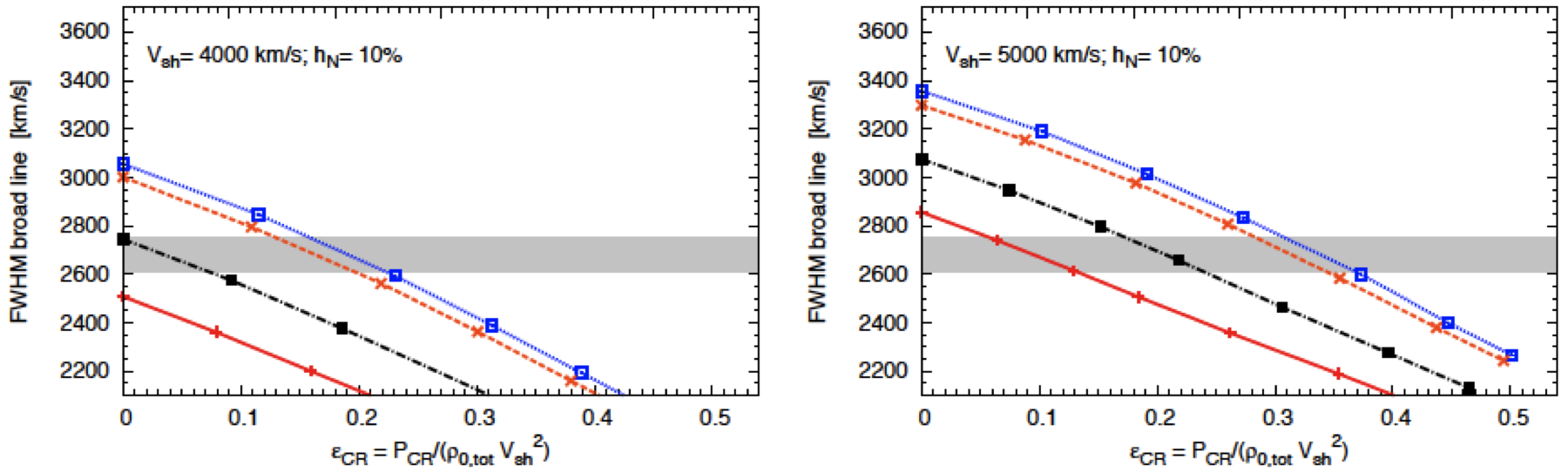}
\caption{FWHM of the broad Balmer line as a function of the CR acceleration efficiency for the SNR 0509-67.5, as calculated by \cite{2013arXiv1306.6762M}, assuming a shock velocity $V_{sh}=4000$ km/s (left panel) and $V_{sh}=5000$ km/s (right panel) and a neutral fraction $h_{N}=10\%$. The lines (from top to bottom) refer to different levels of electron-ion equilibration, $\beta_{down}=0.01,~0.1,~0.5,~1$, The shadowed region is the FWHM with $1\sigma$ error bar, as measured by \cite{Helder:2010p618}.}
\label{fig:SNR0509}      
\end{figure}

In Fig. \ref{fig:SNR0509} (from \cite[]{2013arXiv1306.6762M}) I show the FWHM of the broad Balmer line in the SW rim of SNR 0509-67.5 as a function of the acceleration efficiency, for shock velocity $V_{sh}=4000$ km/s (on the left) and $V_{sh}=5000$ km/s (on the right) and a neutral fraction $h_{N}=10\%$. The shaded area represents the FWHM as measured by \cite{Helder:2010p618,Helder:2011p2386}, with a $1\sigma$ error bar. The curves refer to $\beta_{down}=0.01,~0.1,~0.5,~1$ from top to bottom. For low shock speed and for full electron-ion equilibration ($\beta_{down}=1$) the measured FWHM is still compatible with no CR acceleration. On the other hand, for such fast shocks, it is found that $\beta_{down}\ll 1$ \cite[]{2007ApJ.654L.69G,2013SSRv.tmp.75G}, in which case one can see that acceleration efficiencies of $\sim 10-20\%$ can be inferred from the measured FWHM. 

The case of RCW86 is more complex: the results of a measurement of the FWHM of the broad Balmer line were reported by \cite{Helder:2009p660}, where the authors claimed a FWHM of $1100\pm 63$ km/s with a shock velocity of $6000\pm 2800$ km/s and deduced a very large acceleration efficiency ($\sim 80\%$). In a more recent paper by the same authors \cite[]{2013arXiv1306.3994H}, the results of \cite{Helder:2009p660} were basically retracted: several regions of the SNR RCW86 were studied in detail and lower values of the shock velocity were inferred. Only marginal evidence for particle acceleration was found in selected regions. The morphology of this remnant is very complex and it is not easy to define global properties. Different parts of the SNR shock need to be studied separately. In addition, the uncertainty in the distance to SNR RCW86 is such as to make the estimate of the acceleration efficiency even more difficult. 

Anomalous widths of narrow Balmer lines have been also observed in several SNRs (see, e.g. \cite{Sollerman:2003p615}). The width of such lines is in the 30-50 km/s range, implying a pre-shock temperature around 25,000-50,000 K. If this were the ISM equilibrium temperature there would be no atomic hydrogen, implying that the pre-shock hydrogen is heated by some form of shock precursor in a region that is sufficiently thin so as to make collisional ionization equilibrium before the shock unfeasible. The CR precursor is the most plausible candidate to explain such a broadening of the narrow line.

Most important would be to have measurements of the width of the narrow and broad components (and possibly intermediate component) of the Balmer line at the same location in order to allow for a proper estimate of the CR acceleration efficiency. Co-spatial observation of the thermal X-ray emission would also provide important constraints on the electron temperature. So far, this information is not yet available with the necessary accuracy in any of the astrophysical objects of relevance. 

Recent observations of the Balmer emission from the NW rim of SN1006 \cite[]{Nikolic:2013p3111} have revealed a rather complex structure of the collisionless shock. That part of the remnant acts as a bright Balmer source, but does not appear to be a site of effective particle acceleration, as one can deduce from the absence of non-thermal X-ray emission from that region. This reflects in a width of the broad Balmer line that appears to be compatible with the estimated shock velocity in the same region, with no need for the presence of accelerated particles. The observations of \cite{Nikolic:2013p3111} provide however a rather impressive demonstration of the huge potential of Balmer line observations, not only to infer the CR acceleration efficiency, but also as a tool to measure the properties of collisionless shocks. 

\section{Conclusions}
\label{sec:concl}

The problem of the origin of cosmic rays is a complex one: what we observe at the Earth results from the convolution of acceleration inside sources, escape from the sources and propagation in the Galaxy (or in the Universe, for extragalactic cosmic rays). Each one of these pieces consists of a complex and often non-linear combination of pieces of physics. This intricate chain of physical processes and the fact that wildly different spatial and temporal scales are involved represent the very reasons why we are still discussing of the problem of the origin of cosmic rays, one century after the discovery of their existence. 

Here I summarized the main aspects of the physics of acceleration of CRs in SNRs, emphasizing the progress made in the last decade or so, as well as the numerous loose ends deriving from the comparison between theoretical predictions and observational findings. 

At the time of writing this review, there is enough circumstantial evidence suggesting that SNRs accelerate the bulk of Galactic CRs, so as to introduce the concept of SNR paradigm. This evidence is mainly based on the following pieces of observation: 1) gamma ray measurements, both from the ground and from space, prove that SNRs accelerate particles up to at least $50-500$ TeV \cite[]{2013APh.43.71A,2013arXiv1307.6571B,2013arXiv1307.6572B,2012APh.39.61H}. In some of these cases (for instance in Tycho) one can make the case that the observed gamma ray emission is most likely due to the decay of neutral pions, thereby supporting the hypothesis that CR protons are being accelerated. 2) X-ray spectrum and morphology strongly suggest that magnetic field amplification is taking place at SNR shocks \cite[]{Volk:2005p968}, in virtually all young SNRs that we are aware of, with field strength of order few hundred $\mu G$ \cite[]{Vink:2012p2755}. This phenomenon is most easily explained if accelerated particles induce the amplification of the fields through the excitation of plasma instabilities. In this way, particles scatter on waves that are produced by the same particles that are being accelerated \cite[]{Schure:2012p3068}. 3) In selected SNRs there is evidence for anomalous width of the Balmer lines, that can be interpreted as the result of efficient CR acceleration at SNR shocks \cite[]{Heng:2010p1398}. 

Despite the confidence that SNRs may act as the main sources of the bulk of Galactic CRs, at present there is not yet any evidence of an individual SNR accelerating CRs up to the knee, although, as discussed by \cite{Caprioli:2011p2134,Caprioli:2012p2411}, this may not be surprising, because of the relatively short duration of the phase during which acceleration to the highest energies is expected to take place. More disturbing is the lack of a complete understanding of the physical mechanisms responsible for magnetic field amplification. I discussed here several ideas on how magnetic field amplification may occur and how this phenomenon feeds back on the distribution function of accelerated particles. While it appears that there are several ways of describing the large magnetic fields inferred from X-ray morphology, it seems harder to produce these fields on spatial scales relevant for particle scatterings at the highest energies. In other words, the issue of the highest energy achievable at SNR shocks remains open. Promising results in this direction are however recently arising from numerical investigations of the development of a filamentation instability \cite[]{2012MNRAS.419.2433R,2013ApJ.765L.20C}, that might represent a breakthrough in our understanding of the connection between particle escape from the accelerator and generation of turbulence at the necessary spatial scales. 

Magnetic field amplification and CR dynamical reaction on the accelerator represent the two main ingredients of the non-linear theory of particle acceleration at SNR shocks. The main predictions of the theory are that 1) the spectra of accelerated particles are no longer power laws, being concave in shape and possibly harder than predicted by the test-particle theory of DSA, and 2) that the temperature of the plasma behind the shock is expected to be lower at a SNR shock that is accelerating CRs effectively than it would be in the absence of particle acceleration. 

The spectra of accelerated particles predicted by NLDSA, as well as the test-particle spectra, are at odds with the current observations of gamma ray emission from SNRs and with the anisotropy observed at Earth. The physical reason for this discrepancy is that since the spectra of particles accelerated at SNRs are so hard, the required diffusion coefficient in the Galaxy is a rather steep function of energy, $D(E)\propto E^{0.7}$ at relativistic energies \cite[]{Berezhko:2007p1010}. Such a dependence is known to be incompatible with the measured anisotropy at energies $E\gtrsim 10$ TeV \cite[]{Ptuskin:2006p666,Blasi:2012p2024}. The hard spectra inside the sources also appear to be incompatible with the gamma ray spectra from a sample of SNRs \cite[]{Caprioli:2011p2134}. It is worth recalling that the spectra of particles escaping a SNR are not as concave as the spectra of particles accelerated at any given time at the shock \cite[]{Caprioli:2010p133}, but this effect is not sufficient to solve the anisotropy problem. Several authors \cite[]{Ptuskin:2010p1025,Caprioli:2010p133,Caprioli:2012p2411} suggested that appreciably steeper spectra may be obtained by assuming fast moving scattering centers in the upstream fluid, but this effect appears to be dependent on rather poorly known characteristics of the waves responsible for the scattering. 

A deeper look into the physics of particle acceleration in SNRs will be possible with the upcoming new generation of gamma ray telescopes, most notably the Cherenkov Telescope Array (CTA) \cite[]{Acharya20133}. The increased sensitivity of CTA is likely to lead to the discovery of a considerable number of other SNRs that are in the process of accelerating CRs in our Galaxy. The high angular resolution will allow us to measure the spectrum of gamma ray emission from different regions of the same SNR so as to achieve a better description of the dependence of the acceleration process upon the environment in which acceleration takes place. 

Interestingly, it has recently been realized that the presence of accelerated particles in the shock region of a SNR exploding in a partially ionized medium leads to considerable modification of the acceleration process \cite[]{2013ApJ.768.148M}, as well as to modification of the shape of the Balmer line emission from hydrogen atoms \cite[]{2012p3105,2013ApJ.768.148M}. Measurements of the Balmer emission from SNRs that show evidence of particle acceleration is a unique tool to measure the CR acceleration efficiency. The very high angular resolution of optical observations may, in principle make possible to achieve a detailed investigation of the CR acceleration process in SNRs. 

The general picture that arises from the SNR paradigm inspires some confidence that we may unfold the mechanism responsible for the acceleration of CR protons up to a few PeV, and of nuclei of charge $Z$ to an energy $Z$ times larger. For iron nuclei this implies that the maximum energy should be $\sim 10^{17}$ eV. This energy should also flag the end of the Galactic CR spectrum. The fact that this energy is much lower than the ankle, where traditionally the transition from Galactic to extragalactic CR has been placed, has stimulated a considerable interest in the development of models that may be able to describe at once the CR spectrum in the transition region and the chemical composition observed by different experiments in the relevant energy region (see \cite[]{2012APh.39.129A} for a review). At the time of writing of this review, it is unclear whether the low maximum energy inferred based on the SNR paradigm are compatible with the observed chemical composition and spectra. Recent data collected with the KASCADE-Grande experiment \cite[]{Apel:2013p3150} and ICETOP \cite[]{IceCubeCollaboration:2013p3139,Aartsen:2013p042004} suggest that some additional CR component is needed in the energy region between $10^{17}$ eV and $10^{19}$ eV. The required chemical composition by these data at $10^{18}$ eV is a roughly equal mix of light and heavy nuclei, which does not appear to be in obvious agreement with the chemical composition observed by the Pierre Auger Observatory \cite[]{2010PhRvL104i1101A}, HiRes \cite[]{2007JPhG34401S} and Telescope Array \cite[]{2013EPJWC5206002S}, which find a chemical composition at $10^{18}$ eV that is dominated by a light chemical component. The understanding of the transition region through increasingly more accurate measurements of chemical composition is a crucial step towards figuring out the origin of ultra high energy cosmic rays, which still represents a big unsolved problem. 

\begin{acknowledgements}
The author is grateful to his friends and colleagues in the Arcetri High Energy Astrophysics Group, R. Aloisio, E. Amato, R. Bandiera, N. Bucciantini, G. Morlino, O. Petruk for daily discussions on everything, as well as to D. Caprioli and P.D. Serpico for continuous collaboration and to Tom Gaisser for providing Figure 1. The author is also grateful to Tony Bell for a long discussion at the Aspen Center for Physics. This work was completed while at the Aspen Center for Physics, supported in part by the National Science Foundation under Grant No. PHYS-1066293, by the Simons Foundation and by PRIN INAF 2010. 

\end{acknowledgements}

\bibliographystyle{spbasic}      
\bibliography{crbib}   

%
%

\end{document}